\documentclass[11pt, letterpaper]{article}
\usepackage{setspace}
\usepackage{amsmath,mathrsfs,amsthm,dsfont}
\usepackage{caption}
\usepackage{verbatim}
\usepackage{subcaption}
\usepackage{graphicx,psfrag,epsf}
\usepackage{enumerate}
\usepackage{natbib}
\usepackage{bbm}
\usepackage{bm}
\usepackage{amssymb}
\usepackage[affil-it]{authblk}
\usepackage{verbatim}
\usepackage{extarrows}
\usepackage{framed}
\usepackage[colorlinks=true,citecolor=blue]{hyperref}
\usepackage{algorithm}
\usepackage{algorithmic}
\usepackage{url} 
\usepackage{float}
\usepackage{geometry}
\usepackage{lscape} 
\usepackage{multirow}
\usepackage{booktabs}
\usepackage{pifont}
 \usepackage{graphicx}
 \usepackage{pdfpages}
\usepackage{empheq}
\usepackage{thmtools}       
\usepackage[title]{appendix}
\usepackage[normalem]{ulem}
\usepackage{etoc}
\usepackage{enumitem}
\newcommand{\stkout}[1]{\ifmmode\text{\sout{\ensuremath{#1}}}\else\sout{#1}\fi}
\usepackage{amssymb}

\DeclareMathOperator*{\argmax}{arg\,max}
\DeclareMathOperator*{\argmin}{arg\,min}
\def\X{\boldsymbol{X}}
\def\x{\boldsymbol{x}}
\def\q{\boldsymbol{q}}
\def\d{\boldsymbol{d}}
\def\g{\boldsymbol{g}}

\newcommand{\bfsym}[1]{\ensuremath{\boldsymbol{#1}}}

\def\bgamma{\bfsym \gamma}

\def\bmu{\bfsym { m}}

\def\blambda {\bfsym {\lambda}}

\def\btau{\bfsym {\tau}}

\def\bgamma{\boldsymbol{\gamma}}

\newcommand\indep{\protect\mathpalette{\protect\independenT}{\perp}}
\def\independenT#1#2{\mathrel{\rlap{$#1#2$}\mkern2mu{#1#2}}}
\newcommand{\R}{\ensuremath{\mathbb{R}}}

\newcommand{\E}{\ensuremath{\mathbb{E}}}
\newcommand{\Prob}{\ensuremath{\mathbb{P}}}

\newcommand{\Var}{\text{Var}}

\theoremstyle{definition}
\newtheorem{assumption}{Assumption}

\newtheorem{theorem}{Theorem}
\newtheorem{lemma}{Lemma}

\newtheorem{definition}{Definition}
\newtheorem{corollary}{Corollary}
\newtheorem{proposition}{Proposition}

\title{Minimax Regret Estimation for Generalizing Heterogeneous Treatment Effects with Multisite Data\thanks{We thank Zijian Guo and Zhenyu Wang for engaging conversations and advice about distributionally robust optimization.}}

\author{Yi Zhang\thanks{Independent Researcher. Email: \href{mailto:yizhang0017@gmail.com}{yizhang0017@gmail.com}} \qquad Melody Huang\thanks{Assistant Professor, Department of Political Science and Department of Statistics \& Data Science, Yale University, New Haven, CT 06520, U.S.A. Email: \href{mailto:melody.huang@yale.edu}{melody.huang@yale.edu}} \qquad Kosuke Imai\thanks{Professor, Department of Government and Department of Statistics, Harvard University.  1737 Cambridge Street, Institute for Quantitative Social Science, Cambridge MA 02138, U.S.A.  Email: \href{mailto:imai@harvard.edu}{imai@harvard.edu} URL:
      \href{https://imai.fas.harvard.edu}{https://imai.fas.harvard.edu}} }

\begin{document}

\maketitle

\bigskip
\begin{abstract}
To test scientific theories and develop individualized treatment rules, researchers often wish to learn heterogeneous treatment effects that can be consistently found across diverse populations and contexts.  We consider the problem of generalizing heterogeneous treatment effects (HTE) based on data from multiple sites.  A key challenge is that a target population may differ from the source sites in unknown and unobservable ways. This means that the estimates from site-specific models lack external validity, and a simple pooled analysis risks bias.  We develop a robust CATE (conditional average treatment effect) estimation methodology with multisite data from heterogeneous populations. We propose a minimax-regret framework that learns a generalizable CATE model by minimizing the worst-case regret over a class of target populations whose CATE can be represented as convex combinations of site-specific CATEs. Using robust optimization, the proposed methodology accounts for distribution shifts in both individual covariates and treatment effect heterogeneity across sites. We show that the resulting CATE model has an interpretable closed-form solution, expressed as a weighted average of site-specific CATE models. Thus, researchers can utilize a flexible CATE estimation method within each site and aggregate site-specific estimates to produce the final model. Through simulations and a real-world application, we show that the proposed methodology improves the robustness and generalizability of existing approaches.


\end{abstract}

\clearpage
\section{Introduction}\label{sec:intro}
A central objective of causal inference, shared by both scientific research and industry applications, is to uncover generalizable cause-and-effect relationships. Generalizing heterogeneous treatment effects (HTEs) is crucial in this pursuit, as identifying consistent HTEs across diverse populations can provide deeper insight into the underlying causal mechanisms. Estimating generalizable HTEs can also lead to more efficacious individualized treatment rules across heterogeneous populations, with applications ranging from precision medicine in the health sciences to data-driven decision-making in the social sciences.

In this paper, we study the problem of generalizing HTEs based on data from multiple sites. 
Specifically, we consider the question of how to learn the conditional average treatment effect (CATE) for a target population when site-specific CATEs can be estimated from multiple source sites, but the target CATE may differ from those source CATEs in unknown and unobservable ways. Differences between the target and source CATEs can arise from observed and unobserved unit characteristics of respective populations, the way in which treatment affects the outcome, and the details of what constitutes treatment itself \citep{egami2023elements}. Such problems appear in many practical applications where researchers seek a treatment-effect model that generalizes beyond existing study populations.

For example, in the social sciences, experiments are often conducted in different settings, e.g., in laboratories \citep[e.g.,][]{falk2009lab}, online \citep[e.g.,][]{coppock2020small, kertzer2022experiments, peyton2022generalizability, bassan2025generalizability}, and in the field \citep[e.g.,][]{kalla2020reducing, dunning2019information, green2003getting, meager2019understanding, eskreis2019large}. One may be interested in generalizing these experimental results to develop better policies in a more general setting. Similarly, in medical studies, researchers use electronic health records data to extrapolate the results of small clinical trials to a larger population \citep[e.g.,][]{dahabreh2019generalizing}. Related challenges also arise in observational causal inference, where researchers seek to use existing study data to inform future policy decisions by inferring their potential impacts \citep[e.g.,][]{dugoff2014generalizing,dahabreh2021study}.

To address this challenge, we develop a robust CATE estimation methodology based on multisite data from heterogeneous populations.
Because target and source CATEs can differ in unknown ways, estimates from site-specific models lack external validity, whereas a naive pooled analysis also risks bias.
We propose a {\it minimax-regret} framework to learn a generalizable treatment effect model by minimizing the worst-case regret relative to the oracle predictor over a class of target populations (Section~\ref{sec:main:minimax-regret}).
In the main text, we develop this framework under squared loss, where the oracle predictor admits a direct CATE interpretation. More broadly, however, the same minimax-regret perspective applies to treatment-effect prediction under other loss functions. We discuss the details of this extension in Section~\ref{subsec:multi} and Appendix~\ref{appendix:extensionGLM}. 
Thus, the proposed approach guarantees robust predictive performance of individual treatment effects without specifying the exact relationship between the target and source site populations.
To our knowledge, this is the first work to consider the minimax regret criterion for robust multisite treatment-effect generalization, although it has been explored for traditional statistical learning \citep[e.g.,][]{agarwal2022minimax, mo2024minimax} and policy learning \citep[e.g.,][]{lei2023policy}.

Specifically, we consider a class of target populations whose CATE can be expressed as a convex combination of site-specific CATEs.
We also assume that the marginal distribution of covariates in the target population is identifiable though the distribution can arbitrarily differ from that of any source site population.
Under this broad class of target distributions, we derive a closed-form expression for the minimax-regret CATE model as a weighted average of site-specific CATE models. 
The aggregation weights are interpretable and can be computed directly from estimated site-specific CATE models without access to individual-level data in the source sites.
Thus, our methods are communication efficient and privacy-preserving, while allowing the use of various machine learning models that have recently been developed to flexibly estimate the CATE from a single site \citep[e.g.,][]{hahn2020bayesian,imai2013estimating,kunzel2019metalearners,nie2021covariate,kennedy2023towards,shalit2017estimating,wager2018estimation}.

Our theoretical analysis establishes the convergence rate of the proposed minimax regret CATE estimator, demonstrating that an accurate estimation of site-specific CATE leads to that of the robust target model.
This requirement underscores the importance of using flexible machine learning methods for source sites.
Finally, we compare the proposed minimax regret framework with two alternative objectives: minimax risk, which minimizes the maximum risk, and minimax relative-risk, which minimizes the risk relative to a baseline CATE model (Section~\ref{sec:otherobjective}). 
We show that, relative to the minimax-regret estimator, the minimax risk estimator tends to overfit to target distributions with higher noise levels, while the minimax relative-risk estimator is sensitive to the choice of the baseline CATE model.
Through simulation studies and a real-world empirical application, we demonstrate the robust performance of the proposed methodology across a wide range of target distributions (Sections~\ref{sec:simulation}~and~\ref{sec:application}).

\subsection*{Related Literature} 

Estimation of HTEs with multisite data has gained significant attention in recent years \citep[see][for a recent review]{brantner2023methods}.
The most commonly used approaches are based on parametric models such as mixed-effects models \citep[e.g.,][]{debray2015get,burke2017meta,seo2021comparing}.
Unfortunately, these methods are susceptible to bias caused by model misspecification.
In contrast, the proposed minimax regret approach is nonparametric and places much weaker restrictions on cross-site treatment effect heterogeneity.
Although some have recently developed nonparametric approaches based on machine learning \citep[e.g.,][]{tan2022tree,shyr2023multi}, their focus is the estimation of site-specific CATEs. Instead, our framework aims to generalize CATE to a class of target populations by using data from multiple source sites.

Our work contributes to the growing literature on external validity \citep[e.g.,][]{bareinboim2016causal,curth2021really,egami2023elements,miratrix2021applied,stuart2015assessing}.
A common assumption in this literature is the transportability of the treatment effect model to a target population \citep{kallus2018confounding, dahabreh2019extending, dahabreh2023efficient, yang2023elastic, cheng2021adaptive, hatt2022combining, huang2023leveraging}.
In practice, however, this assumption is often violated because of various observed and unobserved sources of cross-site heterogeneity.
Some have developed sensitivity analyses that address this violation of assumption \citep[e.g.,][]{huang2024overlap, huang2024sensitivity, nguyen2017sensitivity, nie2021covariate}, while others have proposed a sampling strategy to improve generalizability \citep[e.g.,][]{egami2024designing,tipton2023designing}.
These works, however, primarily focus on the average treatment effect (ATE), and none considers the robust estimation of CATE.

A key modeling ingredient of our framework is the convex-hull assumption on the target CATE. This assumption provides a transparent and interpretable way to relate an unobserved target to a collection of observed source sites while avoiding unsupported extrapolation beyond the heterogeneity already present in the data. A closely related idea appears in simplex-constrained weighting methods in causal inference, most prominently in synthetic control, where an unobserved target is represented by a convex combination of observed donor units \citep{abadie2010synthetic,abadie2021using,doudchenko2016balancing}. More broadly, related balancing and weighting approaches in observational studies similarly seek to represent a target population using weighted combinations of observed units while avoiding excessive extrapolation beyond the support of the data \citep{zubizarreta2015stable}. At the same time, our formulation differs from econometric evidence-aggregation approaches that restrict the target parameter through partial identification, distance bounds, or low-dimensional structure \citep{stoye2012minimax,manski2020toward,ishihara2021evidence,yata2021optimal,christensen2022optimal,olea2023decision,olea2024externally,tian2023learning}, as well as from causal approaches that model cross-site heterogeneity through latent site-specific confounding \citep{nie2021covariate,menzel2023transfer,gechter2024generalizing}.

Our proposed robust optimization framework is also related to the growing literature on distributional robustness in causal inference.
Much of this literature has focused on ATE estimation \citep[e.g.,][]{jin2022sensitivity,dorn2024doubly,yadlowsky2018bounds} and policy learning \citep[e.g.,][]{carranza2024robust,ishihara2021evidence,kallus2021minimax,kido2022distributionally,lei2023policy,manski2004statistical,mo2021learning,stoye2012minimax}.
While \citet{kern2024multi} studies CATE estimation, their framework is designed to be robust to unknown covariate shifts in the target distribution.
In contrast, our framework allows the CATE function itself to vary across sites in unknown ways, even after conditioning on covariates. Thus, our setting involves a richer form of cross-site heterogeneity than pure covariate shift and calls for a different methodological framework.

Beyond causal inference, the minimax distributionally robust optimization (DRO) framework has been extensively studied in fields such as optimization \citep{ben2013robust,duchi2021statistics}, machine learning \citep{duchi2021learning,sagawa2019distributionally}, and statistical learning theory \citep{awasthi2023open,zhang2023optimal}.
In the context of standard supervised learning, several minimax objectives have been proposed, including minimizing the worst-case risk \citep{hu2018does,sagawa2019distributionally}, the worst-case regret \citep{mo2024minimax,hastings2024taking,agarwal2022minimax}, and the worst-case explained variance or relative-risk \citep{meinshausen2015maximin,wang2023distributionally}.
For a detailed comparison of these minimax objectives, see \citet{mo2024minimax}.

While \cite{mo2024minimax}, \cite{wang2023distributionally}, and \cite{hastings2024taking} also apply DRO to settings with multiple heterogeneous populations, they focus on traditional prediction problems with directly observed outcomes. These settings can leverage standard iterative optimization techniques to solve the minimax objective efficiently.
In contrast, our work studies robust counterfactual prediction across multiple heterogeneous source sites in a causal inference setting, where the outcomes of interest are not directly observed. Under squared loss, this yields a robust CATE estimator in the main text, but the broader framing is one of minimax-regret treatment-effect prediction. This distinction underscores both the need for and the value of a closed-form characterization in our setting.
Lastly, while our proposed minimax regret framework is closely related to that of \cite{mo2024minimax}, their closed-form results are limited to linear regression models for prediction.
In contrast, we consider a more general model class for CATE estimation, enabling the use of nonparametric machine learning models for flexible estimation. 

\section{Minimax Regret Distributionally Robust Learning}
\label{sec:main:minimax-regret}

In this section, we develop a minimax regret distributionally robust learning framework for multisite treatment-effect prediction. We begin by describing the problem setup, in which we have the CATE estimates from multiple source sites but only know the distribution of covariates in a target population. We focus on the squared-loss formulation where the corresponding oracle predictor admits a direct CATE interpretation. We show that the resulting minimax regret problem has a closed-form solution under the proposed framework, which in turn enables separate and flexible estimation of site-specific CATE models followed by aggregation into a robust target model. In Appendix~\ref{appendix:extensionGLM}, we generalize our minimax regret framework to a broader class of loss functions.

\subsection{Setup}
\label{subsec:setup}

Suppose we have data from $S$ {\it source sites}, indexed by $s\in[S]:=\{1,\ldots,S\}$.
For each source site $s$, let $\X^{(s)}\in \mathcal{X}$ denote the pre-treatment covariates and $A^{(s)} \in \{0,1\}$ denote the binary treatment indicator.
We define $(Y^{(s)}(1),Y^{(s)}(0))$ as the potential outcomes under treatment $A^{(s)}=1$ and control $A^{(s)}=0$ conditions, respectively.
For each unit $i=1,\ldots,n_s$ in source site $s$, we assume $\left(Y_i^{(s)}(1), Y_i^{(s)}(0), \X_i^{(s)}\right)$ is drawn independently
from a {\it site-specific} distribution $P^{(s)} = \left(P^{(s)}_{(Y(1),Y(0))\mid \X}, P^{(s)}_{\X}\right)$, where $P^{(s)}_{(Y(1),Y(0))\mid \X=\x}$ is the conditional distribution of potential outcomes for the covariate value $\x\in\mathcal{X}$, and $P^{(s)}_{\X}$ represents the marginal covariate distribution.
Under standard SUTVA assumptions \citep{rubin1990comment}, the observed outcome variable is defined based on the realized treatment as  $Y_i^{(s)}= A_i^{(s)}\cdot Y_i^{(s)}(1) + (1 - A_i^{(s)})\cdot Y_i^{(s)}(0)$. Thus, for each source site $s$, we obtain $n_s$ independent and identically distributed (i.i.d.) observations, denoted as \( \left\{Y^{(s)}_i, \X^{(s)}_i, A_i^{(s)}\right\}_{i=1}^{n_s} \).

For each source site $s$, we consider the {\it site-specific} conditional average treatment effect (CATE) function with respect to the distribution $P^{(s)}$, which is defined as,
\begin{equation}
\tau^{(s)}(\x) \ := \ \E_{P^{(s)}} \big[Y(1) - Y(0) \mid \X=\x \big]
\end{equation}
for any given $\x\in\mathcal{X}$. 
Throughout the paper, we omit the superscript $^{(s)}$ for random variables when their respective distributions are clear from context.
Given potential site-level differences, we allow $P^{(s)}$ to vary across sites, implying that distributional shifts may occur not only in covariates \( \X \) but also in the heterogeneous treatment effect, i.e., \( Y(1) - Y(0)\) given \(\X \). This means that we allow each source site to have a distinct, non-identical CATE function \( \tau^{(s)}(\x) \).

Our goal is to estimate a generalizable CATE model $$\tau_Q(\x):=\E_Q[Y(1)-Y(0)\mid \X=\x]$$ 
with performance guarantees for a target population $Q=(Q_{(Y(1), Y(0))\mid\X}, Q_{\X})$. 
Under the squared-loss formulation, $\tau_Q(\x)$ is the oracle best predictor of the latent individual treatment effect $Y(1)-Y(0)$ given covariates $\X=\x$.
We consider the setting in which the marginal covariate distribution $Q_{\X}$ is identifiable for the target population, but its conditional distribution of potential outcomes given that the covariates $Q_{(Y(1), Y(0))\mid\X}$ may differ from those of source site populations in unknown ways.  Specifically, we observe an i.i.d. sample of size $n_Q$ from the target distribution where only pre-treatment covariates are observed for each unit, denoted by $\{\X^{Q}_i\}_{i=1}^{n_Q}$. This is a common scenario in practice, as researchers typically have access to pre-treatment covariates, even when treatment has not yet been administered.  

Our method is applicable when researchers have access to a collection of source sites and a target population of interest, so long as the identifying assumptions for CATE (i.e., uncounfoundedness and positivity) hold within each site. This means that researchers can flexibly apply our proposed framework to multisite randomized control trials, multiple observational studies, or a combination of both. Throughout, we assume that researchers have access to the distribution of a common set of baseline covariates $\X$ in each source population and the target population. See Section \ref{sec:concl} for an additional discussion.

\subsection{Multisite Distributionally Robust Estimation Framework}
\label{subsec:multi}

We develop a methodology that leverages the availability of multiple source sites and achieves good overall prediction accuracy of heterogeneous treatment effects across a wide range of target populations.
Here, we focus on the squared-loss setting, in which the corresponding oracle predictor is the target CATE. Thus, we seek a robust CATE model. Because the covariate distribution across the target population is assumed to be known, the main uncertainty concerns the conditional potential outcome distribution given $\X$ (i.e., $Y(1) - Y(0) \mid \X$). With some abuse of terminology, we refer to this as the `target distribution' of interest.

Our method therefore considers an {\it uncertainty set} of plausible target distributions that are related to the source-site distributions. Concretely, we impose a mild restriction on the unknown target CATE function by assuming that $\tau_Q(\cdot)$ can be expressed as a convex combination of the source-site CATE functions, $\{\tau^{(s)}(\cdot)\}_{s=1}^S$.
The formal definition of this multisite uncertainty set is given here.
\begin{definition}[Multisite Uncertainty Set] \label{df:uncertainty_set}
    Define the uncertainty set $\mathcal{C}\left(Q_{\X}\right)$ as:
\begin{equation}\label{equ:Cqx}
\mathcal{C}\left(Q_{\X}\right):=\left\{Q=\left(Q_{\X}, Q_{(Y(1),Y(0))\mid \X}\right) \ \bigg | \ \tau_Q(\cdot)=\sum_{s=1}^S q_s \cdot \tau^{(s)}(\cdot) \quad \text { with } \quad \q \in \Delta_{S-1}\right\}      
\end{equation}
where $\Delta_{S-1}=\left\{\q \in \mathbb{R}^S \mid \sum_{s=1}^S q_s=1, \min_s q_s \geq 0\right\}$ denotes the $(S-1)$-dimensional simplex. 
\end{definition}

The uncertainty set restricts the class of joint distributions over $(Y(1),Y(0),\X)$ to which the target population belongs.  
In particular, we formulate this constraint only on the CATE model $\tau_Q(\cdot)$ rather than directly restricting the entire conditional distribution of potential outcomes $Q_{(Y(1), Y(0))\mid \X}$.
Thus, our multisite uncertainty set spans the entire convex hull of the source CATE functions, ensuring that it is large enough to accommodate a wide range of discrepancies between the target and each source distribution, while avoiding being overly conservative.

This formulation differs from much of the existing literature, which often assumes that the CATE is identical across studies and cross-study differences arise primarily through covariate shift. In contrast, our multisite uncertainty set allows the conditional potential outcome model to vary across source sites even after conditioning on $\X$, thereby accommodating heterogeneity due to both conditional and covariate shifts.

We next develop a distributionally robust optimization (DRO) framework to learn a generalizable CATE model that is robust across all plausible target distributions in this multisite uncertainty set $\mathcal{C}\left(Q_{\X}\right)$.
We begin by observing that a CATE function, $\E[Y(1)-Y(0) \mid \X=\x]$, can be viewed as a {\it counterfactual} prediction model, where the outcome is the individual treatment effect $Y(1)-Y(0)$ and the predictors are the pre-treatment covariates $\X$. 
Our goal then is to derive a CATE model that achieves good prediction accuracy over the range of target populations given in Definition~\ref{df:uncertainty_set}.

The standard approach in the DRO literature is to optimize the {\it worst-case} prediction performance over the uncertainty set.
Formally, for a given model class $\mathcal{F}$ and a general loss function $\ell: \mathbb{R}\times \mathbb{R}\rightarrow \mathbb{R}$,
this minimax approach defines the robust prediction model as the solution to the following optimization problem:
\begin{equation}\label{equ:minimax:generalloss}
    f^\ast_{\text{risk}}(\cdot) \ := \ \underset{f \in \mathcal{F}}{\arg \min } \underset{Q\in  \mathcal{C}\left(Q_{\X}\right)}{\max} 
\E_{Q}[ \ell\left(Y(1)-Y(0),f(\X)\right) ],
\end{equation}
where $f^\ast_{\text{risk}}(\cdot)$ minimizes the {\it worst-case prediction risk} over the uncertainty set $\mathcal{C}\left(Q_{\X}\right)$.
While the minimax approach based on the raw risk objective is conceptually straightforward and has been widely studied in the DRO literature \citep[see, e.g.,][]{ben2013robust, duchi2021learning, zhang2023optimal}, its solutions are often overly pessimistic because the objective is highly sensitive to varying noise levels across target distributions in the uncertainty set \citep[e.g.,][]{agarwal2022minimax, hastings2024taking}; see also Section~\ref{subsec:minimaxrisk}. 

Thus, we propose an alternative {\it minimax regret} objective \citep{savage1951, manski2011choosing}. For a general loss function $\ell$, the regret of a predictor $f(\cdot)$ under a target distribution $Q$ is the excess risk relative to the oracle predictor in the model class $\mathcal{F}$. Below, we focus on the standard squared-loss setting with continuous outcomes, under which the regret of $f(\cdot)$ can be written as
\begin{equation}\label{equ:regretrisk}
\E_Q[(Y(1)-Y(0)-f(\X))^2] - \underset{f' \in \mathcal{F}}{\min} \; \E_Q[(Y(1)-Y(0)-f'(\X))^2].
\end{equation}
In Appendix~\ref{appendix:extensionGLM}, we extend this framework to a broader class of loss functions.

The proposed minimax regret CATE model minimizes the worst-case regret across the range of target distributions in \(\mathcal{C}\left(Q_{\X}\right)\): 
\begin{equation}\label{equ:HTE:obj:regret}
   f_{\text{regret}}^*(\cdot) =\underset{f \in \mathcal{F}}{\arg \min } \max _{Q \in \mathcal{C}\left(Q_{\X}\right)} \left\{\mathbb{E}_Q[(Y(1)-Y(0)-f(\X))^2]- \underset{f' \in \mathcal{F}}\min\ \E_Q[(Y(1)-Y(0)-f'(\X) )^2]\right\}.
\end{equation}
This minimax regret CATE model minimizes the worst-case performance gap relative to the best possible model that achieves the best prediction of individual treatment effect for every choice of the target population $Q$.
This accounts for differences in inherent learning difficulties across distributions, preventing a small number of challenging target populations from dominating the final solution and yielding a model that is more robust to heterogeneous noise levels across distributions.

\subsection{Closed-form Identification Result}\label{subsec:idenresults}

Solving the DRO problem in Equation~\eqref{equ:HTE:obj:regret} is challenging for two reasons.
First, the individual treatment effect $Y(1)-Y(0)$ is a counterfactual quantity and is not directly observable.
Second, the target distribution $Q$ is unknown and may differ from the observed source-site distributions, $\{P^{(s)}\}_{s\in[S]}$.
Under squared loss, however, the prediction risk for the latent individual treatment effect admits the decomposition
\[
\E_Q[(Y(1)-Y(0)-f(\X))^2]
=
\E_{Q_{\X}}[(\tau_Q(\X)-f(\X))^2]
+
\E_Q[\Var_Q(Y(1)-Y(0)\mid \X)].
\]
The second term is an irreducible conditional variance term that does not depend on the candidate predictor $f(\cdot)$. As a result, it cancels in the regret criterion, so that under the squared-loss setting studied in the main text, the minimax regret problem can be reduced to a worst-case CATE prediction problem.

Consequently, we can reformulate the original optimization problem in Equation~\eqref{equ:HTE:obj:regret} as:
\begin{equation}\label{equ:regret:distance}
f_{\text{regret}}^*(\cdot) = \underset{f \in \mathcal{F}}{\arg \min } \max _{Q \in \mathcal{C}\left(Q_{\X}\right)} \mathbb{E}_{Q_{\X}}[(f(\X)-\tau_Q({\X}))^2].
\end{equation}
This reformulation reveals that, under squared loss, the proposed robust predictor directly minimizes the worst-case mean squared error (MSE) of estimating the true target CATE model $\tau_Q$ with respect to the fixed covariate distribution $Q_{\X}$. While this type of MSE objective is widely used as an evaluation criterion in the CATE estimation literature \citep{kunzel2019metalearners,curth2021nonparametric}, here it arises as a special case of a minimax-regret counterfactual-prediction problem. Appendix~\ref{appendix:extensionGLM} shows that a related minimax-regret geometry continues to hold beyond squared loss, although the same variance-cancellation argument no longer applies.

To derive a closed-form solution, we impose an oracle identification condition that the model class $\mathcal{F}$ is large enough to contain the convex hull of the source-site CATE models, $\{\tau^{(s)}(\cdot)\}_{s\in[S]}$. This assumption is used to characterize the population-level minimax regret solution and should be interpreted separately from the practical estimation procedure in Section~\ref{subsec:estimation}, where the site-specific CATEs are estimated flexibly from data and then aggregated.

We first show that the problem in Equation~\eqref{equ:regret:distance}, which optimizes over an infinite set of target distributions in the convex hull, is equivalent to a constrained optimization problem with a finite number of constraints corresponding to the vertices of the convex hull. This intermediate result provides useful intuition for our main identification theorem.
\begin{proposition}[Alternative representation of the minimax regret problem]\label{lemma:euqival:regretOPT}
The minimax problem defined in Equation~\eqref{equ:regret:distance} can be equivalently expressed as:
\begin{equation}
\label{equ:equivalConstrainedOPT}
\{f_{\text{regret}}^*(\cdot), R^*\} \ = \ \underset{f \in \mathcal{F}, R \in \mathbb{R}}{\arg \min}\ R \quad \quad \text{subject to} \quad \mathbb{E}_{Q_{\X}}[(f(\X) - \tau^{(s)}(\X))^2] \leq R \quad \text{for all } s \in [S],
\end{equation}
where $R^*$ denotes the smallest worst-case regret achieved by $f_{\text{regret}}^*(\cdot)$. Furthermore, the solution $\{f_{\text{regret}}^*(\cdot), R^*\}$ exists and satisfies the following conditions:
\begin{equation} 
    \begin{aligned}
      f_{\text{regret}}^*(\cdot) =  \sum_{s=1}^{S} q^*_s \cdot \tau^{(s)}(\cdot) \quad \text{subject to} \quad & q^*_s \cdot \left(\mathbb{E}_{Q_{\X}}[( f_{\text{regret}}^*(\X) - \tau^{(s)}(\X))^2] - R^*\right) = 0 \quad \text{for } s \in [S], \\
      &  \mathbb{E}_{Q_{\X}}[( f_{\text{regret}}^*(\X) - \tau^{(s)}(\X))^2]  \leq R^* \quad \text{for } s \in [S],\\
      &   \sum_{s=1}^{S} q^*_s = 1, \quad  q^*_s \geq 0 \quad \text{for }  s \in [S]. 
    \end{aligned} \label{prop:second}
\end{equation}
\end{proposition}
The proof is provided in Appendix~\ref{app:lemma:equival}. 
Proposition~\ref{lemma:euqival:regretOPT} implies that the minimax CATE $f_{\text{regret}}^*(\cdot) $ will be a convex combination of the CATEs from the source sites.
Specifically, a source site $s$ does not contribute to the final solution (i.e., $q^*_s=0$) if the squared $L_2$-distance between its CATE and $f_{\text{regret}}^*(\cdot)$ is strictly less than $R^*$.
In contrast, a source site with a nonzero weight ($q^*_s>0$) has a CATE exactly at the squared $L_2$-distance of $R^*$ from \( f_{\text{regret}}^*(\cdot) \).
Thus, these source sites define the minimax regret boundary.

To further motivate the minimax regret solution and its relationship with the CATE functions of the source sites, we present a simple example. 
Assume that site-specific CATE models are piecewise linear with an interaction term, \( \tau^{(s)}(\X) = \beta_s \cdot X_1 \mathds{1}(X_1 > 0) + X_1 X_2 \), where $\beta_s \in \mathbb{R}$ controls cross-site heterogeneity. Then, in this setting, the resulting minimax regret CATE model is given by $f_{\text{regret}}^*(\cdot)= \beta^*_{\text{regret}} \cdot X_1 \mathds{1}(X_1>0)+X_1X_2$, which retains the same functional form as the source CATEs. Specifically, the minimax coefficient $\beta^*_{\text{regret}}$ is derived from the optimization problem as $\beta^*_{\text{regret}}=\underset{\beta \in \mathbb{R}}{\arg \min } \ \max_{s\in[S]}\  (\beta-\beta_s)^2 =\frac{1}{2}(\max_{s\in[S]} \beta_s+\min_{s\in[S]} \beta_s)$.
In other words, the minimax regret solution corresponds to the midpoint between the CATE models associated with the two boundary groups that have the highest and lowest values of $\beta_s$.
This midpoint selection minimizes the worst-case MSE across source sites by balancing the maximum deviation from both extremes in the convex hull of site-specific CATEs.

Using Proposition~\ref{lemma:euqival:regretOPT}, we now establish our main identification result for the oracle minimax regret CATE, showing that it can be represented as a weighted combination of the site-specific CATE models. 
\begin{theorem}[Minimax regret identification]\label{thm:ident:regret}
      Define $L_2(Q_{\X})$ as the space of square-integrable measurable functions with respect to $Q_{\X}$.
      Suppose that the model class $\mathcal{F}\subset L_2(Q_{\X})$ is convex and contains all the site-specific CATEs, i.e.,  $\tau^{(s)}(\cdot) \in \mathcal{F}$ for all $s \in[S]$. 
      Then, the oracle minimax regret CATE $f^*_{\text{regret}}(\cdot)$ defined in Equation~\eqref{equ:HTE:obj:regret} can be identified as:
\begin{equation}\label{equ:iden:regret}
    f_{\text{regret}}^*(\cdot) =\sum_{s=1}^S q_s^* \cdot \tau^{(s)}(\cdot) \quad \text { with } \quad \q^*=\underset{\q \in \Delta_{S-1}}{\arg \min } \quad \q^\top \Gamma \q - \q^\top \d,
\end{equation}
where the $(k,l)$-th element of $S\times S$ matrix $\Gamma$ is given by $\Gamma_{k, l}=\mathbb{E}_{Q_{\X}}\left[\tau^{(k)}(\X) \tau^{(l)}(\X)\right]$ for $k, l \in [S]$, and $\d:=\left(\Gamma_{1,1},\ldots,\Gamma_{S,S
} \right)^\top$ is the diagonal vector of  $\Gamma$.
\end{theorem}
The proof is given in Appendix~\ref{app:pf:ident:regret}. The aggregation weights \( \q^* = \{q_s^*\}_{s=1}^S \) determine the contribution of each source site to the final CATE model.
These weights are found by solving the quadratic optimization problem given in Equation~\eqref{equ:iden:regret}, where \( \Gamma \) is, by construction, a symmetric and positive semidefinite matrix.

The objective function for determining the aggregation weights comprises two main components.
The first is a quadratic term, \( \q^\top \Gamma \q = \mathbb{E}_{Q_{\X}}\left[\left(\sum_{s=1}^S q_s \cdot \tau^{(s)}(\X)\right)^2\right] \), representing the $L_2$-distance of the aggregated model from the origin.
The second is a linear term, \( \q^\top \d = \mathbb{E}_{Q_{\X}}\left[\sum_{s=1}^S q_s \cdot \tau^{(s)}(\X)^2\right] \), denoting the weighted average of the second moments of each site-specific CATE, both measured with respect to the target covariate distribution \( Q_{\X} \).
This result implies that if the source-site CATE functions are known, we can directly compute the aggregation weights and construct \( f_{\text{regret}}^*(\cdot) \) using Equation~\eqref{equ:iden:regret}.
This computational efficiency is a major advantage of the proposed methodology over many other existing minimax methods.
Theorem~\ref{thm:ident:regret} concerns the common squared-loss setting, where the oracle predictor coincides with the CATE. Appendix~\ref{appendix:extensionGLM} extends this result to a broader class of GLM-motivated loss functions, showing that a related minimax-regret geometry and closed-form characterization continue to hold more generally.

In particular, the resulting solution is agnostic to the specification of the model class, enabling a wide range of flexible single-site CATE estimation methods proposed in the literature. To our knowledge, we are the first to derive a closed-form solution to the minimax regret optimization problem under a flexible model class $\mathcal{F}$.
While \cite{mo2024minimax} also obtains a closed-form expression in a similar multisite setup, the result is limited to linear regression in the standard supervised learning setting.
In contrast, our result is based on a considerably larger model class, enabling the use of many popular nonparametric estimators and machine learning models to approximate complex CATE functions.
Our framework does not require site-specific CATEs to share the same functional form, assuming only that the model class $\mathcal{F}$ encompasses all site-specific CATEs $\{\tau^{(s)}(\cdot)\}_{s=1}^S$.
In practice, one can estimate these site-specific CATEs separately using different methods before aggregating the resulting CATE models (see Section~\ref{subsec:estimation} for more details).

\subsection{Identification with Additional Restrictions}

If site-specific CATE models are highly heterogeneous, optimizing worst-case performance over the entire uncertainty set may yield an overly conservative solution.
In such cases, additional restrictions on the target population, assuming that they can be justified, lead to more informative inference.
We generalize the proposed methodology so that researchers can incorporate auxiliary information on the target distribution by further restricting the uncertainty set.

Specifically, we assume that the combination weights lie within a convex subset of the simplex, $\mathcal{H} \subseteq \Delta_{S-1}$:
\begin{equation}\label{equ:Cqx:H} 
\mathcal{C}\left(Q_{\X},\mathcal{H}\right)  :=  \left\{Q=\left(Q_{\X}, Q_{(Y(1),Y(0))\mid \X}\right) \ \bigg | \ \tau_Q(\cdot)=\sum_{s=1}^S q_s \cdot \tau^{(s)}(\cdot) \quad \text { with } \quad \q \in  \mathcal{H}\right\},
\end{equation}
where the constraint set $\mathcal{H}$ is chosen by researchers.
Here, $\mathcal{H}$ can incorporate auxiliary information about the target population by restricting the feasible weight vector to a user-chosen convex subset of the simplex. Such restrictions may encode prior knowledge, design considerations, or measurement considerations that bound how far the target CATE (or, more generally, the target conditional potential outcome model) can deviate from the source-site CATEs.
For example, one may impose the maximal value of site-specific weights to limit the influence of any single source site on the final CATE model, i.e.,  $q_s \leq c$ for all $s \in [S]$ where $c \in (1/S,1)$. 

Given the generalized uncertainty set $\mathcal{C}(Q_{\X}, \mathcal{H})$, we can define the minimax regret CATE model over $\mathcal{C}(Q_{\X}, \mathcal{H})$ as in Equation~\eqref{equ:regret:distance}, denoting it by $f_{\operatorname{regret},\mathcal{H}}^*(\cdot)$.
We extend Theorem~\ref{thm:ident:regret} by deriving a closed-form solution under the assumption that $\mathcal{H}$ is a convex polytope within the simplex $\Delta_{S-1}$. 
To derive this result, we leverage the fact that any polytope can be represented as the convex hull of a finite set of points.  The proof is given in Appendix~\ref{app:proof_coro}.

\begin{corollary}[Extension to polytopes $\mathcal{H}\subset\Delta_{S-1}$]\label{corollary:ident:regret:polytopes}
Let $\mathcal{H}$ be a convex polytope within the $(S-1)$ dimensional simplex $\Delta_{S-1}$, represented as the convex hull of a finite set of vertices $\{\g_1, \ldots, \g_N\}$, where each vertex $\g_i$ is a distinct element of the simplex, i.e., $\g_i \in \Delta_{S-1} \subset \mathbb{R}^S$ for $i \in [N]$.
Define the transformation matrix \( G = (\g_1, \ldots, \g_N) \in \mathbb{R}^{S \times N}\).  Under the same conditions as in Theorem~\ref{thm:ident:regret}, $f_{\operatorname{regret},\mathcal{H}}^*(\cdot)$ can be identified as:
\begin{equation}\label{equ:iden:regret:polytope}
    f_{\operatorname{regret},\mathcal{H}}^*(\cdot)= (\q^*)^\top \btau_{\operatorname{poly}}(\cdot)  \quad \text { with } \quad \q^*=\underset{\q \in \Delta_{N-1}}{\arg \min } \left(\q^\top \Gamma_{\operatorname{poly}} \q - \q^\top \d_{\operatorname{poly}}\right),
\end{equation}
where \( \btau_{\operatorname{poly}}(\cdot)= G^\top \btau(\cdot) \), with \( \btau(\cdot) := (\tau^{(1)}(\cdot), \ldots, \tau^{(S)}(\cdot))^\top \) representing the vector of site-specific CATE models, \( \Gamma_{\operatorname{poly}} = G^\top \Gamma G \), and 
\[
\d_{\operatorname{poly}} := \left( \mathbb{E}_{Q_{\X}} \left[ \left( \g_1^\top \btau(\X) \right)^2 \right], \ldots, \mathbb{E}_{Q_{\X}} \left[ \left( \g_N^\top \btau(\X) \right)^2 \right] \right)^\top
\]
representing the diagonal vector of \( \Gamma_{\operatorname{poly}} \).
\end{corollary}

For notational simplicity, we focus on using $\mathcal{C}(Q_{\X})$ in the remainder of the paper, unless otherwise stated.
The proposed estimation method and theoretical results extend naturally to a general uncertainty set \( \mathcal{C}(Q_{\X}, \mathcal{H}) \) introduced here.

\subsection{Flexible Two-step Estimation Procedure}\label{subsec:estimation}

Given the population-level identification result in Theorem~\ref{thm:ident:regret}, we now describe a practical plug-in estimator of the minimax regret CATE model based on the observed data. The oracle characterization in Section~\ref{subsec:idenresults} concerns the target population solution induced by the true source-site CATEs, whereas the estimation procedure replaces those unknown CATEs by flexible site-specific estimators learned from data. Our estimator proceeds in two steps: (1) estimate each site-specific CATE model, \( \tau^{(s)}(\cdot) \) for \( s=1, \ldots, S \) and (2) solve for the optimal aggregation weights \( \q^* \), yielding the final robust CATE model as an ensemble of the estimated site-specific CATE models.

A key advantage of our robust CATE estimator is that it allows the use of any available machine learning algorithm to estimate site-specific CATEs \citep[see, e.g.,][]{hahn2020bayesian,imai2013estimating,kennedy2023towards,kunzel2019metalearners,nie2021quasi,wager2018estimation}.
Each source-site CATE can be estimated independently, potentially using different methods across sites.
In addition, unlike other distributed learning methods, our two-step estimation procedure avoids the need to share individual-level data across sites \citep{sagawa2019distributionally,hu2018does} or employ iterative optimization \citep{deng2020distributionally,mo2024minimax}.
Therefore, our approach is both communication efficient and privacy-preserving. 

Formally, to estimate \( \tau^{(s)}(\cdot) \) using observed data, we impose the standard assumptions of unconfoundedness and positivity in the CATE estimation literature.
\begin{assumption}[Identifiability within a source site]\label{ass:unconfound:overlap}
For each site $s\in[S]$, we assume that 
\begin{enumerate}[label=(\alph*)]
\item \textit{Unconfoundedness}: $Y^{(s)}_i\left(a\right) \indep A^{(s)}_i \mid \X^{(s)}_i=\x$ 
\item \textit{Positivity}: $\exists$ $\epsilon>0$ such that
 $\epsilon<\Prob(A^{(s)}_i = a\mid \X^{(s)}_i=\x)<1-\epsilon$, for $a\in\{0,1\}$
\end{enumerate}
\end{assumption}
\noindent Assumption~\ref{ass:unconfound:overlap} ensures that within each site, the CATE model \( \tau^{(s)}(\cdot) \) is identifiable and can be consistently estimated using its locally observed data alone.

To ensure that the final robust CATE model performs well on the target covariate distribution $Q_{\X}$, we assume that $Q_{\X}$ lies within the support of each site-specific covariate distribution, $P_{\X}^{(s)}$ for $s\in [S]$.
The assumption allows us to aggregate site-specific CATE models, each of which is estimated using local observed data, while avoiding extrapolation.
\begin{assumption}[Covariate overlap]\label{ass:Xsupport}
For each source site $s\in[S]$, we assume that the target covariate distribution $Q_{\X}$ is absolutely continuous with respect to $P_{\X}^{(s)}$, i.e., $\sup_{\x\in\mathcal{X}} d Q_{\X}(\x)/d P_{\X}^{(s)}(\x) <\infty$.
\end{assumption}
Importantly, Assumption \ref{ass:Xsupport} implies \emph{target-to-site} overlap: for each source site $s$, we assume that the target covariate support is contained in the support of each site and the site-specific CATE $\tau^{(s)}(\x)$ is well-defined on $\mathrm{supp}(Q_{\X})$. A separate \emph{site-to-site} overlap assumption is not needed for our identification and characterization results (e.g., Theorem~1), since the key quantities are evaluated under the target distribution $Q_{\X}$ rather than requiring direct comparability between any two source sites.

Assumption~\ref{ass:Xsupport} also helps connect source-domain estimation guarantees to the target-domain error criterion used later in Assumption~\ref{ass:CATE:rate}. In particular, the bounded density-ratio condition implies a standard change-of-measure bound, so that for any square-integrable error function $g$,
\[
\E_{Q_{\X}}[g(\X)^2]
=
\E_{P_{\X}^{(s)}}\!\left[\frac{dQ_{\X}}{dP_{\X}^{(s)}}(\X) g(\X)^2\right]
\le
\left\|\frac{dQ_{\X}}{dP_{\X}^{(s)}}\right\|_{\infty}\E_{P_{\X}^{(s)}}[g(\X)^2].
\]
Thus, source-site $L_2$ guarantees can be transferred to the target covariate domain provided the overlap constant is well-behaved. In practice, the quality of this transfer depends on the degree of source-to-target overlap. One may further improve target-domain performance using target-aware estimation strategies such as importance weighting, overlap diagnostics, or trimming.

Under Assumptions~\ref{ass:unconfound:overlap}~and~\ref{ass:Xsupport}, we can construct the plug-in estimator of $f_{\operatorname{regret}}^*(\cdot)$ based on Theorem~\ref{thm:ident:regret}:
\begin{equation}\label{equ:estor:regret}
   \hat{f}_{\operatorname{regret}}(\cdot) = \sum_{s=1}^S \hat{q}_s \cdot \hat{\tau}^{(s)}(\cdot) \quad \text{with} \quad \hat{\q} = \underset{\q \in \Delta_{S-1}}{\arg \min} \ \q^{\top} \widehat{\Gamma} \q - \q^{\top} \hat{\d},
\end{equation}
where each entry of $\widehat{\Gamma}$ and $\hat{\d}$ is estimated using empirical observations from the target population, $\{\X^{Q}_i\}_{i=1}^{n_Q}$, given by
\[
\begin{aligned}
    & \widehat{\Gamma}_{k, l} = \frac{1}{n_{Q}} \sum_{i=1}^{n_{Q}} \hat{\tau}^{(k)}(\X_i^{Q}) \hat{\tau}^{(l)}(\X_i^{Q}) \quad \text{for} \quad k, l \in [S], \\
    & \hat{d}_{s} = \frac{1}{n_{Q}} \sum_{i=1}^{n_{Q}} \left( \hat{\tau}^{(s)}(\X_i^{Q}) \right)^2 \quad \text{for} \quad s \in [S].
\end{aligned}
\]
We use $\hat{\tau}^{(s)}(\cdot)$ to denote a generic estimator of $\tau^{(s)}(\cdot)$ for each $s \in [S]$.

\subsection{Theoretical Results}\label{sec:theory}

We establish the theoretical guarantees for the proposed minimax regret CATE estimator $\hat{f}_{\operatorname{regret}}(\cdot)$ introduced in Section~\ref{subsec:estimation}.
Specifically, we derive the rate of convergence towards the true minimax regret CATE model $f^*_{\operatorname{regret}}(\cdot)$.

We first introduce several necessary assumptions.
\begin{assumption}[Positive definiteness] \label{ass:finitegroup:pdefinite}
The number of source sites $S$ is fixed and finite. The matrix $\Gamma$ defined in Theorem~\ref{thm:ident:regret} is positive definite with its smallest eigenvalue $\lambda_{\min }(\Gamma)> 0$.
\end{assumption}

While $\Gamma$ is a positive semi-definite matrix by definition, we further assume that it is positive definite.
This ensures that the objective function to determine the optimal aggregation weights in Theorem~\ref{thm:ident:regret} is strictly convex, guaranteeing a unique solution \( \q^* \). 
In practice, this assumption is easily satisfied as long as the site-specific CATE models $\{\tau^{(s)}(\cdot)\}_{s=1}^S$ are linearly independent.

In addition, we assume that the outcome is bounded and that the site-specific CATE estimators are $L_2$ consistent with respect to the target covariate distribution.
Specifically, for a function $f(\cdot)$, we define its $L_q$ norm measured over the target covariate distribution $Q_{\X}$ as $\|f(\cdot)\|_{Q, q}:=$ $\left(\mathbb{E}_{Q_{\X}}\left[f(\X)^q\right]\right)^{1 / q}$ where $q \geq 1$.
\begin{assumption}[Bounded outcome]\label{ass:bdd:outcome}
    We assume that the potential outcomes $Y_i(a)$ are almost surely bounded, i.e., $\exists\  M\geq0$ such that $\lvert Y_i(a)\rvert\leq M$.
\end{assumption}
\begin{assumption}[$L_2$ consistent estimation of site-specific CATE models]\label{ass:CATE:rate}
     Let $n=\min _{s \in[S]} n_s$ denote the smallest sample size among the $S$ source sites. We assume that there exists a positive sequence $\delta_n$ converging to zero as $n$ grows such that  
\[\max _{s \in[S]}\left\|\hat{\tau}^{(s)}(\cdot)-\tau^{(s)}(\cdot)\right\|_{Q, 2}\leq \delta_{n}. \]
\end{assumption} 
Assumption~\ref{ass:CATE:rate} constrains the $L_2$ convergence rate of $\hat{\tau}^{(s)}(\cdot)$ over the target covariate distribution. 
In contrast, standard guarantees for site-specific CATE estimators are often stated with respect to the source-site covariate distribution. As discussed above, the overlap condition provides a change-of-measure argument that can transfer such source-domain guarantees to the target domain, up to the corresponding density-ratio constant.
In the current literature, the convergence rates have been established for different CATE estimators under various conditions (see Appendix~\ref{appendix:learner:rate}).

The following theorem establishes the convergence rate for the proposed minimax regret CATE estimator $\hat{f}_{\operatorname{regret}}(\cdot)$ defined in Equation~\eqref{equ:estor:regret}. 
\begin{theorem}[$L_2$ error for the minimax regret CATE estimator]\label{thm:l2error:regret}
  Suppose that Assumptions~\ref{ass:unconfound:overlap}--\ref{ass:CATE:rate} hold. Then, with probability greater than $1-\frac{1}{t}-\frac{1}{t^2}$ for $t>1$,
  \begin{equation*}
    \left\|\hat{f}_{\operatorname{regret}}(\cdot)-f^*_{\operatorname{regret}}(\cdot)\right\|_{Q, 2}  \leq \sqrt{2 S}\left[\delta_n+2M\left\{\frac{1.5t S^2}{\lambda_{\min }(\Gamma)}\left(\frac{4 M}{\sqrt{t S}}\delta_n+\delta_n^2+\frac{4M^2}{\sqrt{n_Q}} \right)\wedge \rho_{\Delta_{S-1}}  \right\} \right],
  \end{equation*}
  where $\rho_{\Delta_{S-1}}:=\max _{\q, \q^{\prime} \in \Delta_{S-1}}\left\|\q-\q^{\prime}\right\|_2$ is the constant diameter of the simplex $\Delta_{S-1}$. 
\end{theorem}
The proof is given in Appendix~\ref{app:pf:Thm2}.
Theorem~\ref{thm:l2error:regret} characterizes the estimation error of $\hat{f}_{\operatorname{regret}}(\cdot)$ with two main components. The first term, $\delta_n$, originates from the direct estimation error of site-specific CATE models, while the second term, which depends on $\frac{1}{\lambda_{\min }(\Gamma)}\left(\delta_n+\delta_n^2+\frac{1}{\sqrt{n_Q}} \right) $, originates from the estimation error of the aggregated weights $\{\hat{q}_s\}_{s\in[S]}$.
In particular, both the (smallest) sample size of the source site data, $n$, and the sample size of the target covariate data, $n_Q$, affect the accuracy of $\hat{f}_{\operatorname{regret}}(\cdot)$. 
As $\delta_n$ and $\frac{1}{\sqrt{n_Q}}$ approach zero, this error bound shrinks, eventually falling below the naive constant upper bound $\rho_{\Delta_{S-1}}$.
Therefore, accurately recovering the oracle robust CATE model requires that each site-specific CATE model be estimated with reasonable precision. 
Additionally, the linear dependence among site-specific CATE models influences the error bound through the smallest eigenvalue $\lambda_{\min }(\Gamma)$ of the correlation matrix $\Gamma$. As $\lambda_{\min }(\Gamma)\rightarrow 0$, which indicates near-linear dependence of site-specific CATE models $\{\tau^{(s)}\}_{s\in[S]}$, estimating the optimal aggregation weights becomes more challenging.

Theorem~\ref{thm:l2error:regret} is related to recent finite-sample analyses for robust aggregation in multisource supervised learning \citep[e.g.,][]{guo2023statistical,wang2023distributionally}, which study plug-in estimation of a population-level robust optimizer. However, there are important differences. First, the target object is a multisite counterfactual-prediction rule induced by a minimax-regret formulation over CATE models.  In addition, the oracle aggregation weights solve a different quadratic program, involving both $\Gamma$ and $\d$. As a result, the bound must control estimation error from both the first-stage site-specific CATE estimators and the regret-specific weight estimation step.

Using Theorem~\ref{thm:l2error:regret}, we can quantify the MSE between the estimator $\hat{f}_{\operatorname{regret}}(\cdot)$ and a given target CATE model $\tau_Q(\cdot)$.
Suppose that there is a fixed target distribution with covariate distribution $Q_{\X}$ and a true CATE model $\tau_Q(\cdot)$. Then, the MSE between \( \hat{f}_{\operatorname{regret}}(\cdot) \) and  $\tau_Q(\cdot)$ can be bounded by:
\begin{align*}
    \left\|\hat{f}_{\operatorname{regret}}-\tau_Q\right\|_{Q, 2}^2  \leq  \left\|\hat{f}_{\operatorname{regret}}-f^*_{\operatorname{regret}}\right\|_{Q, 2}^2 &+  \left\|\tau_Q-f^*_{\operatorname{regret}}\right\|_{Q, 2}^2 \\&+ 2 \left\|\hat{f}_{\operatorname{regret}}-f^*_{\operatorname{regret}}\right\|_{Q, 2}\left\|\tau_Q-f^*_{\operatorname{regret}}\right\|_{Q, 2}.
\end{align*}
Therefore, the performance of the robust CATE estimator $\hat{f}_{\operatorname{regret}}(\cdot)$ on any fixed target distribution $Q$ depends on two factors: the convergence rate of $\hat{f}_{\operatorname{regret}}(\cdot)$ to its population counterpart, and the inherent distance between the oracle CATE models $\tau_Q(\cdot)$ and $f^*_{\operatorname{regret}}(\cdot)$.
Specifically, when the sample sizes for both the source sites and the target covariate distribution are sufficiently large, the $L_2$ error for the minimax regret estimator, $\left\|\hat{f}_{\operatorname{regret}}-f^*_{\operatorname{regret}}\right\|_{Q, 2}$, will shrink to zero. 
As a result, the distance between the minimax regret CATE estimator and any target CATE will converge to their true oracle gap, $\left\|\tau_Q-f^*_{\operatorname{regret}}\right\|_{Q, 2}$.
However, as the uncertainty set of target distributions becomes larger, the worst-case MSE between the robust CATE estimator and target CATEs increases, indicating that it becomes more challenging to build a generalizable CATE model that performs consistently well across a wide range of heterogeneous distributions.

\section{Comparison with Other Minimax Approaches}\label{sec:otherobjective}

In this section, we investigate the properties of the robust CATE models derived from two alternative objectives --- minimax risk and minimax relative risk ---, and compare their properties with those of the proposed minimax regret CATE model.
Although many of these existing works consider standard supervised learning problems \citep{mo2024minimax,wang2023distributionally,hastings2024taking}, we keep our discussion focused on the CATE estimation in causal inference.

\subsection{Minimax Risk} \label{subsec:minimaxrisk}

As briefly mentioned in Section~\ref{subsec:multi}, a common approach in the multi-group DRO framework is to minimize the maximum risk rather than the regret \citep{duchi2021learning,zhang2023optimal}. For comparison, we consider the robust CATE estimator based on the following minimax squared-error objective:
\begin{equation}\label{equ:HTE:sqobj}
f_{\operatorname{risk}}^* \ = \ \underset{f \in \mathcal{F}}{\arg \min } \max _{Q \in \mathcal{C}\left(Q_{\X}\right)} \mathbb{E}_Q[(Y(1)-Y(0)-f(\X))^2].
\end{equation}

A main problem of this formulation is that the resulting minimax risk CATE estimator can be sensitive to different noise levels across source sites.
To illustrate this, we decompose the objective function using the law of total expectation as follows:
\begin{equation}\label{equ:sqloss:decomp}
    \E_Q[ (\tau_Q(\X)-f(\X) )^2]+\E_Q[\Var_{Q}(Y(1)-Y(0)\mid\X)].
\end{equation}
Compared to the minimax regret problem shown in Equation~\eqref{equ:regret:distance}, the minimax risk problem includes an additional conditional variance term, which represents the noise level in the individual treatment effect conditional on covariates for the target distribution $Q$.
By optimizing this objective, the minimax risk solution will adapt to differences in inherent learning difficulty across populations.
When noise levels vary across target distributions, this minimax risk problem may lead to overfitting to those high-variance distributions that are heavily influenced by a few noisy sites.

In addition, the conditional variance in Equation~\eqref{equ:sqloss:decomp} is generally unidentifiable since we never observe $Y(1)$ and $Y(0)$ at the same time.
This makes it difficult to obtain an identifiable closed-form solution using the minimax risk estimation approach.
In Appendix~\ref{appendix:sqloss}, however, we derive a closed-form expression for the minimax risk estimator in the special case of $S=2$ with a more restrictive uncertainty set.
The result further illustrates the sensitivity of the minimax estimator to a target population with a high variance.

\subsection{Minimax Relative-risk}\label{subsec:relative-risk}

Another minimax objective is based on a relative-risk function, which compares the prediction accuracy of a model \(f(\cdot)\) against a pre-specified baseline model \(f_{\operatorname{base}}(\cdot)\). 
We can then define the minimax relative-risk CATE estimator as:
\begin{equation}\label{equ:HTE:obj:relativebase}
      f_{\operatorname{rel}}^*(\cdot;f_{\operatorname{base}})=\underset{f \in \mathcal{F}}{\arg \min } \max _{Q \in \mathcal{C}\left(Q_{\X}\right)} \mathbb{E}_Q\left[(Y(1)-Y(0)-f(\X))^2-(Y(1)-Y(0)-f_{\operatorname{base}}(\X))^2\right]. 
\end{equation}
This type of objective, which benchmarks a decision rule against a natural baseline, has been used for robust optimization in policy learning \citep{ben2021safe, kallus2021minimax, lei2023policy} and supervised learning \citep{meinshausen2015maximin, wang2023distributionally}.

When $\mathcal{F}$ contains the convex hull of true site-specific CATEs, the above objective function is equivalent to the $L_2$-distance between the model \(f(\cdot)\) and the true CATE, normalized by the distance between the baseline model and the true CATE:
\begin{equation}\label{equ:relativeloss:decomp}
    \E_Q[ (\tau_Q(\X)-f(\X) )^2] - \E_Q[(\tau_Q(\X)-f_{\operatorname{base}}(\X))^2].
\end{equation}
This objective provides a robustness guarantee by ensuring that $f_{\text{rel}}^*(\cdot)$ performs at least as well as \(f_{\operatorname{base}}(\cdot)\).
Unlike the minimax regret objective, which compares a prediction model $f(\cdot)$ to the optimal CATE model within each target distribution $Q$, the relative-risk approach benchmarks $f(\cdot)$ against a common baseline model.
As a result, the properties of \( f_{\operatorname{rel}}^*(\cdot) \) depend heavily on the choice of the baseline model.

In practice, selecting an appropriate baseline model often requires additional prior information.
A common default choice is the zero constant model \( f_{\operatorname{base}} = 0 \), which was originally proposed by \cite{meinshausen2015maximin} in classical linear regression settings \citep[see also][]{guo2023statistical} and later adopted by \cite{wang2023distributionally} for statistical machine learning. 
When $f_{\operatorname{base}} = 0$ and assuming that the treatment effect is centered, i.e., the ATE $\E_{P^{(s)}}[Y(1) - Y(0)] = 0$ for $s = 1, \ldots, S$, the negative value of the objective in Equation~\eqref{equ:HTE:obj:relativebase} can be interpreted as the variance explained by the model $f$.
Thus, the solution \( f_{\operatorname{rel}}^* \) ensures improved performance over the zero null model when predicting individual treatment effects.
In the statistical learning literature, this objective is often referred to as the {\it maximin explained variance} \citep{guo2023statistical, wang2023distributionally}.

Here, we derive a closed-form solution for the minimax relative-risk estimator $f_{\operatorname{rel}}^*$ for any baseline model $f_{\operatorname{base}}$.
In particular, due to the convex-concave structure of the optimization problem in Equation~\eqref{equ:HTE:obj:relativebase}, this closed-form solution holds for the generalized uncertainty set $\mathcal{C}(Q_{\X}, \mathcal{H})$ over any convex subset $\mathcal{H} \subseteq \Delta_{S-1}$.
\begin{proposition}[Minimax relative-risk identification]\label{prop:ident:baselineloss}
    Suppose that the function class $\mathcal{F}$ is convex with $\tau^{(s)}(\cdot) \in \mathcal{F}$ for all $s \in[S]$ and $\mathcal{H}$ is a convex subset of $\Delta_{S-1}$. Then $f_{\operatorname{rel}}^*(\cdot; f_{\operatorname{base}})$ in Equation~\eqref{equ:HTE:obj:relativebase}, when optimized over $\mathcal{C}(Q_{\X}, \mathcal{H})$, can be identified as
\begin{equation}\label{equ:iden:relativebase}
 f_{\operatorname{rel}}^*(\cdot;f_{\operatorname{base}})=\sum_{s=1}^S q_s^* \cdot \tau^{(s)}(\cdot) \quad \text { with } \quad \q^*=\underset{\q \in \mathcal{H}}{\arg \min } \  \E_{Q_{\X}}\left[\sum_{s=1}^S q_s\cdot \tau^{(s)}(\X) - f_{\operatorname{base}}(\X)\right]^2.
\end{equation}
\end{proposition}
The proof is given in Appendix~\ref{app:pf:ident:base}.
According to this proposition, the minimax relative-risk CATE estimator $f_{\operatorname{rel}}^*(\cdot; f_{\operatorname{base}})$ can be expressed as a weighted combination of site-specific CATE models $\{\tau^{(s)}(\cdot)\}_{s=1}^S$.
This is similar to the minimax regret estimator in Theorem~\ref{thm:ident:regret}.
The aggregation weights, however, are determined by minimizing the distance between the aggregated model and the baseline model.
Thus, the performance of $f_{\operatorname{rel}}^*$ depends critically on a reasonable $f_{\operatorname{base}}$.
In particular, if $f_{\operatorname{base}}(\cdot)$ lies within the uncertainty set $\mathcal{C}(Q_{\X}, \mathcal{H})$, the minimax relative-risk CATE estimator $f_{\operatorname{rel}}^*(\cdot;f_{\operatorname{base}})$ will simply reduce to $f_{\operatorname{base}}(\cdot)$.
If \( f_{\operatorname{base}}(\cdot) \) lies outside the uncertainty set, \( f_{\operatorname{rel}}^*(\cdot;f_{\operatorname{base}}) \) will correspond to the point in \( \mathcal{C}(Q_{\X}, \mathcal{H}) \) that is closest to \( f_{\operatorname{base}}(\cdot) \).

In Sections~\ref{sec:simulation}~and~\ref{sec:application}, we compare the empirical performance of our proposed minimax regret CATE model with the minimax relative-risk CATE model under the standard choice of the zero baseline model.
For the construction of the minimax relative-risk CATE estimator, we can similarly use plug-in estimates, as done for the minimax regret estimator (see Section~\ref{subsec:estimation}).

\section{Simulation Studies}\label{sec:simulation}

We conduct simulation studies to evaluate the performance of our proposed minimax regret CATE estimator under various scenarios.
To benchmark its performance, we compare it against the pooled CATE estimator, as well as the minimax relative-risk CATE estimator discussed in Section~\ref{sec:otherobjective}.
We do not consider the minimax risk estimator given its lack of an identifiable closed-form solution.

\subsection{Setup}

We generate simulated data from $S=10$ source sites.
For each source site $s\in[S]$, the data generation process is as follows:
\begin{align*}
    & \X^{(s)}_i \overset{\mathrm{i.i.d.}}{\sim} \mathcal{N}\left(\boldsymbol{0}, \mathbf{I}_5\right)\\
    & A^{(s)}_i \mid \X^{(s)}_i \overset{\mathrm{i.i.d.}}{\sim}  \operatorname{Bernoulli}(0.5) \\
    &  Y_i^{(s)} = \mu_{0}^{(s)}(\X_i^{(s)}) + A_i^{(s)}\cdot \tau^{(s)}(\X_i^{(s)}) +\varepsilon_i^{(s)}, \quad \text{where} \quad \varepsilon^{(s)}_i \overset{\mathrm{i.i.d.}}{\sim} \mathcal{N}(0,1).
\end{align*}
Here, $\mu^{(s)}_{0}(\cdot)$ and $\tau^{(s)}(\cdot) $ denote the conditional outcome function under control and the CATE function for site $s$, respectively.
To introduce distributional shifts in treatment effect heterogeneity across sites, we generate $\mu^{(s)}_{0}(\cdot)$ and $\tau^{(s)}(\cdot)$ with site-specific parameters.
Specifically, we generate the control baseline model as:
\begin{equation*}
\mu_{0}^{(s)}(\X) =  \alpha_s X_1 + \sum_{j=2}^5 X_j,
\end{equation*}
where $\alpha_s$ denotes the site-specific parameter to be specified later.
For the CATE functions, we consider two distinct settings:
\begin{align*}
    & \text{(A)}: \quad \tau^{(s)}(\X) =  \beta_s X_1 \mathds{1}(X_1>0)+ 0.2 (X_1X_2+X_2X_3), \\
    & \text{(B)}: \quad 
    \tau^{(s)}(\X) = 
    \begin{cases} 
        \beta_s\cdot 0.6+ \frac{2}{1+\exp(-12(X_1 - 1/2))}\cdot\frac{2}{1+\exp(-12(X_5 - 1/2))}, & \text{if } s \in \{1, 2,3\}, \\
           \beta_s \cdot X_1 \mathds{1}(X_1>0) + 0.2 (X_1X_2+X_2X_3), & \text{if } s \in \{4,5,6\},   \\
       \beta_s \cdot  0.5X_2^2 + 0.3(X_3+X_4), & \text{if } s \in \{7, 8,9,10\}. \\
    \end{cases}
\end{align*}
In both settings, the coefficient $\beta_s$ varies across sites to introduce site-specific heterogeneity. Setting~A assumes a consistent nonlinear functional form for $\tau^{(s)}$ across all sites, while Setting~B considers a more complex scenario with varying functional forms of $\tau^{(s)}$ across sites, creating a more challenging setting, particularly for the pooling approach.
This simulation setup is partially adapted from \citet{brantner2024comparison}.
The site-specific parameters $\alpha_s$ and $\beta_s$ are sampled i.i.d from a mixture of normal distributions: $\alpha_s, \beta_s \overset{\mathrm{i.i.d.}}{\sim} 0.7\times \mathcal{N}(0,0.75^2)+0.3\times\mathcal{N}(3,0.75^2) \).

For a given total sample size $n_{\operatorname{total}}$, each source $s\in[S]$ is allocated a local sample size determined by the sampling weight $q_{\text{mix}}\in \Delta_{S-1}$. Unless stated otherwise, we assume a default sample size of $n_{\operatorname{total}}=5,000$ and a balanced sample allocation with  $q_{\text{mix}}=(\frac{1}{S},\ldots,\frac{1}{S})$.
Additionally, we generate target covariate data $\{\X^{Q}_i\}_{i=1}^{n_Q}$ with a fixed sample size of $n_Q=10,000$. For simplicity, we assume no covariate shifts between the target and source sites, generating $\X^{Q}_i \overset{\mathrm{i.i.d.}}{\sim}  \mathcal{N}\left(\boldsymbol{0}, \mathbf{I}_5\right) $.

For site-specific CATE estimation, we employed multiple methods, including the R-learner and X-learner \citep{nie2021quasi, kunzel2019metalearners} implemented via the R package \texttt{rlearner}, and the causal forest \citep{wager2018estimation} using the R package \texttt{grf}.
For the proposed minimax regret and minimax relative-risk estimators, we first computed the aggregation weights by solving the respective QP problems using the R package \texttt{CVXR} and then ensembled the source-site CATE estimates to derive the final CATE models.
Additionally, we included a pooled CATE estimator, which directly pools all source data together and trains a single CATE model.

To evaluate the robustness of our approach, we will compare the empirical regret of different methods under two settings: (1) the performance on each individual source site distribution, and (2) the {\it worst-case} regret across all possible target distributions in $\mathcal{C}\left(Q_{\X}\right)$. 
Under the convex-hull assumption, this worst-case target distribution is attained at one of the source sites. Therefore, in our simulation setup, the worst-case regret over $\mathcal{C}\left(Q_{\X}\right)$ is equivalent to the maximum regret across the $S$ source sites, with the corresponding site-specific CATE serving as the oracle target CATE.
According to Equation~\eqref{equ:regret:distance}, the empirical regret is measured as the MSE between the estimated CATE model, $\hat{f}(\cdot)$, and a given target CATE, $\tau_Q(\cdot)$, over the target covariate observations: $\frac{1}{n_Q}\sum_{i=1}^{n_Q}(\tau_Q(\X^{Q}_i)-\hat{f}(\X^{Q}_i))^2$.
For each setup, we conduct 1,000 Monte Carlo simulations.

\subsection{Results}

Before presenting the main results, we note that in the current settings, all base single-site CATE estimation methods (R-learner, X-learner, and causal forest) perform reasonably well and yield similar final multisite CATE estimates. Therefore, we report the results based on the R-learner in the main text, with additional results for the X-learner and causal forest provided in Appendix~\ref{append:add:sim}.

\begin{figure}[t]
    \centering
    \includegraphics[width=1\linewidth]{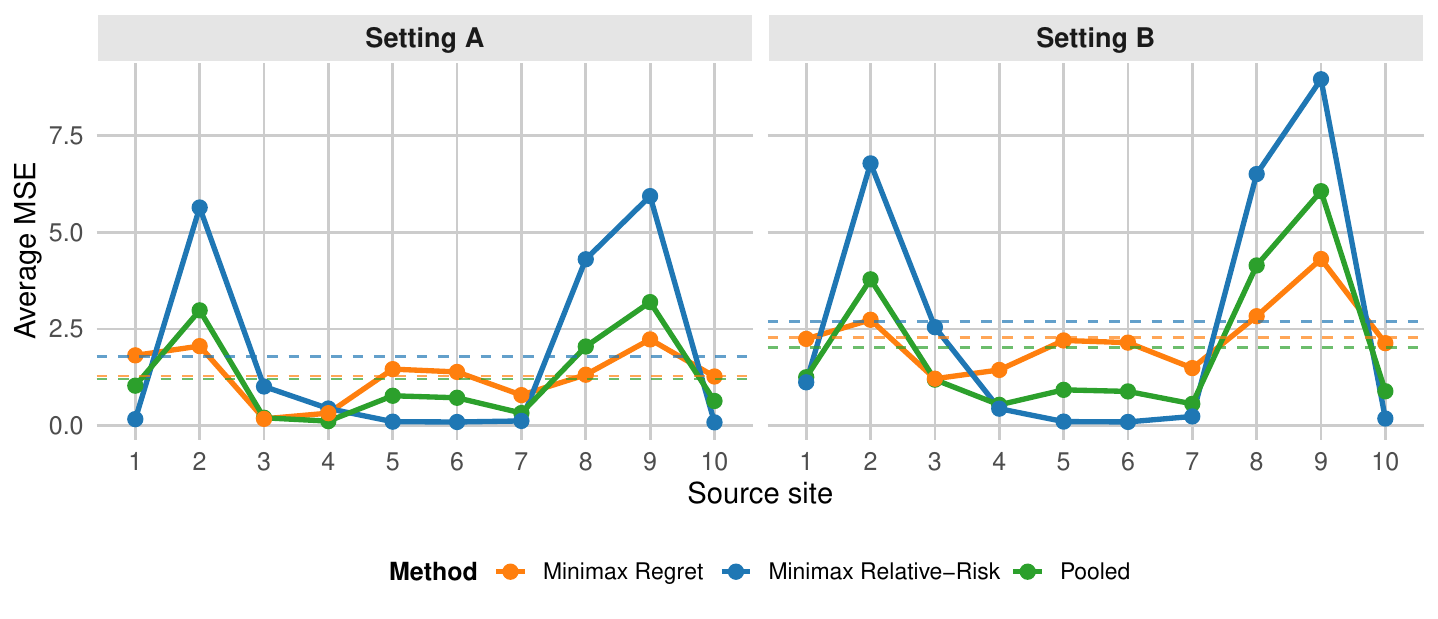} 
    \caption{Average MSE of multisite CATE estimates from three methods across 1,000 simulations, evaluated over different source sites, using R-learner as the site-specific CATE estimation method. Dashed horizontal lines indicate the corresponding average MSE across source sites for each method.}
    \label{fig:sim:varytarget}
\end{figure}

We first examine the performance of different CATE estimators on each individual source site.
Figure~\ref{fig:sim:varytarget} presents the regret averaged across simulations for each method under both Settings~A~and~B.
Our proposed minimax regret estimator (orange) demonstrates robust performance across diverse source sites, when compared to the minimax relative-risk (blue) and pooled (green) estimators.
In particular, the minimax regret estimator tends to perform especially well on the source sites (e.g., sites~2,~8,~and~9) that are most influential in the worst-case optimization problem.  These ``boundary'' sites bind in the minimax optimization and determine the robust solution; their squared $L_2$ distance from the oracle minimax regret solution attains the worst-case value in Proposition~\ref{lemma:euqival:regretOPT}. Because the minimax regret estimator is designed to control the maximum error over such sites, it can trade off some accuracy on more interior sites in order to improve worst-case robustness. This helps explain why some intermediate sites can occasionally have somewhat larger error under the minimax regret estimator, even though its overall worst-case performance is best. 
In contrast, the minimax relative-risk estimator is pulled toward the conservative baseline and therefore performs poorly on several sites. Moreover, the pooled estimator lacks worst-case robustness guarantees and can perform badly when the target distribution deviates from the pooled source distribution.

Setting~B introduces greater heterogeneity across sites by allowing varying functional forms for the CATE, in contrast to the consistent functional forms in Setting~A.
This increased heterogeneity amplifies the difficulty of learning a generalizable CATE model across diverse target distributions, resulting in higher worst-case regret for all methods. While the relative performance ranking remains the same as before, the pooled and relative-risk estimators show greater vulnerability to this functional heterogeneity. The performance of the pooled estimator, in particular, becomes worse as the single pooled model struggles to perform well when functional forms differ significantly across sites.

In comparison, the minimax regret approach maintains robust performance, with smaller variation and better error control across sites than the other methods.
These findings emphasize the adaptability of the proposed estimator to complex cross-site heterogeneity while providing strong worst-case guarantees. As a result, the minimax regret approach offers a consistent advantage in addressing the challenges posed by heterogeneous settings.
We also consider a covariate-shift variant of this first simulation setup, in which the target covariate distribution is shifted relative to the source sites. The corresponding results are reported in Appendix~\ref{appendix:sim:covshift}.

We also evaluate the impact of imbalanced sample sizes of source sites on the performance of each method.
We simulate three scenarios with a total sample size of \( n_{\operatorname{total}} = 5000 \): (1) ``Balanced'': all sites have an equal sample size, 
(2)	``Half-and-half'': half of the sites have three times the sample size of the others, with larger samples either in the first 
or the second half, 
and (3) ``One Large'': one site has a sample size 10 times larger than the others, with the largest site being either the 1st 
or the 5th site. 

\begin{figure}[t]
    \centering
    \includegraphics[width=1\linewidth]{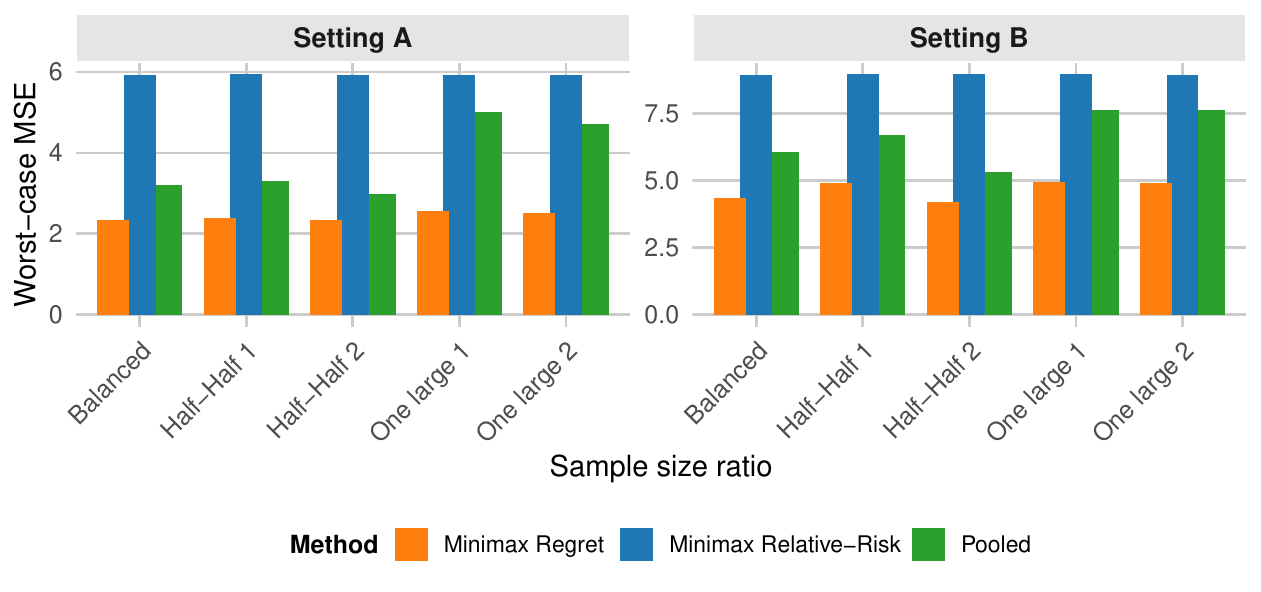}
   \caption{Worst-case MSE averaged across 1,000 iterations under varying sample size ratios: (1) {\bf Balanced}: all sites have equal sample sizes, (2) {\bf Half-and-Half}: half of the sites have three times the sample size of the others, and (3) {\bf One Large}: one site has a sample size 10 times larger than the others. For scenarios (2) and (3), we evaluate two configurations, each with a different subset of large sites. We use R-learner for site-specific CATE estimation.}
    \label{fig:sim:siteimbalance}
\end{figure}

As shown in Figure~\ref{fig:sim:siteimbalance}, the pooled estimator is highly sensitive to the imbalances of site sample sizes, with its performance heavily influenced by the large sites. This sensitivity arises because the pooled method weights sites proportionally to their sample sizes.
When site sample sizes vary significantly, the pooled estimator becomes dominated by the larger sites, leading to biased estimates and poor generalization to smaller sites or target distributions.

By contrast, the minimax regret and minimax relative-risk estimators demonstrate robust performance across all configurations. These methods remain stable even in the presence of substantial sample size imbalances, though the minimax relative-risk estimator is too conservative and performs worst in terms of MSE. This robustness comes from the ability to balance contributions from all source sites, mitigating the disproportionate influence of large sites. 
Each site contributes appropriately to the final CATE estimate based on how much its CATE differs from others, as reflected in the robust optimization objective, as long as the site has enough samples to reliably estimate its local CATE.
The results highlight the advantage of the minimax regret estimator in scenarios with heterogeneous sample sizes, as it avoids the pitfalls of oversampling bias inherent in the pooled approach.

\section{Empirical Application: Microcredit Experiments}\label{sec:application}

\subsection{Background and Setup}

To illustrate our proposed method, we turn to a set of microcredit experiments. These experiments were carried out in different geographic regions around the world -- Morocco, Bosnia and Herzegovina, Mexico, and Mongolia -- with the aim of evaluating whether microcredit fosters entrepreneurship among poor households, as measured in increases in profit \citep[e.g.,][]{roodman2012due}. Within each experiment, participants were randomly provided with increased access to credit from a Microfinance Institution (MFI). However, because the geographic regions in which these experiments took place were so different and because the specifics of each experimental design varied, there is a great degree of heterogeneity across the different experiments \citep{meager2019understanding}.

For example, the experiment in Bosnia \citep{augsburg2015impacts} randomized whether individuals who would otherwise be marginally rejected were provided loans. In contrast, other studies chose random communities to open a branch of an MFI for easier access. This implies that, in addition to differences in context across sites, the treatment itself was implemented differently across experiments. The studies also varied in the MFI's relationship with the local community. For example, in the study conducted by \citet{augsburg2015impacts}, the MFI was already operating locally before the experiment. In contrast, the experiments conducted in \citet{angelucci2015microcredit,attanasio2015impacts,crepon2015estimating} considered MFIs that were not previously operating locally. Our framework does not require these treatment implementations to be identical across sites; instead, such differences are absorbed into the site-specific potential outcome distributions and CATE functions. We summarize the details of each experimental site in Appendix Table~\ref{tbl:microcredit_summary}. 

We leverage the multi-site nature of the microcredit experiments to evaluate the performance of our proposed minimax regret approach in two ways. First, we synthetically generate different target populations to mimic plausible forms of distributional shift in practice. Second, we conduct a leave-one-out evaluation, in which each of the four experimental sites is treated in turn as the target population and the remaining three sites are used as sources.

For both sets of evaluations, we fit a CATE model using Causal Forest within each source site \citep{wager2018estimation}. The source CATEs range from -\$0.52 (in the site of Mongolia) to \$150 (in the site of Morocco). 

\begin{table}[ht]
\centering
\begin{tabular}{lcccccc}
  \toprule
Site & Min. & 1st Quantile & Median & Mean & 3rd Quantile & Max. \\ 
  \midrule
Mexico & 9.57 & 21.10 & 26.01 & 27.45 & 35.35 & 49.20 \\ 
  Mongolia & -0.52 & 0.07 & 0.19 & 0.18 & 0.35 & 0.62 \\ 
  Bosnia & 29.25 & 47.78 & 53.92 & 53.77 & 60.64 & 77.85 \\ 
  Morocco & 72.51 & 95.41 & 102.07 & 104.20 & 111.54 & 149.59 \\ 
   \bottomrule
\end{tabular}
\caption{Summary statistics for the different CATEs across the four sites.}
\end{table}

We compare the performance of the minimax regret approach against benchmark methods, including the minimax relative-risk and pooled approaches.
Overall, the minimax regret approach ensures the best worst-case performance across the various scenarios. Additionally, even when evaluating individual scenarios separately, the minimax regret approach generally outperforms both the pooling and minimax relative-risk methods. 

\subsection{Synthetically generated target populations} \label{subsec:synth_target}
We consider four scenarios that synthetically generate target populations to mimic the types of distributional shifts researchers may encounter in practice. Specifically, empirical target distributions are constructed by resampling from the original source site data, with the total number of target observations held constant across all scenarios. We treat the oracle CATE as the weighted average of source-site CATEs. The four scenarios are summarized below. 

\begin{enumerate} 
\item \textit{Pooled source distribution}. The target distribution is identical to the pooled source distribution, with mixture weights proportional to the sample sizes of the original source sites.
\item \textit{Equally weighted distribution}. The target distribution is a mixture of all source distributions with equal weights.
\item \textit{Single site dominated distribution}. 
The target distribution is primarily dominated by a single source site, mimicking a scenario in which the target population closely resembles subsets of the source sites. The sampling weights are set as $\left(\frac{5}{8}, \frac{1}{8}, \frac{1}{8}, \frac{1}{8}\right)$, with one site contributing $\frac{5}{8}$ of the target population and the remaining three sites contributing $\frac{1}{8}$ each.
\item \textit{Two sites dominated distribution.} The target distribution is dominated by two source sites, each contributing 40\% of the weight, while the remaining two sites contribute 10\% each.
\end{enumerate}


\begin{table}[t!]
\centering
\begin{tabular}{lccc}
  \toprule
 & Pooled & Regret & Relative-risk \\ \midrule 
 Scenario 1:\\
  ~Equivalent to Source Distribution & \textbf{190.07} & 403.57 & 5019.01 \\ 
\\ \midrule 
     Scenario 2: \\ 
~Equal Weight & 296.42 & \textbf{35.01} & 2025.88  \\ 
\midrule 
Scenario 3:\\ 
  ~Bosnia & 240.48 & \textbf{32.64} & 2501.27 \\ 
  ~Mexico & 517.94 & \textbf{229.34} & 1308.08 \\ 
  ~Mongolia & \textcolor{red}{1101.91} & \textcolor{red}{687.03} & \textbf{497.21} \\ 
  ~Morocco & \textbf{214.41} & 514.85 & \textcolor{red}{5417.91} \\ \midrule 
  Scenario 4: \\
  ~Bosnia and Morocco &\textbf{ 108.25} & 230.11 & 4254.59 \\ 
  ~Mexico and Bosnia & 371.78 & \textbf{81.15} & 1823.94 \\ 
  ~Mexico and Mongolia & 891.85 & \textbf{541.66} & 684.97 \\ 
  ~Mexico and Morocco & 170.66 & \textbf{43.61} & 3062.59 \\ 
  ~Mongolia and Bosnia & 629.47 & \textbf{228.33} & 1175.04 \\ 
  ~Mongolia and Morocco & 223.83 & \textbf{5.96} & 2284.56 \\ 
   \bottomrule
\end{tabular}
\caption{The results in terms of MSE for each method. Bolded entries indicate the best-performing method with the lowest error within each scenario (i.e., minimum value for each row), while \textcolor{red}{red} entries highlight the worst-case error for each method across \textit{all} scenarios (i.e., maximum value for each column). }
\label{tbl:microcredit_performance}
\end{table}

Table~\ref{tbl:microcredit_performance} presents the results in terms of MSE.
In Scenario~1 (Pooled source distribution), as expected, the pooled approach achieves the lowest MSE by perfectly aligning with the target distribution. In Scenario~2 (Equally weighted distribution), however, due to the imbalance in the original source sample sizes, the pooled approach performs worse than the minimax regret method.
The minimax relative-risk approach performs poorly in both scenarios because its solution aims to minimize the distance to the conservative baseline model, causing it to overweight the CATE model to the most extreme source distribution.

In Scenario 3 (Single site dominated distribution) across most settings, the pooled approach performs poorly because its implicit weighting reflects the observed sample, which does not align well with the target distribution. The only exception is the case where Morocco dominates the target distribution. This is because Morocco has the largest sample size in the source data, making this specific setting closer to the complete pooling distribution.

In general, the minimax regret approach also outperforms the relative-risk approach in most cases. The exception occurs in the setting where the target distribution is dominated by Mongolia, the most extreme source distribution with a CATE closest to zero baseline. In this case, the minimax relative-risk approach performs better, as its conservativeness ensures good performance in this extreme setting. We observe a similar pattern in Scenario~4, in which the target distribution is dominated by two of the sites. 

Additionally, we evaluate how well the estimated models adapt to the different target population distributions. Specifically, we compare the aggregation weights in the final minimax CATE model estimated by each method across the different scenarios. For comparison, we report the weights for the pooled approach as the original source sample size ratios, which remain fixed regardless of the target distribution. Although both the minimax relative-risk and minimax regret approaches theoretically produce aggregation weights that depend on the target covariate distribution, empirically, the weights do not vary significantly across different target scenarios. Specifically, the estimated weights for the minimax relative-risk approach are consistently dominated by a single site (Mongolia), whereas the minimax regret approach assigns non-zero weights across two different source distributions (Mongolia and Morocco). See Appendix Figure \ref{fig:scenario_weights} for visualization of the weights.

In summary, our proposed minimax regret approach demonstrates both robustness and adaptability in real-world multisite settings. It guarantees robust performance across a broad range of target distributions, outperforming the naive pooled method, especially in scenarios with significant heterogeneity across sites, while being less conservative than the minimax relative-risk approach, which is heavily influenced by the pre-specified baseline.

\subsection{Leave-one-out Evaluation}

While Section \ref{subsec:synth_target} demonstrated the benefits of our minimax regret approach across a variety of different target populations, the target populations were synthetically generated. We now evaluate the performance of our proposed approach in a setting where we do not know how the target population is necessarily constructed. We leverage the multi-site nature of the experiments to conduct a leave-one-out evaluation. We hold out each of the experimental sites and treat it as the target population, and use the other three sites as the source sites.

Examining the distribution of the estimated CATEs across sites, we find that two sites are substantially more extreme than the others. Morocco's CATEs are much larger, ranging from \$72 to \$149, whereas Mongolia's are much smaller, ranging from $-$\$0.52 to \$0.6. In contrast, the CATEs in Mexico and Bosnia range from roughly \$10 to \$78. As a result, the convex-hull assumption is likely less credible when Morocco or Mongolia is treated as the target population. Notably, in \textit{both} of these settings, the minimax regret approach still results in the smallest mean squared error, relative to the pooled and relative-risk approaches.
Overall, these results suggest that the proposed minimax regret estimator remains competitive even when the target site is not constructed to satisfy the maintained convex-hull assumption exactly. In particular, it delivers the best worst-case performance across the held-out sites, which is consistent with its design goal of controlling error under heterogeneous and only partially known target populations.

\begin{table}[ht]
\centering
\begin{tabular}{lccc}
  \toprule
Target & Pooled & Regret & Relative-risk \\ 
  \midrule
Mexico & 251.71 & 407.75 & 4768.19 \\ 
  Mongolia & 138.58 & 43.56 & 1586.07 \\ 
  Bosnia & 113.76 & 441.29 & 4969.64 \\ 
  Morocco & 2278.28 & 1806.26 & 5193.96 \\ 
  \midrule 
  Worst-case Error &  2278.28 & \textbf{1806.26} & 5193.96 \\ 
   \bottomrule
\end{tabular}
\end{table}

\section{Concluding Remarks} \label{sec:concl}

In this paper, we develop a minimax regret framework to learn a generalizable CATE model across heterogeneous multisite data. We formulate the problem as a minimax counterfactual prediction task and leverage the regret criterion under the assumption that an unknown target CATE model can be modeled as a convex mixture of source site CATE models.  We demonstrate that the proposed minimax regret CATE model is robust to unknown distributional shifts in treatment effect and provides performance guarantees across a wide range of target populations.
We further show that the robust CATE model admits an interpretable closed-form solution.
This approach is computationally efficient, enables the use of flexible off-the-shelf CATE estimation methods to obtain site-specific estimates, and avoids sharing individual-level data across sites when computing the robust CATE model. 
Although the main text focuses on the squared-loss setting, where the oracle predictor has a direct CATE interpretation, Appendix~\ref{appendix:extensionGLM} shows that the same minimax-regret perspective extends to a broader class of loss functions. 

More broadly, our framework allows researchers to leverage data from multiple heterogeneous sites to construct a treatment-effect model that generalizes to a target population without imposing strong parametric restrictions on cross-site heterogeneity. Because treatment effects may vary across studies and across individuals, results from any single study may not transport well to the population where decisions are ultimately made. By synthesizing evidence across heterogeneous studies while explicitly targeting the covariate distribution of the target population, our approach yields CATE estimates that are both more stable than site-specific estimates alone and more externally relevant than naive pooling. In turn, a generalizable CATE can support downstream decisions such as prioritizing outreach, tailoring eligibility criteria, or conducting cost-benefit analyses for individuals or subgroups in the target population.

An important direction for future work is interpretability. While the resulting robust CATE model may be combined with downstream tools such as variable-importance measures \citep{benard2023variable} or subgroup distillation methods \citep{huang2025distilling}, developing principled methods for extracting stable and interpretable effect moderators from the proposed multisite estimator remains an open problem.

Furthermore, our approach implicitly assumes that all sites share the same set of covariates for the estimation of CATE.
In practice, however, the measurement of the same constructs can differ across sites while some sites may completely lack key covariates, creating challenges such as discordant measurements or systematic missing data \citep{brantner2023methods}. More involved settings arise when the availability of covariates differs across source sites and/or between sources and the target.
In principle, one could also incorporate sites with partially observed covariates by introducing additional structure, such as imputation models that map site-specific covariates into the common $\X$ representation, or by explicitly modeling effect modification through latent or proxy variables. These extensions require extra assumptions and careful sensitivity analysis, and we view them as possible directions for future work. 

Another interesting avenue for future research is to consider how to use multiple source sites to help mitigate potential concerns about overlap violations between the sources and the target population. 

\bigskip

\bibliographystyle{chicago}
\bibliography{ref.bib}

\newpage
\appendix
\renewcommand\thefigure{\thesection.\arabic{figure}}    
\setcounter{figure}{0}
\numberwithin{figure}{section}
\renewcommand\thetable{\thesection.\arabic{table}}    
\setcounter{table}{0} 
\numberwithin{table}{section}
\renewcommand\theequation{\thesection.\arabic{equation}}    
\setcounter{equation}{0}
\numberwithin{equation}{section}
\renewcommand\theproposition{\thesection.\arabic{proposition}}    
\setcounter{proposition}{0}
\numberwithin{proposition}{section}
\setcounter{lemma}{0} 
\renewcommand{\thelemma}{\thesection.\arabic{lemma}}
\numberwithin{lemma}{section}

\renewcommand\thetheorem{\thesection.\arabic{theorem}}
\setcounter{theorem}{0}
\numberwithin{theorem}{section}

\renewcommand\theassumption{\thesection.\arabic{assumption}}
\setcounter{assumption}{0}
\numberwithin{assumption}{section}

\begin{center}
 \huge Supplementary Appendix 
\end{center}

\etocdepthtag.toc{mtappendix}
\etocsettagdepth{mtchapter}{none}
\etocsettagdepth{mtappendix}{subsection}
\tableofcontents

\section{Closed-Form Minimax Squared-error CATE Estimator}\label{appendix:sqloss}

We derive a closed-form solution for the minimax estimator based on the squared-error objective, under the special case where $S = 2$ and an additional assumption on the target distribution. Specifically, we restrict the target distribution $Q$ to the following uncertainty set:
\begin{equation}
\widetilde{\mathcal{C}}\left(Q_{\X}\right) := \left\{Q = \left(Q_{\X}, Q_{(Y(1),Y(0)) \mid \X}\right) \ \bigg| \ Q_{Y(1) - Y(0) \mid \X} = \sum_{s=1}^S q_s \cdot P_{Y(1) - Y(0) \mid \X}^{(s)} \quad \text{with} \quad \q \in \Delta_{S-1}\right\}.
\end{equation}
Notably, this restricted uncertainty set $\widetilde{\mathcal{C}}\left(Q_{\X}\right)$ is a subset of the uncertainty set $\mathcal{C}\left(Q_{\X}\right)$ considered in the main text.
It further constrains the entire target conditional treatment effect distribution $Q_{Y(1) - Y(0) \mid \X}$ to be a convex combination of the source site distributions $\left\{P_{Y(1) - Y(0) \mid \X}^{(s)}\right\}_{s=1}^S$.

Now, we derive a closed-form solution for $f_{\operatorname{risk}}^*$ in Equation~\eqref{equ:HTE:sqobj} when optimizing over the restricted uncertainty set $\widetilde{\mathcal{C}}\left(Q_{\X}\right)$.
\begin{proposition}\label{prop:sqloss}
  Suppose $S=2$. Define $\varepsilon_i^{(s)}:=Y_i^{(s)}(1)-Y_i^{(s)}(0)-\tau^{(s)}(\X_i^{(s)})$ for $s=1,2$. Assume $\tau^{(1)} \neq \tau^{(2)}, \E_{P^{(1)}}\left[\left(\varepsilon_i^{(1)}\right)^2 \mid \X_i^{(1)}\right]=\sigma_1^2$ and $\E_{P^{(2)}}\left[\left(\varepsilon_i^{(2)}\right)^2  \mid \X_i^{(2)}\right]=$ $\sigma_2^2$, and $\varepsilon_i^{(s)}$ being independent of $\X_i^{(s)}$ for $s\in[S]$.  Then, $f_{\operatorname{risk}}^*$ in Equation~\eqref{equ:HTE:sqobj}, optimized over $\widetilde{\mathcal{C}}\left(Q_{\X}\right)$, can be written as:
$$
f_{\operatorname{risk}}^*=q^*_1 \tau^{(1)}+\left(1-q^*_1\right) \tau^{(2)} \quad \text { with } \quad q^*_1=0 \vee\left(\frac{1}{2}+\frac{\sigma_1^2-\sigma_2^2}{2\left\|\tau^{(1)}-\tau^{(2)}\right\|_{Q_{\X}, 2}^2}\right) \wedge 1,
$$
where $\left\|\cdot\right\|_{Q_{\X}, 2}$ is defined as the $L_2$-norm measured over the target covariate distribution $Q_{\X}$.
\end{proposition}
The proof is omitted because the result follows from Proposition~2 of \cite{wang2023distributionally} by applying their standard statistical learning framework to the current causal context.
Proposition~\ref{prop:sqloss} implies that \( f_{\operatorname{risk}}^* \) can be expressed in closed form as a convex combination of the site-specific CATEs, $\tau^{(1)}$ and $\tau^{(2)}$.
The estimator, however, is less robust than our proposed minimax regret estimator \( f_{\operatorname{regret}}^* \).
Specifically, when noise levels are homogeneous between the two sites, i.e., $\sigma_1^2 = \sigma_2^2$, $f_{\operatorname{risk}}^*$ is the equal average of $\tau^{(1)}$ and $\tau^{(2)}$, which equals \( f_{\operatorname{regret}}^* \).
In contrast, in the case of heterogeneous noise levels, $f_{\operatorname{risk}}^*$ becomes sensitive to the noisier site. 
For example, when one site has a significantly higher level of noise than the other, i.e., $\sigma_1^2 \gg \sigma_2^2$, $f_{\operatorname{risk}}^*$ is dominated by $\tau^{(1)}$.
This result aligns with the intuition based on Equation~\eqref{equ:sqloss:decomp}, where \( f_{\operatorname{risk}}^* \) primarily optimizes prediction performance for the site with higher noise.

\section{Convergence Rates for CATE Estimators}\label{appendix:learner:rate}

We characterize $\delta_n$ for specific examples of CATE estimators. For clarity, we omit the superscript site indicator $s$ in the following examples and let $n$ denote the generic sample size of a given site.
Define a function as \( p \)-smooth if it has \( p \) continuous and bounded derivatives. Assume the following smoothness conditions for each function:
$$
\begin{aligned}
\mu_a(\x) & := \mathbb{E}[Y(a) \mid \X = \x] \text{ is } p_{\mu_a}\text{-smooth,} \\
\pi(\x)  &:= \mathbb{P}(A = 1 \mid \X = \x) \text{ is } p_{\pi}\text{-smooth,} \\
\tau(\x) & := \mu_1(\x) - \mu_0(\x) \text{ is } p_{\tau}\text{-smooth.}
\end{aligned}
$$
Assume further that each of these functions is estimable at a minimax rate of \( n^{\frac{-p_i}{2 p_i + d}} \), where \( p_i \) corresponds to the smoothness level of the respective function, and $d$ is the dimension of covariates.
Table~\ref{table:deltarate} gives the error bound of some commonly used meta-learners.

\begin{table}[H]
\centering \singlespacing
\begin{tabular}{@{}ll@{}}
\toprule 
Estimator & $\delta_n$ rate \\ 
\midrule
\textbf{T-learner} \citep{kunzel2019metalearners} & 
  $O\left(n^{\frac{- p_{\mu_0}}{2 p_{\mu_0} + d}} + n^{\frac{- p_{\mu_1}}{2 p_{\mu_1} + d}}\right)$ \\ 
\midrule
\textbf{X-learner} \citep{kunzel2019metalearners} & 
  $O\left(n^{\frac{- p_{\tau}}{2 p_{\tau} + d}} + n^{\frac{- p_{\mu_0}}{2 p_{\mu_0} + d}} + n^{\frac{- p_{\mu_1}}{2 p_{\mu_1} + d}}\right)$ \\
\midrule
\textbf{DR-learner} \citep{kennedy2023towards} & 
  $O\left(n^{\frac{-p_{\tau}}{2 p_{\tau} + d}} + n^{-\left(\min_a \frac{p_{\mu_a}}{2 p_{\mu_a} + d} + \frac{p_{\pi}}{2 p_{\pi} + d}\right)}\right)$ \\
\bottomrule
\end{tabular}
\caption{Examples of $\delta_n$ for Different CATE Estimators. Note: If $\min_a \frac{p_{\mu_a}}{2 p_{\mu_a} + d} + \frac{p_{\pi}}{2 p_{\pi} + d} > \frac{p_{\tau}}{2 p_{\tau} + d}$, the DR-learner achieves the oracle rate $n^{\frac{- p_{\tau}}{2 p_{\tau} + d}}$.} \label{table:deltarate}
\end{table}

\section{Extension to a Broader Class of Loss Functions}\label{appendix:extensionGLM}

We generalize our minimax-regret framework beyond squared error loss by considering a broader
class of loss functions. In particular, we use the likelihood-motivated losses based on generalized linear models (GLMs) with canonical links. We also allow for a general prediction class
\(\mathcal{F}\subset L_2(Q_{\X})\). We demonstrate that the geometric structure for the minimax solution we obtained in the squared-loss setting can be generalized to this case.

\subsection{GLM-motivated loss and risk in function space}

Let \(\mathcal{F}\subset L_2(Q_{\X})\) be a nonempty convex class of measurable
functions \(f:\mathcal{X}\to\R\). For an outcome \(y\in\R\), covariates
\(\x\in\mathcal{X}\), and a predictor \(f\in\mathcal{F}\), a canonical-link GLM
has log-likelihood
\[
\log p_f(y\mid \x)
\propto y\,f(\x) - \psi\big(f(\x)\big),
\]
up to the dispersion parameter and additive terms that do not depend on
\(f\), where \(\psi:\R\to\R\) is the log-partition function. We assume
\(\psi\) is strictly convex and smooth.

Motivated by this representation, we define the loss as the negative
log-likelihood
\[
\ell_f(y,\x)
:= \psi\big(f(\x)\big) - y\,f(\x),
\qquad f\in\mathcal{F}.
\]
For a target distribution $Q$, the associated population risk for predicting the
individual treatment effect $\Delta Y:=Y(1)-Y(0)$ is
\begin{equation}
\label{eq:risk-GLM-general}
\mathrm{Risk}(f,Q)
:= \E_Q\!\left[\ell_f(\Delta Y,\X)\right]
= \E_Q\!\left[\psi\big(f(\X)\big) - \Delta Y\,f(\X)\right].
\end{equation}
Let
\[
\tau_Q(\x) := \E_Q[\Delta Y\mid \X=\x],
\qquad
\langle g,h\rangle := \E_{Q_{\X}}[g(\X)\,h(\X)]
\]
denote the conditional treatment effect function and the $L_2(Q_{\X})$ inner product,
respectively. Then, Equation~\eqref{eq:risk-GLM-general} can be written as
\begin{align}
\mathrm{Risk}(f,Q)
&= \E_{Q_{\X}}\big[\psi(f(\X))\big]
   \;-\; \langle \tau_Q, f\rangle \notag\\
&=: \Psi(f) - \langle \tau_Q, f\rangle,
\label{C.2}
\end{align}
where $\Psi(f) := \E_{Q_{\X}}[\psi(f(\X))]$ is convex in $f$.

A notable special case arises when \(\psi(\eta) = \eta^2/2\), corresponding to
Gaussian linear regression. In this case,
\[
\ell_f(y,\x)
= \tfrac12\big(f(\x) - y\big)^2 \quad \text{(up to a constant in \(y\))},
\]
so \(\mathrm{Risk}(f,Q)\) reduces to the squared-error risk considered in the
main text. Another classical example is logistic regression for binary outcomes
\(y\in\{0,1\}\), where \(\psi(\eta)=\log(1+\mathrm{e}^{\eta})\) and
\[
\ell_f(y,\x)
= \log\big(1+\mathrm{e}^{f(\x)}\big) - y\,f(\x)
\]
is the standard logistic negative log-likelihood.

\subsection{Regret and Bregman divergence in function space}

Next, we define the regret objective as the excess risk relative to the best
predictor in \(\mathcal{F}\). For each \(Q\), let
\[
f^*(Q) \in \argmin_{f\in\mathcal{F}} \ \mathrm{Risk}(f,Q)
\]
denote the (unique) oracle minimizer of \(\mathrm{Risk}(\cdot,Q)\). The regret of a
predictor \(f\in\mathcal{F}\) under \(Q\) is then defined as 
\[
\mathrm{Regret}(f,Q)
:= \mathrm{Risk}(f,Q) - \mathrm{Risk}\big(f^*(Q),Q\big).
\]
To obtain a clean characterization of $f^*(Q)$, we impose the following regularity condition on $\Psi$.  
\begin{assumption}\label{ass:Psi-legendre-functional}
\(\Psi:\mathcal{F}\to\R\) is Fr\'echet differentiable, strictly convex and
lower semicontinuous, and its convex conjugate
\[
\Psi^*(g)
:= \sup_{f\in\mathcal{F}}\{\langle g,f\rangle - \Psi(f)\},
\qquad g\in L_2(Q_{\X}),
\]
is finite and strictly convex on \(L_2(Q_{\X})\). In particular, \(\Psi\) is of
Legendre type \citep[Chap.~26]{rockafellar2015convex}, so that \(\nabla\Psi\) and \(\nabla\Psi^*\) are mutual inverses on
their domains.
\end{assumption}

The following proposition shows that the regret can be written as the Bregman
divergence generated by \(\Psi\) on the function space \(\mathcal{F}\). For
\(u,v \in \mathcal{F}\), define
\[
D_{\Psi}(u \,\|\, v)
:= \Psi(v) - \Psi(u)
   - \big\langle \nabla \Psi(u),\,v-u\big\rangle,
\]
where \(\langle\cdot,\cdot\rangle\) is the \(L_2(Q_{\X})\) inner product  and
\(\nabla\Psi(u)\in L_2(Q_{\X})\) denotes the Fr\'echet gradient of \(\Psi\) at
\(u\). 
\begin{proposition}[Regret as Bregman divergence in function space]
\label{prop:regret-bregman-functional}
Suppose Assumption~\ref{ass:Psi-legendre-functional} holds and that, for a given
target distribution \(Q\), the conditional effect function
\(\tau_Q\) lies in the range of \(\nabla\Psi\). Then:
\begin{enumerate}
\item[(i)] \emph{(Oracle characterization)} The risk \(\mathrm{Risk}(\cdot,Q)\)
has a unique minimizer \(f^*(Q)\in\mathcal{F}\), characterized by
\[
\nabla\Psi\big(f^*(Q)\big) = \tau_Q,
\qquad\text{equivalently}\qquad
f^*(Q) = \nabla\Psi^*(\tau_Q),
\]
where \(\Psi^*\) is the convex conjugate of \(\Psi\).

\item[(ii)] \emph{(Regret as Bregman divergence)} For any \(f\in\mathcal{F}\),
\[
\mathrm{Regret}(f,Q)
= D_{\Psi}\big(f^*(Q)\,\|\,f\big).
\]
\end{enumerate}
\end{proposition}

We now return to the multisite minimax-regret problem. Recall that
\(P^{(1)},\ldots,P^{(S)}\) denote the source populations.\footnote{At the population level, we do not need to impose a ``no covariate shift'' assumption. All risk and Bregman quantities
are defined with respect to the target base measure \(Q_{\X}\), and the
results below rely only on the structure of target CATE in the uncertainty set, not on how the site marginals \(P_{\X}^{(s)}\) relate to
\(Q_{\X}\). Covariate shift matters for estimation (i.e., how we recover \(\tau_s\)
from data), but not for the oracle minimax geometry.}
For each site \(s\), let
\[
\tau_s(\x) := \E_{P^{(s)}}[\Delta Y \mid \X=\x],
\qquad
f_s^* := f^*\big(P^{(s)}\big),
\]
so that \(f_s^*\) is the site-specific oracle predictor from
Proposition~\ref{prop:regret-bregman-functional}. As in the squared-loss case,
we consider an uncertainty set \(\mathcal{C}(Q_{\X})\) such that, for any
\(Q\in\mathcal{C}(Q_{\X})\), the corresponding CATE function satisfies
\[
\tau_Q(\x)
= \sum_{s=1}^S q_s\,\tau_s(\x)
\quad\text{for some } q=(q_1,\dots,q_S)\in\Delta_{S-1},
\]
so that \(\tau_Q\) lies in the convex hull of \(\{\tau_s\}_{s=1}^S\) in
\(L_2(Q_{\X})\). We then define the minimax-regret predictor as
\[
f^*_{\mathrm{regret}}
\in \argmin_{f\in\mathcal{F}}\;
\max_{Q\in\mathcal{C}(Q_{\X})} \mathrm{Regret}(f,Q).
\]

It is convenient to also introduce the Bregman divergence generated by the
conjugate \(\Psi^*\). For \(a,b\in L_2(Q_{\X})\), define
\[
D_{\Psi^*}(a \,\|\, b)
:= \Psi^*(b) - \Psi^*(a)
   - \big\langle \nabla\Psi^*(a),\,b-a\big\rangle.
\]
By standard Bregman conjugacy identities, for any
\(u,v\in\mathcal{F}\) with \(g_u := \nabla\Psi(u)\) and \(g_v := \nabla\Psi(v)\),
one has
\[
D_{\Psi}(u \,\|\, v)
= D_{\Psi^*}(g_v \,\|\, g_u).
\]
Applying this to \(u=f^*(Q)\), \(v=f\), and using
\(\nabla\Psi\big(f^*(Q)\big)=\tau_Q\) from
Proposition~\ref{prop:regret-bregman-functional}, we obtain
\[
\mathrm{Regret}(f,Q)
= D_{\Psi}\big(f^*(Q)\,\|\,f\big)
= D_{\Psi^*}\big(\theta \,\|\, \tau_Q\big),
\qquad
\theta := \nabla\Psi(f).
\]
Thus, for any fixed \(f\in\mathcal{F}\), the regret depends on \(Q\) only
through its conditional treatment effect function \(\tau_Q\), and in the dual space
it is measured by the divergence \(D_{\Psi^*}(\theta\|\tau_Q)\).

Since \(\tau_Q\) ranges over the convex hull of \(\{\tau_s\}_{s=1}^S\) as
\(Q\) varies over \(\mathcal{C}(Q_{\X})\), the worst-case regret over
\(\mathcal{C}(Q_{\X})\) can be reduced to a finite maximization over the sites:
\begin{lemma}[Vertex reduction in dual (effect) space]
\label{lem:vertex-reduction-functional}
Fix \(f\in\mathcal{F}\) and let \(\theta := \nabla\Psi(f)\). Under the
construction of \(\mathcal{C}(Q_{\X})\) above,
\[
\max_{Q\in\mathcal{C}(Q_{\X})} \mathrm{Regret}(f,Q)
= \max_{1\le s\le S} D_{\Psi^*}\big(\theta \,\|\, \tau_s\big).
\]
Equivalently,
\[
\max_{Q\in\mathcal{C}(Q_{\X})} \mathrm{Regret}(f,Q)
= \max_{1\le s\le S} D_{\Psi}\big(f_s^* \,\|\, f\big).
\]
\end{lemma}

Lemma~\ref{lem:vertex-reduction-functional} reduces the multisite minimax
problem to a finite maximization over the \(S\) sites. In particular, the
minimax-regret predictor can be written as
\[
f^*_{\mathrm{regret}}
\in \argmin_{f\in\mathcal{F}}
\max_{1\le s\le S} D_{\Psi}\big(f_s^* \,\|\, f\big),
\]
or, in the dual space,
\[
\theta^*_{\mathrm{regret}}
\in \argmin_{\theta\in L_2(Q_{\X})}
\max_{1\le s\le S} D_{\Psi^*}(\theta \,\|\, \tau_s),
\qquad
\theta^*_{\mathrm{regret}} := \nabla\Psi\big(f^*_{\mathrm{regret}}\big).
\]
In the next step, we show that this dual minimax problem admits a Bregman
Chebyshev-center representation, with the site-specific CATE functions
\(\{\tau_s\}_{s=1}^S\) playing the role of ``points" in the dual space.

We now characterize the solution to this dual minimax problem in terms of a Bregman Chebyshev center of the site-specific effect functions
\(\{\tau_s\}_{s=1}^S\).

\begin{theorem}[Bregman Chebyshev center in conjugate space]
\label{thm:bregman-center-functional}
Consider the minimax problem
\begin{equation}
\label{equ:conjugate-minimax-functional}
    \min_{\theta\in L_2(Q_{\X})}\;\max_{1\le s\le S} D_{\Psi^*}(\theta \,\|\, \tau_s).
\end{equation}
Equivalently, in epigraph form,
\[
\min_{\theta\in L_2(Q_{\X}),\,R\ge 0} R
\quad\text{subject to}\quad
D_{\Psi^*}(\theta\|\tau_s)\le R,\;\; s=1,\dots,S.
\]

For \(\bgamma=(\gamma_1,\dots,\gamma_S)\in\Delta_{S-1}\), define
\begin{equation}
\label{equ:conjugate-weights-functional}
\mathcal{H}(\bgamma)
:= \sum_{s=1}^S \gamma_s \Psi^*(\tau_s)
   - \Psi^*\!\left(\sum_{s=1}^S \gamma_s \tau_s\right).
\end{equation}
Then the following hold:
\begin{enumerate}
\item[(i)] \emph{(Existence of optimal weights)} There exists at least one
\[
\bgamma^* \in \argmax_{\bgamma\in\Delta_{S-1}} \mathcal{H}(\bgamma).
\]

\item[(ii)] \emph{(Closed-form dual center)} Any such \(\bgamma^*\) induces a
dual minimax center
\[
\theta^* := \sum_{s=1}^S \gamma_s^* \tau_s
\]
that solves \eqref{equ:conjugate-minimax-functional}, i.e.
\[
\theta^* \in \argmin_{\theta\in L_2(Q_{\X})}
\max_{1\le s\le S} D_{\Psi^*}(\theta \,\|\, \tau_s).
\]

\item[(iii)] \emph{(Optimal worst-case regret)} The optimal worst-case regret
value in dual space is
\[
R^*
:= \min_{\theta\in L_2(Q_{\X})}\;\max_{1\le s\le S} D_{\Psi^*}(\theta \,\|\, \tau_s)
= \mathcal{H}(\bgamma^*).
\]

\item[(iv)] \emph{(Map back to predictor space)} The minimax-regret predictor in
\(\mathcal{F}\) is
\[
f^*_{\mathrm{regret}} = \nabla\Psi^*(\theta^*),
\]
and satisfies
\[
f^*_{\mathrm{regret}}
\in \argmin_{f\in\mathcal{F}}
\max_{1\le s\le S} D_{\Psi}\big(f_s^* \,\|\, f\big).
\]

\item[(v)] \emph{(Supporting sites)} If \(\gamma_s^*>0\) for some \(s\), then
site \(s\) is active at the optimum:
\[
\gamma_s^*>0 \;\Longrightarrow\;
D_{\Psi^*}(\theta^*\|\tau_s) = R^*.
\]
\end{enumerate}
\end{theorem}

Theorem~\ref{thm:bregman-center-functional} admits a natural geometric
interpretation. In the dual space, each site-specific
conditional effect function \(\tau_s\) is an element of \(L_2(Q_{\X})\), and
the divergence \(D_{\Psi^*}(\theta\|\tau_s)\) measures the excess risk
associated with using \(\theta\) instead of the site effect \(\tau_s\). The
minimax problem in Equation~\eqref{equ:conjugate-minimax-functional} therefore
seeks a dual element \(\theta^*\) that serves as the center of a
smallest-radius Bregman ball (under \(D_{\Psi^*}\)) containing all
\(\{\tau_s\}_{s=1}^S\). The theorem shows that this center lies in the convex
hull of the site effects, with the explicit representation
\(\theta^* = \sum_{s=1}^S \gamma_s^* \tau_s\) for some
\(\bgamma^*\in\Delta_{S-1}\) maximizing the dispersion functional
\(\mathcal{H}(\bgamma)\).

Mapping back to the original prediction space via
\(f^*_{\mathrm{regret}} = \nabla\Psi^*(\theta^*)\), the minimax-regret predictor
can be viewed as the GLM predictor whose dual representation is the Bregman
Chebyshev center of the site-specific effect functions in the \(L_2(Q_{\X})\)-Bregman geometry induced by \(\Psi\). Moreover, the
supporting-site property
\(\gamma_s^*>0 \Rightarrow D_{\Psi^*}(\theta^*\|\tau_s)=R^*\) implies that only
a subset of sites is binding at the optimum: these sites lie on the boundary of
the optimal Bregman ball and determine the worst-case regret, while the
remaining sites are strictly interior and do not affect the value of \(R^*\).

\subsection{Connection to the special case of squared loss}

The functional GLM-based framework above nests the squared-error setting as a
special case. To see this, consider the canonical Gaussian GLM with identity
link, corresponding to
\[
\psi(\eta) = \tfrac12 \eta^2, \qquad \eta\in\R.
\]
In this case, for any \(f\in\mathcal{F}\subset L_2(Q_{\X})\),
\[
\Psi(f)
= \E_{Q_{\X}}\big[\psi\big(f(\X)\big)\big]
= \tfrac12\,\E_{Q_{\X}}\big[f(\X)^2\big],
\]
so \(\Psi\) is simply the quadratic functional induced by the \(L_2(Q_{\X})\)
norm. The associated convex conjugate \(\Psi^*\) coincides (up to a constant) with
\(\tfrac12\|\cdot\|_{L_2(Q_{\X})}^2\), and the Bregman divergence
\(D_{\Psi^*}(\theta\|\tau)\) reduces to
\[
D_{\Psi^*}(\theta\|\tau)
= \tfrac12\,\|\theta-\tau\|_{L_2(Q_{\X})}^2,
\]
that is, the squared \(L_2(Q_{\X})\) distance between \(\theta\) and \(\tau\).
Thus, the dual minimax problem in
Equation~\eqref{equ:conjugate-minimax-functional} reduces to finding a smallest
enclosing ball (with respect to L2 norm) that covers the site-specific
moment vectors. In this case, \(\Psi\) and \(\Psi^*\) are both quadratic and
\(\nabla\Psi = \nabla\Psi^* \) equals the identity operator on $L_2(Q_{\X})$, so the conjugate-space solution
\(\theta^*\) coincides with the primal minimax-regret predictor
\(f^*_{\mathrm{regret}}\), and hence with the minimax-MSE estimator obtained in
the squared-loss setting. Thus, Theorem~\ref{thm:bregman-center-functional} can be viewed as a direct extension
of the squared-error minimax-regret characterization to general GLM-type losses
and general function classes \(\mathcal{F}\).

\begin{proof}[Proof of Proposition~\ref{prop:regret-bregman-functional}]
Fix \(Q\) and abbreviate \(\tau_Q\) by \(\tau\) and \(f^*(Q)\) by \(f^*\).
By definition,
\[
\mathrm{Risk}(f,Q)
= \Psi(f) - \langle \tau, f\rangle,
\]
where \(\Psi\) is convex and Fr\'echet differentiable on the convex set
\(\mathcal{F}\subset L_2(Q_{\X})\). Under Assumption~\ref{ass:Psi-legendre-functional},
\(\mathrm{Risk}(\cdot,Q)\) is strictly convex and admits a unique minimizer
\(f^*\). The first-order optimality condition is
\[
\nabla\Psi(f^*) - \tau = 0,
\]
which yields \(\nabla\Psi(f^*)=\tau\). By Legendre duality,
\(\nabla\Psi\) and \(\nabla\Psi^*\) are mutual inverses on their domains, so
\(f^* = \nabla\Psi^*(\tau)\), establishing (i).

For (ii), compute
\begin{align*}
\mathrm{Regret}(f,Q)
&= \mathrm{Risk}(f,Q) - \mathrm{Risk}(f^*,Q) \\
&= \big[\Psi(f) - \langle\tau,f\rangle\big]
   - \big[\Psi(f^*) - \langle\tau,f^*\rangle\big] \\
&= \Psi(f) - \Psi(f^*)
   - \langle\tau, f - f^*\rangle.
\end{align*}
Using \(\tau = \nabla\Psi(f^*)\) from (i), we obtain
\[
\mathrm{Regret}(f,Q)
= \Psi(f) - \Psi(f^*)
  - \big\langle\nabla\Psi(f^*),\,f - f^*\big\rangle
= D_{\Psi}\big(f^*(Q)\,\|\,f\big),
\]
which is precisely the Bregman divergence induced by \(\Psi\).
\end{proof}

\begin{proof}[Proof of Lemma~\ref{lem:vertex-reduction-functional}]
Fix \(f\in\mathcal{F}\) and write \(\theta := \nabla\Psi(f)\). We have 
\[
\mathrm{Regret}(f,Q)
= D_{\Psi^*}(\theta \,\|\, \tau_Q)
\quad\text{for each } Q\in\mathcal{C}(Q_{\X}).
\]
Hence
\[
\max_{Q\in\mathcal{C}(Q_{\X})} \mathrm{Regret}(f,Q)
= \max_{Q\in\mathcal{C}(Q_{\X})} D_{\Psi^*}(\theta \,\|\, \tau_Q).
\]

For fixed \(\theta\), define
\[
g(b) := D_{\Psi^*}(\theta\|b), \qquad b\in L_2(Q_{\X}).
\]
Using the definition
\[
D_{\Psi^*}(\theta\|b)
= \Psi^*(b) - \Psi^*(\theta)
  - \big\langle\nabla\Psi^*(\theta),\,b-\theta\big\rangle,
\]
we see that \(g\) is the sum of the convex function \(\Psi^*(b)\), a constant in
\(b\), and an affine term in \(b\), hence \(g\) is convex.

By construction of \(\mathcal{C}(Q_{\X})\), any \(Q\in\mathcal{C}(Q_{\X})\) has
\(\tau_Q\) in the convex hull of \(\{\tau_s\}_{s=1}^S\), i.e.
\[
\tau_Q = \sum_{s=1}^S q_s\,\tau_s
\quad\text{for some } q=(q_1,\dots,q_S)\in\Delta_{S-1}.
\]
Applying Jensen’s inequality to \(g\),
\[
g(\tau_Q)
= g\Big(\sum_{s=1}^S q_s\tau_s\Big)
\le \sum_{s=1}^S q_s\,g(\tau_s)
\le \max_{1\le s\le S} g(\tau_s).
\]
Taking the maximum over all \(Q\in\mathcal{C}(Q_{\X})\) gives
\[
\max_{Q\in\mathcal{C}(Q_{\X})} D_{\Psi^*}(\theta\|\tau_Q)
\le \max_{1\le s\le S} D_{\Psi^*}(\theta\|\tau_s).
\]

For the reverse inequality, note that each extreme point \(P^{(s)}\) of the
mixture set belongs to \(\mathcal{C}(Q_{\X})\) and satisfies
\(\tau_{P^{(s)}} = \tau_s\). Thus
\[
\max_{Q\in\mathcal{C}(Q_{\X})} D_{\Psi^*}(\theta\|\tau_Q)
\ge D_{\Psi^*}(\theta\|\tau_s)
\quad\text{for all } s,
\]
and hence
\[
\max_{Q\in\mathcal{C}(Q_{\X})} D_{\Psi^*}(\theta\|\tau_Q)
\ge \max_{1\le s\le S} D_{\Psi^*}(\theta\|\tau_s).
\]
Combining the two inequalities yields
\[
\max_{Q\in\mathcal{C}(Q_{\X})} D_{\Psi^*}(\theta\|\tau_Q)
= \max_{1\le s\le S} D_{\Psi^*}(\theta\|\tau_s),
\]
which is the first claimed equality. The equivalent representation in terms of
\(D_{\Psi}\) follows from the conjugacy identity
\(D_{\Psi}(f_s^*\|f)=D_{\Psi^*}(\theta\|\tau_s)\) applied to
\(u=f_s^*(P^{(s)}), v=f\).
\end{proof}

\begin{proof}[Proof of Theorem~\ref{thm:bregman-center-functional}]
Consider the constrained form
\[
\min_{\theta,R} R
\quad\text{s.t.}\quad D_{\Psi^*}(\theta\|\tau_s)\le R,\; s=1,\dots,S.
\]
Introduce Lagrange multipliers \(\lambda_s\ge 0\) for the constraints
\(D_{\Psi^*}(\theta\|\tau_s)\le R\). The Lagrangian is
\[
\mathcal{L}(\theta,R;\blambda)
= R + \sum_{s=1}^S \lambda_s\big(D_{\Psi^*}(\theta\|\tau_s) - R\big)
= R\Big(1-\sum_{s=1}^S \lambda_s\Big)
  + \sum_{s=1}^S \lambda_s D_{\Psi^*}(\theta\|\tau_s).
\]
At any optimal solution we must have \(\sum_{s=1}^S \lambda_s = 1\); otherwise
the infimum over \(R\) would be \(-\infty\). Thus we may restrict to
\(\blambda \in \R_+^S\) with \(\sum_s \lambda_s = 1\), i.e.\ \(\blambda\in\Delta_{S-1}\).
Writing \(\gamma_s := \lambda_s\) and \(\bgamma=(\gamma_1,\dots,\gamma_S)\), the
Lagrangian reduces to
\[
\mathcal{L}(\theta,R;\blambda)
= \sum_{s=1}^S \gamma_s D_{\Psi^*}(\theta\|\tau_s).
\]
By standard saddle-point arguments for convex problems,
\[
R^*
= \min_{\theta\in L_2(Q_{\X})}\max_{1\le s\le S} D_{\Psi^*}(\theta\|\tau_s)
= \min_{\theta\in L_2(Q_{\X})}\max_{\bgamma\in\Delta_{S-1}}
   \sum_{s=1}^S \gamma_s D_{\Psi^*}(\theta\|\tau_s),
\]
and the order of min and max can be exchanged:
\[
R^*
= \max_{\bgamma\in\Delta_{S-1}} \min_{\theta\in L_2(Q_{\X})}
   \sum_{s=1}^S \gamma_s D_{\Psi^*}(\theta\|\tau_s).
\]

For fixed \(\bgamma\), expand the inner objective:
\begin{align*}
\sum_{s=1}^S \gamma_s D_{\Psi^*}(\theta\|\tau_s)
&= \sum_{s=1}^S \gamma_s\Big[
      \Psi^*(\tau_s) - \Psi^*(\theta)
      - \big\langle\nabla\Psi^*(\theta),\,\tau_s-\theta\big\rangle
   \Big] \\
&= \sum_{s=1}^S \gamma_s \Psi^*(\tau_s)
   - \Psi^*(\theta)
   - \Big\langle\nabla\Psi^*(\theta),\,
               \sum_{s=1}^S \gamma_s\tau_s - \theta\Big\rangle.
\end{align*}
Define the mixture effect function
\[
\overline{\tau}(\bgamma) := \sum_{s=1}^S \gamma_s\tau_s.
\]
Then the term in brackets can be recognized as
\[
\Psi^*(\theta)
+ \big\langle\nabla\Psi^*(\theta),\,\overline{\tau}(\bgamma)-\theta\big\rangle
= \Psi^*(\overline{\tau}(\bgamma)) + D_{\Psi^*}(\theta\|\overline{\tau}(\bgamma)),
\]
so
\[
\sum_{s=1}^S \gamma_s D_{\Psi^*}(\theta\|\tau_s)
= \sum_{s=1}^S \gamma_s \Psi^*(\tau_s)
  - \Psi^*(\overline{\tau}(\bgamma))
  - D_{\Psi^*}(\theta\|\overline{\tau}(\bgamma)).
\]
For fixed \(\bgamma\), the last term is minimized at
\(\theta = \overline{\tau}(\bgamma)\) with minimum value zero, yielding
\[
\min_{\theta\in L_2(Q_{\X})} \sum_{s=1}^S \gamma_s D_{\Psi^*}(\theta\|\tau_s)
= \sum_{s=1}^S \gamma_s \Psi^*(\tau_s)
  - \Psi^*\big(\overline{\tau}(\bgamma)\big)
= \mathcal{H}(\bgamma).
\]
Hence
\[
R^* = \max_{\bgamma\in\Delta_{S-1}} \mathcal{H}(\bgamma),
\]
which establishes (i) and (iii), and shows that any maximizer \(\bgamma^*\)
induces an optimal center
\[
\theta^* = \overline{\tau}(\bgamma^*)
= \sum_{s=1}^S \gamma_s^* \tau_s,
\]
proving (ii). Part (iv) follows from the relation
\(f^*_{\mathrm{regret}} = \nabla\Psi^*(\theta^*)\) and the equivalence between
the dual and primal formulations of the minimax problem established earlier.
Finally, complementary slackness in the KKT conditions implies that if
\(\gamma_s^*>0\), then the corresponding constraint is tight, that is
\(D_{\Psi^*}(\theta^*\|\tau_s)=R^*\), proving (v).
\end{proof}

\section{Proofs of Results in the Main Text}\label{appendix:proof}

\paragraph{Additional Notation.}
The $\ell_q$ norm of a $p$-dimensional vector $v$ is defined as $\|v\|_q=\left(\sum_{l=1}^p\left|v_l\right|^q\right)^{\frac{1}{q}}$ for $q \geq 0$ and $\|v\|_{\infty}=\max _{1 \leq l \leq p}\left|v_l\right|$. For a matrix $A$, we use $\lambda_j(A),\|A\|_F$, $\|A\|_2$ and $\|A\|_{\infty}$ to denote its $j$-th largest singular value, Frobenius norm, spectral norm, and element-wise maximum norm, respectively. 

\subsection{Proof of Proposition~\ref{lemma:euqival:regretOPT}}
\label{app:lemma:equival}

The proof of Equation~\eqref{prop:second} is included in Appendix~\ref{app:pf:ident:regret} as an intermediate result for proving Theorem~\ref{thm:ident:regret}. Here, we prove Equation~\eqref{equ:equivalConstrainedOPT}.

\begin{proof}
By definition of minimax optimization, Equation~\eqref{equ:regret:distance} is equivalent to 
\begin{equation}\label{equ:equival:infinite:constrained}
    \begin{aligned}
        (f_{\operatorname{regret}}^*, R^*) := \ &\underset{f \in \mathcal{F}, R\in\mathbb{R}}{\arg \min}\ R \\
        &\text{subject to} \quad \mathbb{E}_{Q_{\X}}[(f(\X) - \tau_Q(\X))^2] \leq R \quad \text{for all } Q \in \mathcal{C}\left(Q_{\X}\right).
    \end{aligned}
\end{equation}
Since each target CATE can be expressed as a weighted average of site-specific CATEs, we only need to show that for any given $R$ and $f$,
\begin{equation*}
    \begin{aligned}
    & \left \{ \mathbb{E}_{Q_{\X}}\left[\left(f(\X) - \sum_{s=1}^S q_s \cdot\tau^{(s)}(\X)\right)^2\right] \leq R ,  \quad \q \in  \Delta_{S-1}\right\}\\
    = \ & 
    \left \{ \mathbb{E}_{Q_{\X}}[(f(\X) - \tau^{(s)}(\X))^2] \leq R , \quad s \in [S] \right\}.
    \end{aligned}
\end{equation*}
It then suffices to show that, for any $ \q \in  \Delta_{S-1}$, the following holds:
\[
\mathbb{E}_{Q_{\X}}\left[\left(f(\X) - \sum_{s=1}^S q_s \cdot\tau^{(s)}(\X)\right)^2\right] \leq \max_{s\in[S]}\  \mathbb{E}_{Q_{\X}}[(f(\X) - \tau^{(s)}(\X))^2].
\]
We can show this as follows:
\begin{equation*}
    \begin{aligned}
        \mathbb{E}_{Q_{\X}}\left[\left(f(\X) - \sum_{s=1}^S q_s \cdot\tau^{(s)}(\X)\right)^2\right]  & =  \mathbb{E}_{Q_{\X}}\left[f^2(\X)-2f(\X)\sum_{s=1}^S q_s \cdot\tau^{(s)}(\X)\right]+\mathbb{E}_{Q_{\X}}\left[\left(\sum_{s=1}^S q_s \cdot\tau^{(s)}(\X)\right)^2\right] \\
        & \leq \mathbb{E}_{Q_{\X}}\left[f^2(\X)-2f(\X)\sum_{s=1}^S q_s \cdot\tau^{(s)}(\X)\right]+ \mathbb{E}_{Q_{\X}}\left[\sum_{s=1}^S q_s \cdot\left(\tau^{(s)}(\X)\right)^2\right] \\
        & = \sum_{s=1}^S q_s\cdot\mathbb{E}_{Q_{\X}}[(f(\X) - \tau^{(s)}(\X))^2] \\
        & \leq \max_{s\in[S]}\  \mathbb{E}_{Q_{\X}}[(f(\X) - \tau^{(s)}(\X))^2] ,
    \end{aligned}
\end{equation*}
where the inequality in the second line is due to $\q \in  \Delta_{S-1}$ and the Cauchy-Schwarz inequality, which ensures that $\left(\sum_{s=1}^S q_s \cdot \tau^{(s)}(\X)\right)^2 \leq \left(\sum_{s=1}^S q_s\right) \left(\sum_{s=1}^S q_s \cdot \left(\tau^{(s)}(\X)\right)^2\right)$ holds pointwise for every value of $\X$.
Thus, we have shown that Equation~\eqref{equ:equival:infinite:constrained} is equivalent to Equation~\eqref{equ:equivalConstrainedOPT}.
\end{proof}

\subsection{Proof of Theorem~\ref{thm:ident:regret}}\label{app:pf:ident:regret}

Since the model class $\mathcal{F}$ contains the true CATE models $\tau_Q(\cdot)$ for all $Q\in\mathcal{C}\left(Q_{\X}\right)$, as noted at the beginning of Section~\ref{subsec:idenresults}, the original minimax regret optimization problem in Equation~\eqref{equ:HTE:obj:regret} can be equivalently reformulated as the minimax MSE optimization problem in Equation~\eqref{equ:regret:distance}.
Therefore, by the first part of Proposition~\ref{lemma:euqival:regretOPT}, we can focus on solving the optimization problem given in Equation~\eqref{equ:equivalConstrainedOPT}, which we refer to as the primal problem. 

To complete the proof, we leverage the results from \cite{elzinga1972minimum}, which provide a closed-form solution for the special case where $\mathcal{F}$ is a linear model class.
We extend their proof by using functional analysis so that we can accommodate a more general model class $\mathcal{F}$ within the $L_2(Q_{\X})$ Hilbert space.

\begin{proof}
We begin by noting that Equation~\eqref{equ:equivalConstrainedOPT} is a convex optimization problem. 
To see this clearly, we consider $\lambda\in[0,1]$ and any $f_1,f_2\in\mathcal{F}$. We can verify that $\mathbb{E}_{Q_{\X}}[(\lambda f_1(\X)+(1-\lambda)f_2(\X) - \tau^{(s)}(\X))^2] \leq \lambda \mathbb{E}_{Q_{\X}}[(f_1(\X) - \tau^{(s)}(\X))^2]+(1-\lambda)\mathbb{E}_{Q_{\X}}[(f_2(\X) - \tau^{(s)}(\X))^2]$ for all $s \in [S]$.
Moreover, the Slater constraint qualification is clearly satisfied since there exist feasible $f_0\in\mathcal{F}$ and $R_0\in\mathbb{R}$ that strictly satisfy the inequality constraint.

Therefore, the Karush-Kuhn-Tucker conditions give a satisfying characterization of both necessary and sufficient optimality conditions for Equation~\eqref{equ:equivalConstrainedOPT}.
Specifically, there exist multipliers $q^*_1,\ldots,q_S^*$ such that
\begin{align}
    \sum_{s=1}^{S} q^*_s & = 1 \label{eq:stationarity_sum}\\
    \sum_{s=1}^{S} q^*_s \cdot (f_{\text{regret}}^*(\cdot) - \tau^{(s)}(\cdot)) & = 0 \label{eq:stationarity_expectation} \\
    q^*_s \cdot (\mathbb{E}_{Q_{\X}}[(f_{\text{regret}}^*(\X) - \tau^{(s)}(\X))^2] - R^*) & = 0 \quad \text{for} \ s \in[S] \label{eq:complementary_slackness}\\
    \mathbb{E}_{Q_{\X}}[(f_{\text{regret}}^*(\X) - \tau^{(s)}(\X))^2]-R^* & \leq  0 \quad \text{for} \ s \in[S]\label{eq:primal_feasibility}\\
    q^*_s & \geq 0 \quad \text{for} \ s \in [S]\label{eq:dual_feasibility}
\end{align}
where Equation~\eqref{eq:stationarity_expectation} is obtained by setting the functional derivative of the Lagrangian associated with Equation~\eqref{equ:equivalConstrainedOPT} with respect to $f(\cdot)$ to zero.
These conditions indicate that $f_{\text{regret}}^*(\cdot)$ will be a convex combination of $\{\tau^{(s)}(\cdot)\}_{s=1}^S$, and $q^*_s=0$ for any $s\in[S]$ such that $\left\|f_{\text{regret}}^*(\X)-\tau^{(s)}(\X)\right\|^2_{Q_{\X},2}$ is strictly smaller than $R^*$. 
Since $\mathcal{F}$ is convex and contains $\{\tau^{(s)}(\cdot)\}_{s=1}^S$, the optimal solution $(f_{\text{regret}}^*(\cdot),R^*)$ exists.

To solve the primal problem in Equation~\eqref{equ:equivalConstrainedOPT}, we write out its Wolfe dual problem \citep{wolfe1961duality}, which is given by:
\begin{align}
    \underset{R\in\mathbb{R},f\in\mathcal{F},\q }{\max }  & \quad R+\sum_{s=1}^S q_s \cdot\left( \mathbb{E}_{Q_{\X}}[(f(\X) - \tau^{(s)}(\X))^2]-R\right), \label{equ:wolfe:obj}\\
    \text{subject to} \quad 
    & \sum_{s=1}^S q_s = 1, \label{equ:wolfe:sumq}\\
    & \sum_{s=1}^{S} q_s \cdot (f(\cdot) - \tau^{(s)}(\cdot))  = 0\label{equ:wolfe:f} \\
    & q_s \geq 0 \quad \text{for} \quad s = 1, \ldots, S.
\end{align}
We will then show that this Wolfe dual problem is equivalent to a quadratic programming (QP) problem.
First, rewrite the objective of the Wolfe dual problem as,
\begin{equation*}
    \begin{aligned}
    & \sum_{s=1}^S q_s \cdot \mathbb{E}_{Q_{\X}}[(f(\X) - \tau^{(s)}(\X))^2]\\
    = \ & \sum_{s=1}^S q_s \cdot \mathbb{E}_{Q_{\X}}[f^2(\X)] - 2\mathbb{E}_{Q_{\X}}\left[f(\X) \sum_{s=1}^S q_s \tau^{(s)}(\X)\right] +\sum_{s=1}^S q_s \cdot \E_{Q_{\X}}[\tau^{(s)}(\X)^2]\\
    = \ & \sum_{s=1}^S q_s\cdot\E_{Q_{\X}}[\tau^{(s)}(\X)^2] -\E_{Q_{\X}}\left[ \left(\sum_{s=1}^S q_s\tau^{(s)}(\X)\right)^2\right]
    \end{aligned}
\end{equation*}
where the second equality follows from Equations~\eqref{equ:wolfe:sumq}~and~\eqref{equ:wolfe:f}.
Since the objective function in Equation~\eqref{equ:wolfe:obj} is independent of $R$ for any feasible $\q$ due to the constraint in Equation~\eqref{equ:wolfe:sumq}, we may choose $R$ arbitrarily.
Thus, the above Wolfe dual problem is equivalent to the following QP problem:
\begin{align}
   \underset{\q}{\max} \quad &
\sum_{s=1}^S q_s \cdot \mathbb{E}_{Q_{\X}}[\tau^{(s)}(\X)^2] 
- \mathbb{E}_{Q_{\X}}\left[ \left(\sum_{s=1}^S q_s \tau^{(s)}(\X)\right)^2 \right], \label{eq:QP_objective} \\
    \text{subject to} \quad 
    & \sum_{s=1}^{S} q_s = 1, \label{eq:QP_constraint_sum} \\
    & q_s \geq 0 \quad \text{for} \quad s = 1, \ldots, S, \label{eq:QP_non_negativity} \\
    \text{with} \quad & f(\cdot) = \sum_{s=1}^{S} q_s \cdot \tau^{(s)}(\cdot), \label{eq:QP_function_f} \\
    \qquad \qquad & R = \sum_{s=1}^{S} q_s \cdot \mathbb{E}_{Q_{\X}}[(f(\X) - \tau^{(s)}(\X))^2]. \label{eq:QP_function_R}
\end{align}
The particular choice in Equation~\eqref{eq:QP_function_R} will later ensure that solving the QP problem solves the primal problem.
This QP optimization problem is concave because the objective can be written as $-(\q^\top \Gamma \q - \q^\top \d)$ and $\Gamma$ is positive semidefinite.

Next, we prove that the solution to the above QP problem is also the solution to the primal problem given in Equation~\eqref{equ:equivalConstrainedOPT}.
Suppose $(\q^*,f^*,R^*)$ solves the QP problem. 
It suffices to show that $(\q^*,f^*,R^*)$ satisfies the KKT conditions of the primal problem.
Equations~\eqref{eq:stationarity_sum},~\eqref{eq:stationarity_expectation}~and~\eqref{eq:dual_feasibility} are directly satisfied by Equations~\eqref{eq:QP_constraint_sum},~\eqref{eq:QP_function_f}~and~\eqref{eq:QP_non_negativity}, respectively.
We leverage the KKT conditions for the QP problem, which guarantees the existence of an unconstrained multiplier $w^*$ and constrained multipliers $\{u^*_s\}_{s=1}^S$ such that:
\begin{align}
  2\E\left[\sum_{s=1}^S q^*_s \tau^{(s)}(\X)\cdot\tau^{(s')}(\X)\right] - \E\left[\tau^{(s')}(\X)^2\right] + w^* - u^*_{s'} & = 0 \quad \text{for} \quad s' = 1, \ldots, S, \label{eq:QPdual_stationarity} \\
  q^*_{s'} u^*_{s'} & = 0 \quad \text{for} \quad s' = 1, \ldots, S, \label{eq:QPdual_complementary_slackness} \\
  u^*_{s'} & \geq 0 \quad \text{for} \quad s' = 1, \ldots, S. \label{eq:QPdual_dual_feasibility}
\end{align}

Using Equation~\eqref{eq:QP_function_f}, we rewrite Equation~\eqref{eq:QPdual_stationarity} as,
\begin{equation}\label{eq:QPdual1}
   \left\|f^*(\X)-\tau^{(s')}(\X)\right\|^2_{Q_{\X},2} -\E\left[f^*(\X)^2\right] - w^* + u^*_{s'} = 0 \quad \text{for} \quad s' = 1, \ldots, S.
\end{equation}
In addition, after multiplying Equation~\eqref{eq:QPdual_stationarity} by $q^*_{s'}$ and summing it over $s'$, we use  
Equation~\eqref{eq:QP_function_R} to obtain,
\begin{equation}\label{eq:QPdual2}
    w^*= R^*-\E\left[f^*(\X)^2\right].
\end{equation}
Combining Equations~\eqref{eq:QPdual1}~and~\eqref{eq:QPdual2}, we obtain,
\begin{equation}
    u^*_{s'}= R^*-    \left\|f^*(\X)-\tau^{(s')}(\X)\right\|^2_{Q_{\X},2}, \quad s' = 1, \ldots, S.
\end{equation}
Due to Equations~\eqref{eq:QPdual_complementary_slackness}~and~\eqref{eq:QPdual_dual_feasibility}, the primal KKT conditions, i.e., Equations~\eqref{eq:complementary_slackness}~and~\eqref{eq:primal_feasibility} hold.
Therefore, if $(\q^*,f^*,R^*)$ solves the QP problem, then  $(f^*,R^*)$ solves the primal problem given in Equation~\eqref{equ:equivalConstrainedOPT}.
\end{proof}

\subsection{Proof of Corollary~\ref{corollary:ident:regret:polytopes}} \label{app:proof_coro}

By definition, a polytope can be represented as the convex hull of a finite set of extreme points, corresponding to its vertices.
Therefore, the uncertainty set $\mathcal{C}\left(Q_{\X}, \mathcal{H}\right)$ can be expressed as a convex hull of $N$ synthetic CATEs: $\{ \g_1^\top \btau, \ldots, \g_N^\top \btau \}$, where each CATE is a convex combination of the original $S$ source sites CATEs.
The result then follows directly from the proof of Theorem~\ref{thm:ident:regret} by treating the $N$ synthetic CATEs as the new set of source CATEs.
\qed

\subsection{Proof of Theorem~\ref{thm:l2error:regret}}\label{app:pf:Thm2}

\begin{proof}
According to the definition of $\hat{f}_{\operatorname{regret}}$ in Equation\eqref{equ:estor:regret}, we have that
\begin{equation*}
    \begin{aligned}
       \left( \hat{f}_{\operatorname{regret}}(\cdot)-f^*_{\operatorname{regret}}(\cdot) \right)^2 & = \left( \sum_{s=1}^S \hat{q}_s \cdot \hat{\tau}^{(s)}(\cdot) -  \sum_{s=1}^S q^*_s \cdot {\tau}^{(s)}(\cdot)  \right)^2 \\
       & = \left( \sum_{s=1}^S \hat{q}_s \cdot \Delta^{(s)}(\cdot) -  \sum_{s=1}^S (q^*_s-\hat{q}_s) \cdot {\tau}^{(s)}(\cdot)  \right)^2 
    \end{aligned}
\end{equation*}
where $\Delta^{(s)}(\cdot):=\hat{\tau}^{(s)}(\cdot)-\tau^{(s)}(\cdot)$ denotes the estimation error of $\hat{\tau}^{(s)}$. It follows that 
\begin{equation*}
    \left( \sum_{s=1}^S \hat{q}_s \cdot \Delta^{(s)}(\cdot) -  \sum_{s=1}^S (q^*_s-\hat{q}_s) \cdot {\tau}^{(s)}(\cdot)  \right)^2  \leq 2\left(\left(\sum_{s=1}^S \hat{q}_s \cdot \Delta^{(s)}(\cdot) \right)^2+\left( \sum_{s=1}^S (q^*_s-\hat{q}_s) \cdot {\tau}^{(s)}(\cdot) \right)^2\right).
\end{equation*}
We then apply the Cauchy-Schwarz inequality to both terms on the right-hand side of the above inequality and establish that
$$
\left(\sum_{s=1}^S \hat{q}_s \cdot \Delta^{(s)}(\cdot) \right)^2 \leq \sum_{s=1}^S \hat{q}_s^2 \sum_{s=1}^S \left(\Delta^{(s)}(\cdot)\right)^2 \leq \sum_{s=1}^S \left(\Delta^{(s)}(\cdot)\right)^2,
$$
where the last inequality holds because $\{\hat{q}_s\}_{s=1}^S \in \Delta_{S-1}$. Furthermore:
$$
\left( \sum_{s=1}^S (q^*_s-\hat{q}_s) \cdot {\tau}^{(s)}(\cdot) \right)^2 \leq
\left\|\hat{\q}-\q^*\right\|_2^2 \sum_{s=1}^S  \left({\tau}^{(s)}(\cdot)\right)^2 .
$$
Then, we can quantify the error of $\left\|\hat{f}_{\operatorname{regret}}-f^*_{\operatorname{regret}}\right\|_{Q, 2}$ as follows,
\begin{equation}\label{equ:proof:ineqmain}
 \begin{aligned}
\left\|\hat{f}_{\operatorname{regret}}-f^*_{\operatorname{regret}}\right\|_{Q, 2} & =\sqrt{\mathbb{E}_{Q}\left(\hat{f}_{\operatorname{regret}}(\X)-f^*_{\operatorname{regret}}(\X)\right)^2} \\
& \leq
\sqrt{2\left(\sum_{s=1}^S\E_{Q}\left[\Delta^{(s)}(\X)^2\right]+\left\|\hat{\q}-\q^*\right\|_2^2 \sum_{s=1}^S \E_{Q}\left[{\tau}^{(s)}(\X)^2\right]\right)} \\
& \leq \sqrt{2 S\left(\delta_n^2+\left\|\hat{\q}-\q^*\right\|_2^2 4M^2\right)} \\
& \leq \sqrt{2 S}\left(\delta_n+\left\|\hat{\q}-\q^*\right\|_2 2M \right),
\end{aligned}   
\end{equation}
where the second inequality is due to Assumptions~\ref{ass:bdd:outcome}~and~\ref{ass:CATE:rate}, and the last inequality holds because both $M$ and $\delta_n$ are non-negative.

To upper bound the term $\left\|\hat{\q}-\q^*\right\|_2$, we will use the following lemma, which is adapted from Lemma 6 in \cite{guo2023statistical} and Lemma 1 in \cite{wang2023distributionally}.
For completeness, we include its proof in Appendix~\ref{app:lemma:proof}.
\begin{lemma}[Estimated weights for minimax-regret]\label{lemma:q:regret}
Define the diameter of the simplex $\Delta_{S-1}$ as $\rho_{\Delta_{S-1}}=\max _{\q, \q^{\prime} \in \Delta_{S-1}}\left\|\q-\q^{\prime}\right\|_2$.
Recall the definitions of  $\q^*=\argmin_{\q \in \Delta_{S-1}} \; \q^\top \Gamma \q - \q^\top \d$ in Equation~\eqref{equ:iden:regret} and $\hat{\q}=\argmin_{\q \in \Delta_{S-1}} \; \q^{\top} \widehat{\Gamma} \q-\q^{\top} \hat{\d}$ in Equation~\eqref{equ:estor:regret}.
If $\lambda_{\min }(\Gamma)>0$, then
$$
\left\|\hat{\q}-\q^*\right\|_2 \leq \frac{S(\|\widehat{\Gamma}-\Gamma\|_{\infty}+\frac{1}{2}\|\hat{\d}-\d\|_{\infty} ) }{\lambda_{\min }(\Gamma)} \wedge \rho_{\Delta_{S-1}},
$$
where $\|\widehat{\Gamma}-\Gamma\|_{\infty}$ is the element-wise maximum norm of a matrix, and $\|\hat{\d}-\d\|_{\infty}$ is the $\ell_\infty$ norm of a vector.
\end{lemma} 

By Lemma~\ref{lemma:q:regret}, we only need to bound the terms $\|\widehat{\Gamma}-\Gamma\|_{\infty}$ and $\|\hat{\d}-\d\|_{\infty}$.
For fixed $k, l \in[S]$, we decompose the error of $\widehat{\Gamma}$ into two parts as follows:
$$
\begin{aligned}
& (\widehat{\Gamma}-\Gamma)_{k, l} \\
= \ &  \frac{1}{n_Q} \sum_{i=1}^{n_Q} \hat{\tau}^{(k)}(\X_i)\hat{\tau}^{(l)}(\X_i)-\mathbb{E}_{Q_{\X}}\left[\tau^{(k)}(\X) \tau^{(l)}\left(\X\right) \right] \\
= \ & \underbrace{\frac{1}{n_Q} \sum_{i=1}^{n_Q}\left(\hat{\tau}^{(k)}(\X_i)\hat{\tau}^{(l)}(\X_i)-{\tau}^{(k)}(\X_i){\tau}^{(l)}(\X_i)\right)}_{(\mathrm{I})_{k, l}} +
\underbrace{\frac{1}{n_Q} \sum_{i=1}^{n_Q}\left({\tau}^{(k)}(\X_i){\tau}^{(l)}(\X_i)-\mathbb{E}_{Q_{\X}}\left[\tau^{(k)}(\X) \tau^{(l)}\left(\X\right) \right]\right) }_{(\mathrm{II})_{k, l}}. 
\end{aligned}
$$

We will upper-bound each term.
For the term $(\mathrm{I})_{k, l}$, which represents the estimation error of $\tau^{(k)}$ and $\tau^{(l)}$, we further decompose it into the following components,
\begin{equation}\label{equ:biasdecom2}
    \begin{aligned}
   \left|(\mathrm{I})_{k, l}\right| & = \left|\frac{1}{n_Q} \sum_{i=1}^{n_Q}{\tau}^{(k)}(\X_i)\Delta^{(l)}(\X_i)
     +    \frac{1}{n_Q} \sum_{i=1}^{n_Q} {\tau}^{(l)}(\X_i)\Delta^{(k)}(\X_i)
    + \frac{1}{n_Q} \sum_{i=1}^{n_Q}\Delta^{(k)}(\X_i)\Delta^{(l)}(\X_i)\right|\\
    & \leq  \left|\frac{1}{n_Q} \sum_{i=1}^{n_Q}{\tau}^{(k)}(\X_i)\Delta^{(l)}(\X_i)\right|
     +  \left|   \frac{1}{n_Q} \sum_{i=1}^{n_Q} {\tau}^{(l)}(\X_i)\Delta^{(k)}(\X_i)\right|
    + \left| \frac{1}{n_Q} \sum_{i=1}^{n_Q}\Delta^{(k)}(\X_i)\Delta^{(l)}(\X_i)\right|\\
    & \leq 2M\sqrt{\frac{1}{n_Q} \sum_{i=1}^{n_Q} \left(\Delta^{(l)}(\X_i)\right)^2}
    +2M\sqrt{\frac{1}{n_Q} \sum_{i=1}^{n_Q} \left(\Delta^{(k)}(\X_i)\right)^2}\\
    & \qquad + \sqrt{\frac{1}{n_Q} \sum_{i=1}^{n_Q} \left(\Delta^{(l)}(\X_i)\right)^2} \sqrt{\frac{1}{n_Q} \sum_{i=1}^{n_Q} \left(\Delta^{(k)}(\X_i)\right)^2}.
    \end{aligned}
\end{equation}
where the first inequality is due to the triangle inequality, and the second inequality is derived from the Cauchy-Schwarz inequality. Next, we apply Markov's inequality to further upper-bound the terms in the above inequality. For any $t>1$, the following inequality holds due to Assumption~\ref{ass:CATE:rate}:
\begin{equation*}
\mathbb{P}\left\{\frac{1}{n_Q} \sum_{i=1}^{n_Q} \left(\Delta^{(l)}(\X_i)\right)^2 \geq t S \delta_n^2\right\} \leq \frac{\mathbb{E}_{Q}\left[\Delta^{(l)}(\X)^2\right]}{ t S \delta_n^2} \leq \frac{1}{ t S} .
\end{equation*}
Then, by union bound, we have 
\begin{equation*}
\mathbb{P}\left(\max_{l\in[S]}\left\{ \frac{1}{n_Q} \sum_{i=1}^{n_Q} \left(\Delta^{(l)}(\X_i)\right)^2 \right\} \leq t S \delta_n^2\right) \geq 1-S\cdot\frac{1}{ t S}=1-\frac{1}{t}.
\end{equation*}
Therefore, with probability at least $1-\frac{1}{t}$, we have,
\begin{equation}\label{equ:proof:term1}
\max _{k, l}\left|(\mathrm{I})_{k, l}\right| \leq 4 M \sqrt{t S}\cdot\delta_n+tS\cdot\delta_n^2 .
\end{equation}

Next, for the term $(\mathrm{II})_{k, l}$, we apply Chebyshev's inequality to upper-bound the term. For any $c>0$,
\begin{equation}\label{equ:proof:cheb}
\begin{aligned}
  \Prob\left(  \left|\frac{1}{n_Q} \sum_{i=1}^{n_Q}{\tau}^{(k)}(\X_i){\tau}^{(l)}(\X_i)-\mathbb{E}_{Q_{\X}}\left[\tau^{(k)}(\X) \tau^{(l)}\left(\X\right) \right]\right|\geq c\right)   &  \leq \frac{\Var\left[\frac{1}{n_Q} \sum_{i=1}^{n_Q}{\tau}^{(k)}(\X_i){\tau}^{(l)}(\X_i)\right]}{c^2}  \\
  & \leq \frac{\E\left[\left({\tau}^{(k)}(\X_i){\tau}^{(l)}(\X_i)\right)^2\right]}{n_Q c^2}\leq \frac{16M^4}{n_Q c^2}.
\end{aligned}
\end{equation}
For $t>1$, we take $c=  4M^2t S / \sqrt{n_Q}$, then the above inequality becomes:
$$
 \Prob\left(  \left|\frac{1}{n_Q} \sum_{i=1}^{n_Q}{\tau}^{(k)}(\X_i){\tau}^{(l)}(\X_i)-\mathbb{E}_{Q_{\X}}\left[\tau^{(k)}(\X) \tau^{(l)}\left(\X\right) \right]\right| \geq 4t S \frac{M^2}{\sqrt{n_Q}}\right) \leq \frac{1}{t^2 S^2} .
$$
Then, by union bound, with probability at least $1-\frac{1}{t^2}$,
\begin{equation}\label{equ:proof:term2}
 \max _{k, l}\left|(\mathrm{II})_{k, l}\right| \leq \frac{4t S M^2}{\sqrt{n_Q}}.
\end{equation}
Combining the inequalities in Equations~\eqref{equ:proof:term1}~and~\eqref{equ:proof:term2}, we establish that, with probability at least $1-\frac{1}{t}-\frac{1}{t^2}$,
\begin{equation*}
   \|\widehat{\Gamma}-\Gamma\|_{\infty} \leq \left(\max _{k, l}\left|(\mathrm{I})_{k, l}\right|  + \max _{k, l}\left|(\mathrm{II})_{k, l}\right| \right) \leq  
t S\left(\frac{4 M \delta_n}{\sqrt{t S}}+\delta_n^2  +\frac{4M^2}{\sqrt{n_Q}}\right).
\end{equation*}
In the meantime, the following inequality also holds:
\begin{equation*}
     \|\hat{\d}-\d\|_{\infty}= \max_{s\in [S]} \left|\widehat{\Gamma}_{s,s}-\Gamma_{s,s}  \right| \leq
 t S\left(\frac{4 M \delta_n}{\sqrt{t S}}+\delta_n^2  +\frac{4M^2}{\sqrt{n_Q}}\right).
\end{equation*}



Combining the above inequalities with Lemma~\ref{lemma:q:regret}, we obtain that, with probability at least $1-\frac{1}{t}-\frac{1}{t^2}$,
\begin{equation*}
 \begin{aligned}
\left\|\hat{f}_{\operatorname{regret}}-f^*_{\operatorname{regret}}\right\|_{Q, 2} \leq  \sqrt{2 S}\left(\delta_n+ \left(\frac{1.5 tS^2\left(\frac{4 M \delta_n}{\sqrt{t S}}+\delta_n^2  +\frac{4M^2}{\sqrt{n_Q}}\right)}{\lambda_{\min }(\Gamma)} \wedge \rho_{\mathcal{H}} \right) 2M \right).
\end{aligned}   
\end{equation*}
\end{proof}

\subsection{Proof of Proposition~\ref{prop:ident:baselineloss}}\label{app:pf:ident:base}

We generalize the proof of Theorem~1 of \cite{wang2023distributionally}, which considers a special case of $f_{\operatorname{base}}=0$.

\begin{proof}
For a target distribution $Q \in \mathcal{C}\left(Q_{\X},\mathcal{H}\right)$, define the population relative-risk function for the function $f(\cdot)$ as:
\[
\begin{aligned}
  R_Q(f):&=  \mathbb{E}_Q[(Y(1)-Y(0)-f(\X))^2-(Y(1)-Y(0)-f_{\operatorname{base}}(\X))^2 ]\\
  &=\mathbb{E}_Q[-2(Y(1)-Y(0))(f(\X)-f_{\operatorname{base}}(\X))+f(\X)^2-f_{\operatorname{base}}(\X)^2 ]. 
\end{aligned}
\]
By the definition of $Q$, the target CATE is a weighted average of source CATEs, characterized by $\tau_Q(\cdot)=\sum_{s=1}^S q_s \cdot \tau^{(s)}(\cdot)$ for $\q\in\mathcal{H}$. 
Therefore,
Equation~\eqref{equ:HTE:obj:relativebase} can be written as: 
\begin{equation}\label{equ:pf:adv:originobj}
  \begin{aligned}
   f_{\operatorname{rel}}^*(\cdot;f_{\operatorname{base}}) & =   \underset{f \in \mathcal{F}}{\arg \min } \max _{Q \in \mathcal{C}\left(Q_{\X},\mathcal{H}\right)} R_Q(f)\\
   & = \underset{f \in \mathcal{F}}{\arg \min } \max _{\q \in \mathcal{H}} \ \sum_{s=1}^S q_s \cdot \mathbb{E}_{Q_{\X}}\left[-2\tau^{(s)}(\X)(f(\X)-f_{\operatorname{base}}(\X))+f(\X)^2-f_{\operatorname{base}}(\X)^2 \right].
\end{aligned}
\end{equation}
Suppose that we can swap the order of min and max, which will be justified later.
Then, the objective can be rewritten as,
\begin{equation}\label{equ:pf:adv:swapobj}
\underset{\q \in \mathcal{H}}{\arg \max}\    \underset{f \in \mathcal{F}}{\min }\ \sum_{s=1}^S q_s \cdot \mathbb{E}_{Q_{\X}}\left[-2\tau^{(s)}(\X)(f(\X)-f_{\operatorname{base}}(\X))+f(\X)^2-f_{\operatorname{base}}(\X)^2 \right].
  \end{equation}
For any given $\q\in\mathcal{H}$, the solution to the inner minimization is given by,
\begin{equation}\label{equ:pf:fq}
\begin{aligned}
     f^{\q}(\cdot) \ =  \ &   
      \underset{f \in \mathcal{F}}{\arg\min }\  \mathbb{E}_{Q_{\X}}\left[-2\sum_{s=1}^S q_s \cdot \tau^{(s)}(\X)f(\X)+f(\X)^2 \right] \\
     = \ &  \underset{f \in \mathcal{F}}{\arg\min }\  \mathbb{E}_{Q_{\X}}\left[f(\X)-\sum_{s=1}^S q_s \cdot \tau^{(s)}(\X) \right]^2\\
     = \ & \sum_{s=1}^S q_s \cdot \tau^{(s)}(\cdot),
\end{aligned}
\end{equation}
where the second equality follows because the function class $\mathcal{F}$ contains all the $\{\tau^{(s)}(\cdot)\}_{s=1}^S$ and is convex.
Thus, the solution to the optimization problem in Equation~\eqref{equ:pf:adv:swapobj} is given by:
\begin{equation*}
    \q^*= \underset{\q \in \mathcal{H}}{\arg \min } \ \mathbb{E}_{Q_{\X}}\left[\sum_{s=1}^S q_s \cdot \tau^{(s)}(\X) - f_{\operatorname{base}}(\X)\right]^2
\end{equation*}
and 
\begin{equation}\label{equ:pf:f*swap}
       f_{\operatorname{swap}}^*(\cdot)= \sum_{s=1}^S q^*_s \cdot \tau^{(s)}(\cdot).
\end{equation}

Next, we provide a justification for interchanging max and min by showing that $f_{\operatorname{swap}}^*(\cdot)$ is also the solution to the original objective given in Equation~\eqref{equ:pf:adv:originobj}.
For any $t\in[0,1]$ and $\q\in\mathcal{H}$, we use the fact that $\q^* + t(\q-\q^*) \in\mathcal{H}$ and by the definition of $\q^*$, we have
\begin{equation*}
\mathbb{E}_{Q_{\X}}\left[\sum_{s=1}^S (q^*_s+t(q_s-q^*_s)) \cdot \tau^{(s)}(\X)  - f_{\operatorname{base}}(\X)\right]^2 \geq \mathbb{E}_{Q_{\X}}\left[\sum_{s=1}^S q^*_s \cdot \tau^{(s)}(\X)  - f_{\operatorname{base}}(\X)\right]^2.
\end{equation*}
The above inequality further implies
\begin{equation*}
t^2\cdot \mathbb{E}_{Q_{\X}}\left[\sum_{s=1}^S (q_s-q^*_s)\tau^{(s)}(\X) \right]^2 +2t\cdot \mathbb{E}_{Q_{\X}}\left[ \left(\sum_{s=1}^S q^*_s \cdot \tau^{(s)}(\X)-f_{\operatorname{base}}(\X)\right) 
\left(\sum_{s=1}^S(q_s-q^*_s)\tau^{(s)}(\X)\right)\right] \geq 0 
\end{equation*}
for all $t\in[0,1]$. 

By taking $t\rightarrow 0+$, we have,
\begin{equation*}
  \mathbb{E}_{Q_{\X}}\left[ \left(\sum_{s=1}^S q^*_s \cdot \tau^{(s)}(\X)-f_{\operatorname{base}}(\X)\right) 
\left(\sum_{s=1}^S(q_s-q^*_s)\tau^{(s)}(\X)\right)\right]\geq0,
\end{equation*}
which is equivalent to, for any $\q\in\mathcal{H}$,
\begin{equation}\label{equ:pf:quadineq}
\begin{aligned}
     \mathbb{E}_{Q_{\X}}&\left[ \left(\sum_{s=1}^S q^*_s \cdot \tau^{(s)}(\X)-f_{\operatorname{base}}(\X)\right)  
\left(\sum_{s=1}^S q_s \cdot \tau^{(s)}(\X)\right)\right] \\
& \geq  \mathbb{E}_{Q_{\X}}\left[ \sum_{s=1}^S q^*_s\cdot \tau^{(s)}(\X)\right]^2- \mathbb{E}_{Q_{\X}}\left[ \sum_{s=1}^S q^*_s\tau^{(s)}(\X)\cdot f_{\operatorname{base}}(\X)\right].
\end{aligned}
\end{equation}

Given the original objective in Equation~\eqref{equ:pf:adv:originobj}, we have the following inequality,
\begin{equation}\label{equ:pf:1srdir}
\begin{aligned}
   & \underset{f \in \mathcal{F}}{\min } \max _{\q \in \mathcal{H}} \  \mathbb{E}_{Q_{\X}} \left[-2\sum_{s=1}^S q_s \cdot \tau^{(s)}(\X)(f(\X)-f_{\operatorname{base}}(\X))+f(\X)^2 -f_{\operatorname{base}}(\X)^2\right] \\
    \leq & \max _{\q \in \mathcal{H}} \ \mathbb{E}_{Q_{\X}}\left[-2\sum_{s=1}^S q_s \cdot \tau^{(s)}(\X)(f_{\operatorname{swap}}^*(\X)-f_{\operatorname{base}}(\X))+f_{\operatorname{swap}}^*(\X)^2 -f_{\operatorname{base}}(\X)^2\right] \\
   =& \max _{\q \in \mathcal{H}} \ \mathbb{E}_{Q_{\X}}\left[-2\left(\sum_{s=1}^S q_s \cdot \tau^{(s)}(\X)\right)\left(\sum_{s=1}^S q^*_s \cdot \tau^{(s)}(\X)-f_{\operatorname{base}}(\X)\right) 
+\left(\sum_{s=1}^S q^*_s \cdot \tau^{(s)}(\X)\right) ^2 -f_{\operatorname{base}}(\X)^2\right]\\
\leq & \ -\mathbb{E}_{Q_{\X}}\left[\sum_{s=1}^S q^*_s \cdot \tau^{(s)}(\X) - f_{\operatorname{base}}(\X)\right]^2,
\end{aligned}
\end{equation}
where the first inequality follows because $f^*_{\operatorname{swap}}\in\mathcal{F}$, and the last inequality holds due to Equation~\eqref{equ:pf:quadineq}. 
Moreover, since $\q^*\in\mathcal{H}$, we have, 
\begin{equation} \label{equ:pf:2nddir}
\begin{aligned}
     & \underset{f \in \mathcal{F}}{\min } \max _{\q \in \mathcal{H}} \ \mathbb{E}_{Q_{\X}} \left[-2\sum_{s=1}^S q_s \cdot \tau^{(s)}(\X)(f(\X)-f_{\operatorname{base}}(\X))+f(\X)^2 -f_{\operatorname{base}}(\X)^2\right] \\
      \geq  &\  \underset{f \in \mathcal{F}}{\min } \ \mathbb{E}_{Q_{\X}} \left[-2\sum_{s=1}^S q^*_s \cdot \tau^{(s)}(\X)(f(\X)-f_{\operatorname{base}}(\X))+f(\X)^2 -f_{\operatorname{base}}(\X)^2\right] \\
      \geq  & -\mathbb{E}_{Q_{\X}}\left[\sum_{s=1}^S q^*_s \cdot \tau^{(s)}(\X) - f_{\operatorname{base}}(\X)\right]^2,
\end{aligned}
\end{equation}
where the last inequality follows from the fact that by definition of $f^{\q}(\cdot) $ in Equation~\eqref{equ:pf:fq}, the minimizer of the right hand side of the above inequality is exactly $f_{\operatorname{swap}}^*(\cdot)$ defined in Equation~\eqref{equ:pf:f*swap}.

Combining Equations~\eqref{equ:pf:1srdir}~and~\eqref{equ:pf:2nddir}, we have shown that the original optimization problem achieves the optimal value $\mathbb{E}_{Q_{\X}}\left[\sum_{s=1}^S q^*_s \cdot \tau^{(s)}(\X) - f_{\operatorname{base}}(\X)\right]^2$.
Since $f_{\operatorname{swap}}^*(\cdot)\in\mathcal{F}$ and  $\q^*\in \mathcal{H}$, we have established that $f_{\operatorname{swap}}^*(\cdot)$ is the optimizer to the original objective.
\end{proof}

\subsection{Proof of Lemma~\ref{lemma:q:regret}}\label{app:lemma:proof}

\begin{proof}
By the definition of $\q^*$, for any $t \in(0,1)$, we have
$$
\left(\q^*\right)^{\top} \Gamma \q^* - (\q^*)^{\top} \d \leq\left[\q^*+t\left(\hat{\q}-\q^*\right)\right]^{\top} \Gamma\left[\q^*+t\left(\hat{\q}-\q^*\right)\right]  - (\q^*+t\left(\hat{\q}-\q^*\right))^{\top} \d  .
$$
By taking $t \rightarrow 0+$, we have
\begin{equation}\label{equ:lemma2:equ1}
2\left(\q^*\right)^{\top} \Gamma\left(\hat{\q}-\q^*\right) - \left(\hat{\q}-\q^*\right)^{\top} \d  \geq 0 .
\end{equation}
Similarly, by the definition of $\hat{\q}$, for any $t \in(0,1)$, we have
$$
(\hat{\q})^\top \widehat{\Gamma} \hat{\q} - (\hat{\q})^{\top} \hat{\d}  \leq\left[\hat{\q}+t\left(\q^*-\hat{\q}\right)\right]^{\top} \widehat{\Gamma}\left[\hat{\q}+t\left(\q^*-\hat{\q}\right)\right]  - (\hat{\q} +t\left(\q^*-\hat{\q}\right))^{\top} \hat{\d} 
$$
and therefore, 
\[
t^2\left(\q^*-\hat{\q}\right)^{\top}\widehat{\Gamma}\left(\q^*-\hat{\q}\right) + 2t\left(\q^*-\hat{\q}\right)^{\top} \widehat{\Gamma}\hat{\q} -t\left(\q^*-\hat{\q}\right)^{\top} \hat{\d}  \geq 0.
\]
Dividing by $t>0$ results in the following inequality:
$$
2\left(\q^*\right)^{\top} \widehat{\Gamma}\left(\q^*-\hat{\q}\right)+(t-2)\left(\q^*-\hat{\q}\right) \widehat{\Gamma}\left(\q^*-\hat{\q}\right)- \left(\q^*-\hat{\q}\right)^{\top}\hat{\d}   \geq 0 .
$$
Since $2-t>0$, we have
\begin{equation}\label{equ:lemma2:equ2}
\left(\q^*-\hat{\q}\right)^{\top} \widehat{\Gamma}\left(\q^*-\hat{\q}\right)  \leq \frac{1}{2-t} \left[ 2\left(\q^*\right)^{\top} \widehat{\Gamma}\left(\q^*-\hat{\q}\right) -\left(\q^*-\hat{\q}\right)^{\top} \hat{\d} \right].
\end{equation}
Furthermore, we have,
$$
\begin{aligned}
   & 2 \left(\q^*\right)^{\top} \widehat{\Gamma}\left(\q^*-\hat{\q}\right)- \left(\q^*-\hat{\q}\right)^{\top} \hat{\d} \\
   = \ & 2\left(\q^*\right)^{\top} (\widehat{\Gamma}-\Gamma)\left(\q^*-\hat{\q}\right)+2\left(\q^*\right)^{\top} \Gamma\left(\q^*-\hat{\q}\right) - \left(\q^*-\hat{\q}\right)^{\top} \d- \left(\q^*-\hat{\q}\right)^{\top} (\hat{\d}-\d) \\
    \leq\ & 2\left(\q^*\right)^{\top} (\widehat{\Gamma}-\Gamma)\left(\q^*-\hat{\q}\right)- \left(\q^*-\hat{\q}\right)^{\top} (\hat{\d}-\d) .
\end{aligned}
$$
where the inequality follows from Equation~\eqref{equ:lemma2:equ1}.
Combining this with Equation~\eqref{equ:lemma2:equ2}, we obtain that 
\begin{equation}
    \left(\q^*-\hat{\q}\right)^{\top}  \widehat{\Gamma}\left(\q^*-\hat{\q}\right) \leq 
    \frac{1}{2-t}\left[2\left(\q^*\right)^{\top} (\widehat{\Gamma}-\Gamma)\left(\q^*-\hat{\q}\right)- \left(\q^*-\hat{\q}\right)^{\top} (\hat{\d}-\d)\right].
\end{equation}
Since the $\hat{\q}$ and $\q^*$ are symmetric, we switch the roles of $\{\widehat{\Gamma}, \hat{\d}, \hat{\q}\},\{\Gamma, \d, \q^*\}$ and establish:
\begin{equation}
    \begin{aligned}
    \left(\q^*-\hat{\q}\right)^{\top} {\Gamma}\left(\q^*-\hat{\q}\right) \leq & \frac{1}{2-t}\left[ 2\left(\hat{\q}\right)^{\top} ({\Gamma}-\widehat{\Gamma})\left(\hat{\q}-\q^*\right)- \left(\hat{\q}-\q^*\right)^{\top} (\d-\hat{\d})\right] \\
    \leq &  \frac{1}{2-t}\left[2\|\hat{\q}\|_2\|\widehat{\Gamma}-\Gamma\|_2\left\|\q^*-\hat{\q}\right\|_2 + \left\|\q^*-\hat{\q}\right\|_2 \|\d-\hat{\d}\|_2\right].
    \end{aligned}
\end{equation}
Since $ \lambda_{\min }(\Gamma)\left\|\q^*-\hat{\q}\right\|_2^2\leq  \left(\q^*-\hat{\q}\right)^{\top} {\Gamma}\left(\q^*-\hat{\q}\right) $, we obtain,
$$
\lambda_{\min }(\Gamma)\left\|\q^*-\hat{\q}\right\|_2^2 \leq \frac{1}{2-t}\left[2\|\hat{\q}\|_2\|\widehat{\Gamma}-\Gamma\|_2\left\|\q^*-\hat{\q}\right\|_2 + \left\|\q^*-\hat{\q}\right\|_2 \|\d-\hat{\d}\|_2\right],
$$
where $\|\widehat{\Gamma}-\Gamma\|_2$ denotes the spectral norm of the matrix, $\widehat{\Gamma}-\Gamma$.

Finally, if $\lambda_{\min }(\Gamma) > 0$, we can establish the result by letting $t \to 0+$:
$$
\left\|\q^*-\hat{\q}\right\|_2 \leq \frac{1}{\lambda_{\min }(\Gamma)} \left[ \|\widehat{\Gamma}-\Gamma\|_2 + \frac{1}{2} \|\d-\hat{\d}\|_2\right] 
 \leq \frac{S (\|\widehat{\Gamma}-\Gamma\|_{\infty}+\frac{1}{2}\|\hat{\d}-\d\|_{\infty} )}{\lambda_{\min }(\Gamma)},
$$
where the last inequality is due to the fact that $\|\widehat{\Gamma}-\Gamma\|_2\leq\|\widehat{\Gamma}-\Gamma\|_F\leq S\|\widehat{\Gamma}-\Gamma\|_{\infty}$.
\end{proof}

\section{Additional Tables and Figures}

\subsection{Additional Simulation Results}\label{append:add:sim}

\begin{figure}[H]
    \centering
    \includegraphics[width=1\linewidth]{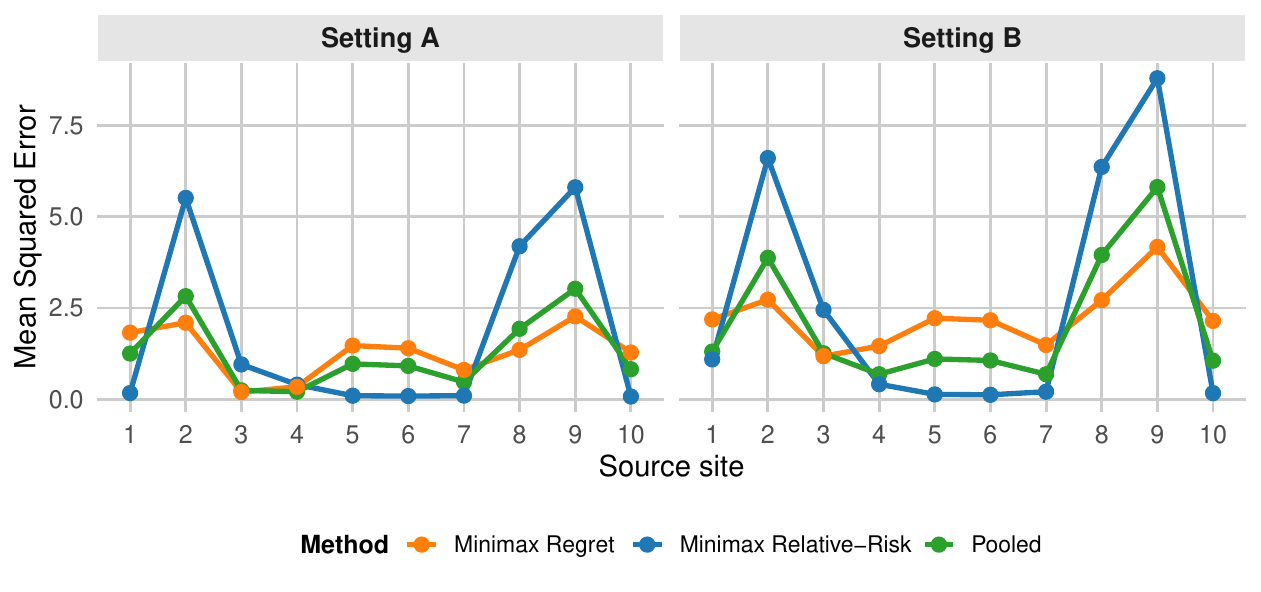} 
       \caption{Average MSE of multisite CATE estimates from three methods across 1,000 simulations, evaluated over different source sites, using {\bf X-learner} as the site-specific CATE estimation method.}
\end{figure}

\begin{figure}[H]
    \centering
    \includegraphics[width=1\linewidth]{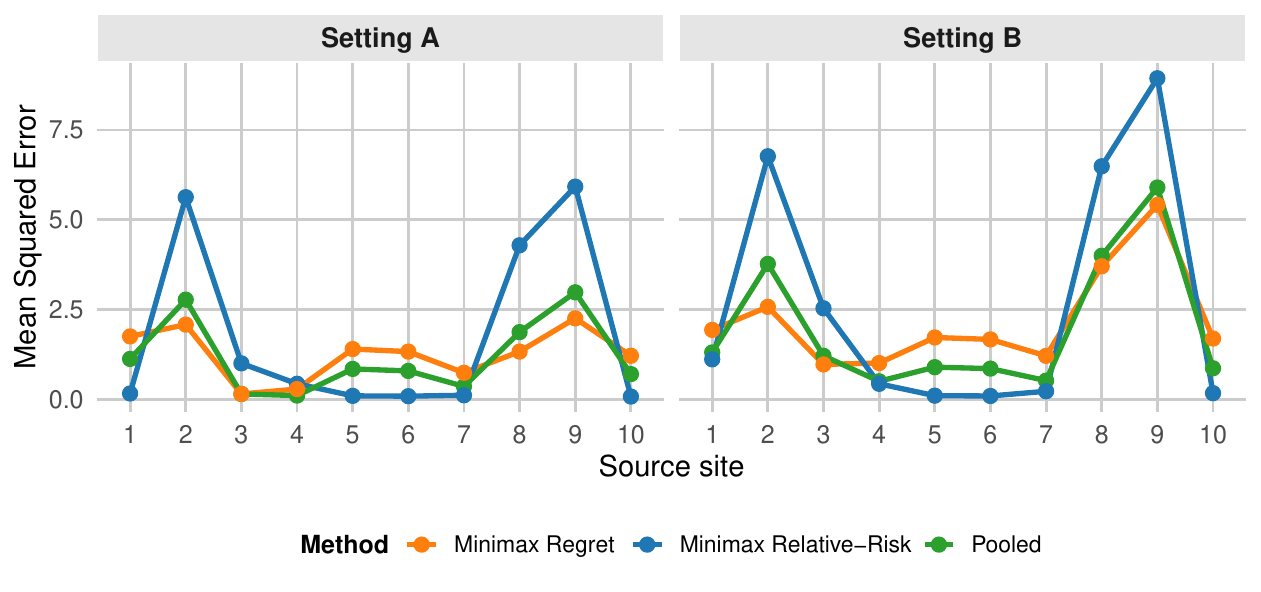} 
    \caption{Average MSE of multisite CATE estimates from three methods across 1,000 simulations, evaluated over different source sites, using {\bf causal forest} as the site-specific CATE estimation method.}
\end{figure}





\begin{figure}[H]
    \centering
    \includegraphics[width=1\linewidth]{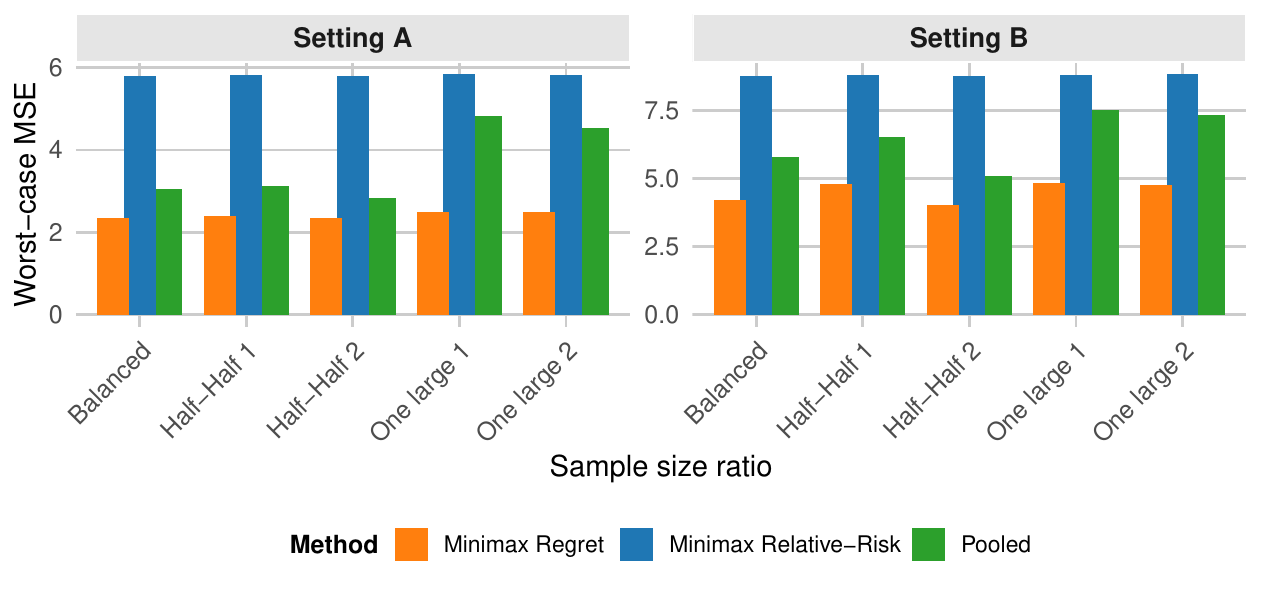}
   \caption{Worst-case MSE averaged across 1,000 iterations under varying sample size ratios: (1) {\bf Balanced}: all sites have equal sample sizes, (2) {\bf Half-and-Half}: half of the sites have three times the sample size of the others, and (3) {\bf One Large}: one site has a sample size 10 times larger than the others. For scenarios (2) and (3), we evaluate two configurations, each with a different subset of large sites. We use {\bf X-learner} for site-specific CATE estimation.}
\end{figure}

\begin{figure}[H]
    \centering
    \includegraphics[width=1\linewidth]{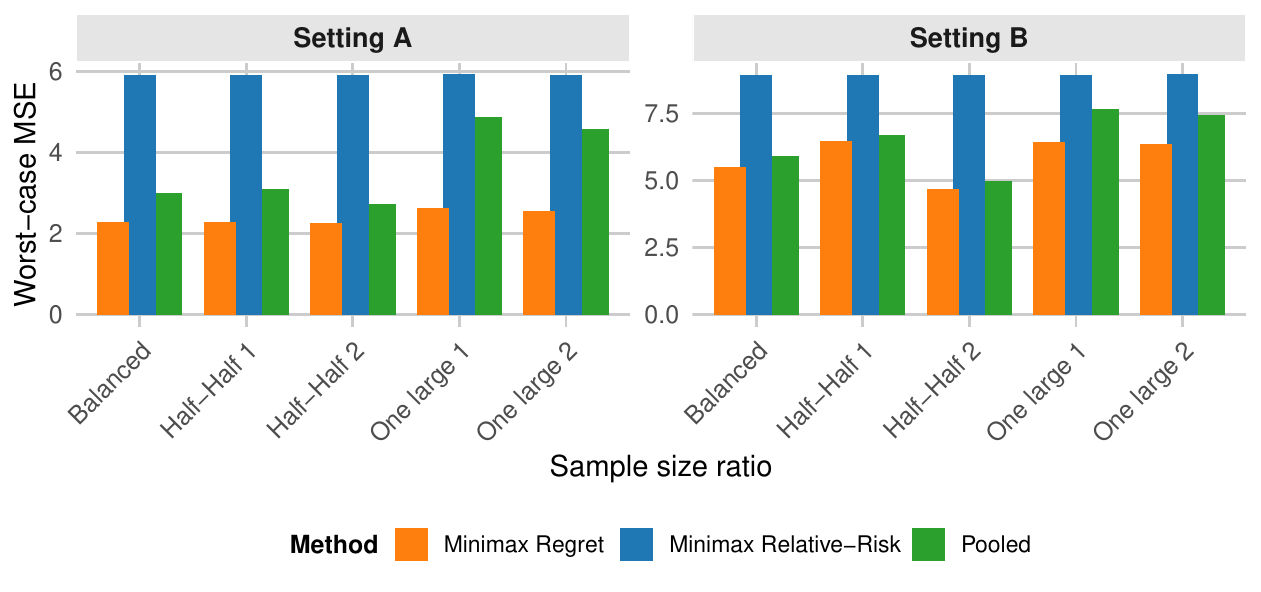}
   \caption{Worst-case MSE averaged across 1,000 iterations under varying sample size ratios: (1) {\bf Balanced}: all sites have equal sample sizes, (2) {\bf Half-and-Half}: half of the sites have three times the sample size of the others, and (3) {\bf One Large}: one site has a sample size 10 times larger than the others. For scenarios (2) and (3), we evaluate two configurations, each with a different subset of large sites. We use {\bf causal forest} for site-specific CATE estimation.}
\end{figure}

\subsection{Additional simulation under covariate shift}
\label{appendix:sim:covshift}

To examine the robustness of the proposed approach under covariate shift, we consider an additional simulation variant based on the same data-generating mechanism as in the main text, but with a shifted target covariate distribution. Specifically, we keep the source-site data-generating process unchanged, while generating the target covariate distribution
\[
\X_i^Q \overset{\mathrm{i.i.d.}}{\sim} \mathcal{N}(\bmu_Q, \mathbf{I}_5),
\qquad
\bmu_Q=(1,-1,0.5,0,0)^\top.\]
We use the same treatment assignment mechanism, baseline outcome model, and site-specific CATE specifications as in Settings~A and~B of Section~\ref{sec:simulation}. This induces a mean shift in the target covariate distribution while preserving overlap between the target and each source site. 

We report results using the same single-site learners as in the main text and evaluate performance using the mean squared error between the estimated CATE model and the oracle target CATE over the shifted target covariate sample.

\begin{figure}[t]
    \centering
    \includegraphics[width=1\linewidth]{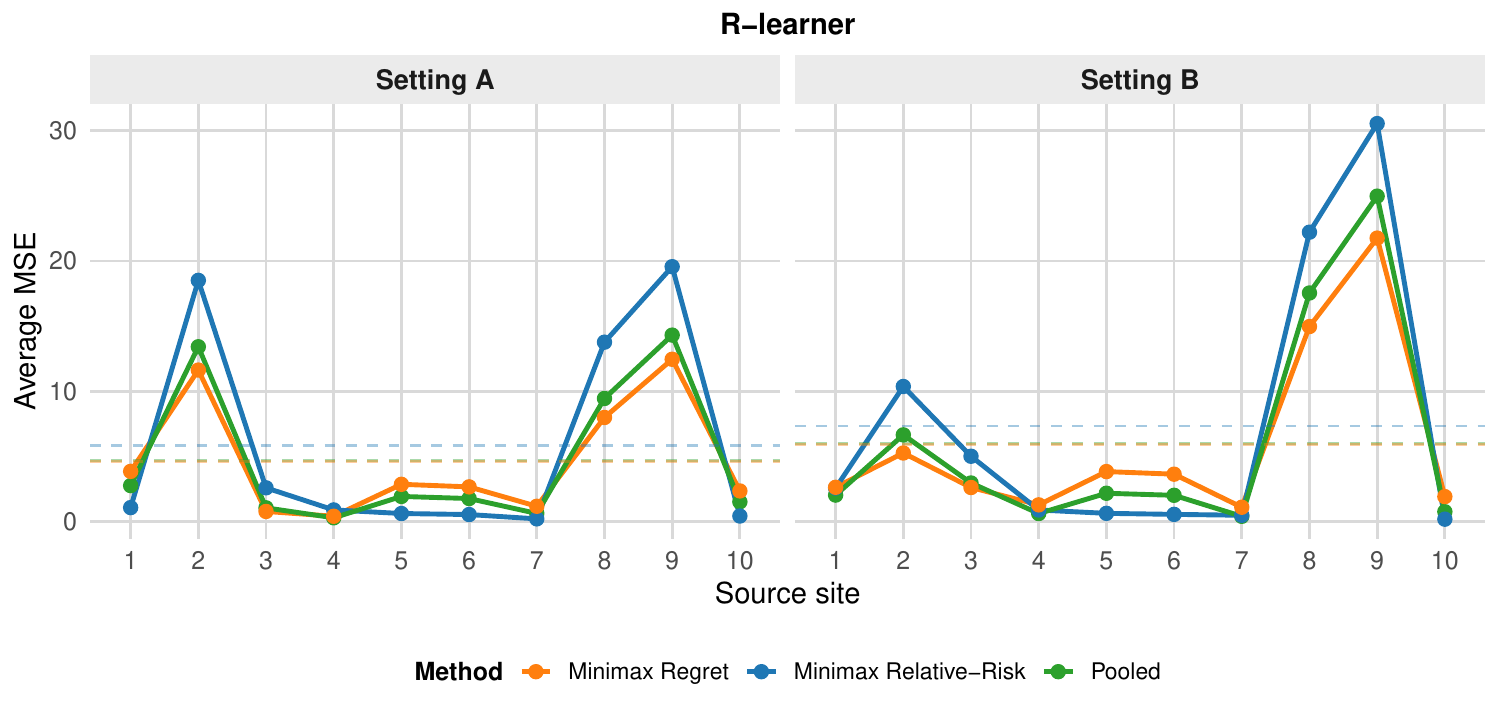}
    \caption{Average MSE of multisite CATE estimates from three methods across 1,000 simulations under covariate shift, evaluated over different source sites, using R-learner as the site-specific CATE estimation method. Dashed horizontal lines indicate the corresponding average MSE across source sites for each method.}
    \label{fig:sim:covshift:R}
\end{figure}

Figure~\ref{fig:sim:covshift:R} reports the site-by-site average MSE under covariate shift using the R-learner as the site-specific CATE estimation method, for both Settings~A and~B. Dashed horizontal lines indicate the corresponding average MSE across source sites for each method.

\begin{figure}[t]
    \centering
    \includegraphics[width=1\linewidth]{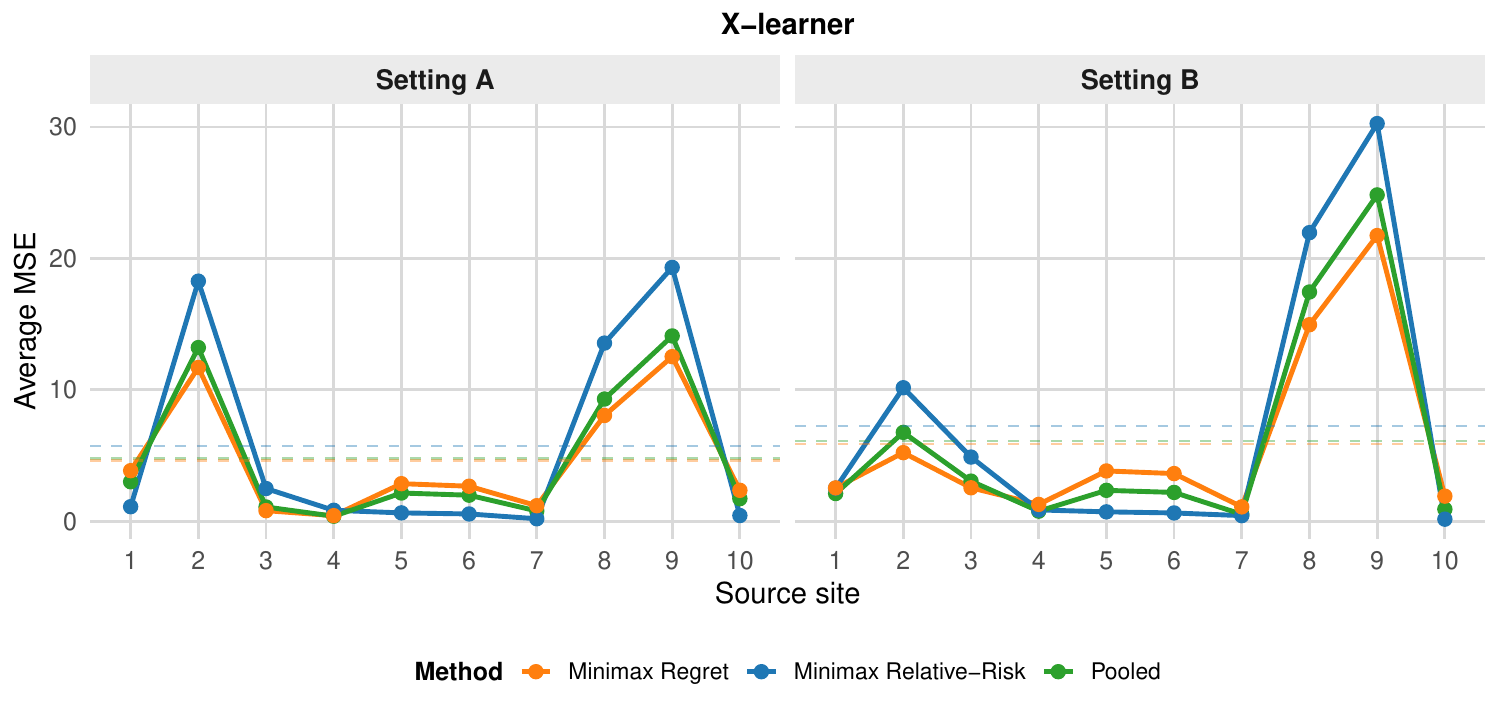}
    \caption{Average MSE of multisite CATE estimates from three methods across 1,000 simulations under covariate shift, evaluated over different source sites, using X-learner as the site-specific CATE estimation method. Dashed horizontal lines indicate the corresponding average MSE across source sites for each method.}
    \label{fig:sim:covshift:X}
\end{figure}

\begin{figure}[t]
    \centering
    \includegraphics[width=1\linewidth]{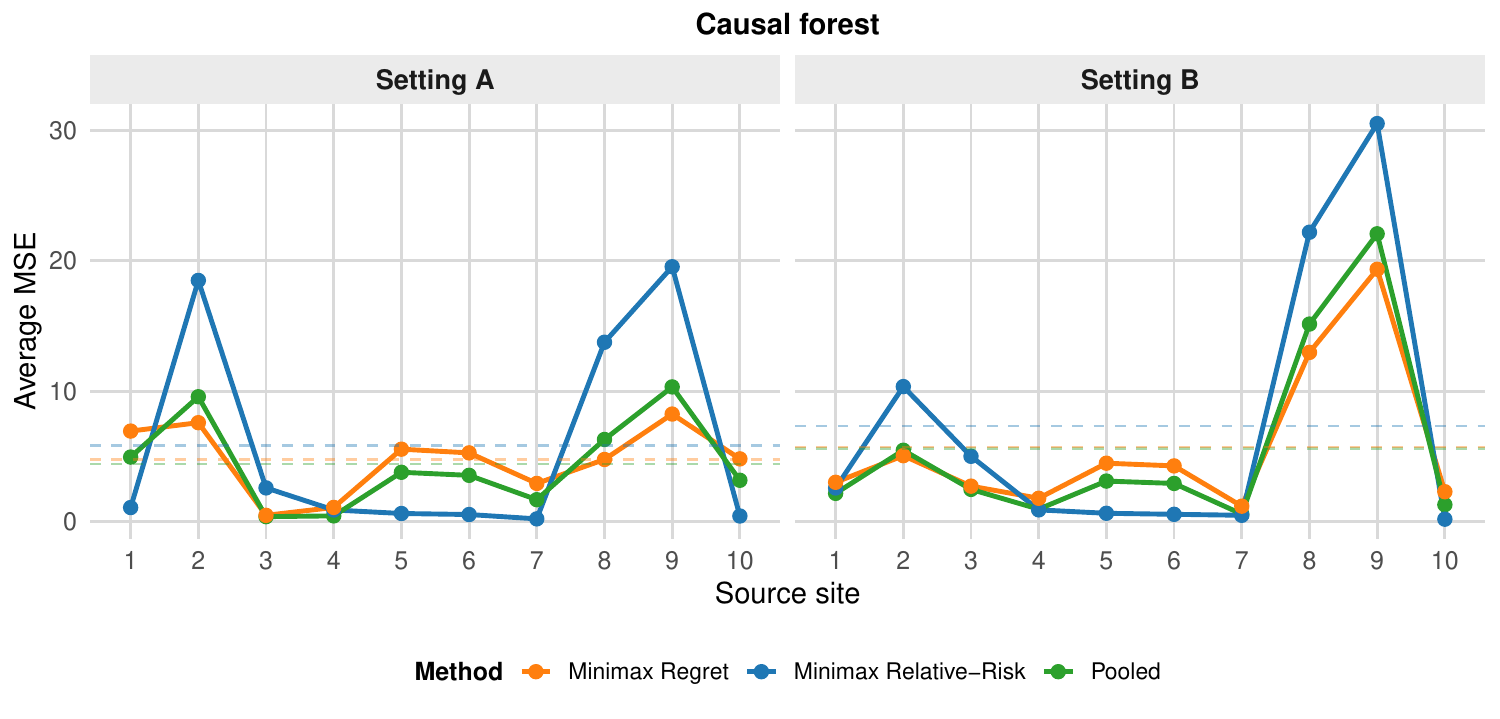}
    \caption{Average MSE of multisite CATE estimates from three methods across 1,000 simulations under covariate shift, evaluated over different source sites, using causal forest as the site-specific CATE estimation method. Dashed horizontal lines indicate the corresponding average MSE across source sites for each method.}
    \label{fig:sim:covshift:forest}
\end{figure}

For completeness, we also report the corresponding results using the X-learner and causal forest in Figures~\ref{fig:sim:covshift:X} and~\ref{fig:sim:covshift:forest}. The overall qualitative conclusions remain similar to those in the baseline simulations: the minimax regret estimator continues to provide the strongest worst-case robustness, while the pooled estimator remains more sensitive to distributional mismatch between the source and target populations.

\subsection{Additional Empirical Application Results}\label{append:add:realapp}

\begin{landscape} 
\small 
\begin{table} 
\begin{tabular}{p{0.2\textwidth}p{0.25\textwidth}p{0.25\textwidth}p{0.25\textwidth}p{0.25\textwidth}}\toprule 
Study & Bosnia \& Herzegovina & Mexico & Mongolia & Morocco\\
& \citet{augsburg2015impacts} & \citet{angelucci2015microcredit} & \citet{attanasio2015impacts} & \citet{crepon2015estimating}\\\midrule 
Treatment & Provide loans to marginally rejected borrowers & Open branches, promote loans & Open branches, target likely borrowers & Open branches \\ 
Treatment Prop.  & 0.55 & 0.37 & 0.73 & 0.49 \\ 
Randomization Level & Individual & Community & Community & Community \\ 
Urban/Rural & Both & Both & Rural & Rural \\ 
Target Women? & No & Yes & Yes & No\\
MFI already operates locally? & Yes & No & No & No \\ 
Microloan Liability Type & Individual & Group & Both & Group \\ 
Collateralized & Yes & No & Yes & No \\ 
Other competing MFI's & Yes & Yes & Yes & No \\ 
Avg. interest rate & 22\% APR & 100\% APR & 24\$ APR & 13.5\% APR \\ 
Sampling Frame & Marignal applicants & Women between 18-60 who own businesses (or wish to start one) & Women who registered interest in loans and met eligibility criteria & Random sample with likely borrowers \\ 
Study Duration & 14 months & 16 months & 19 months & 24 months \\ 
\underline{\textbf{Demographic Covariates}}\\ 
  Female & 0.41 & 1.00 & 1.00 & 0.15 \\ 
  Age & 37.78 & 37.37 & 39.96 & 47.50 \\ 
  Baseline Profit & 3321.91 & -16.51 & 19.98 & 341.39 \\ 
  Baseline Expenditures & 1167.80 & 1104.16 & 53.00 & 470.88 \\ 
  Baseline Loan Amount  & 4049.07 & 276.63 & 0.00 & 234.38 \\ 
  Existing Business & 0.62 & 0.48 & 0.60 & 0.12 \\ \bottomrule 
\end{tabular} 
\caption{Summary of the different sites. Akin to \citet{meager2019understanding}, we emphasize differences across the experiments in the sampling frame, as well as treatment implementation.}
\label{tbl:microcredit_summary}
\end{table} 
\end{landscape} 
\begin{figure}[H]
\centering 
\textbf{(a) Scenario 3}\\ 
    \includegraphics[width=\textwidth]{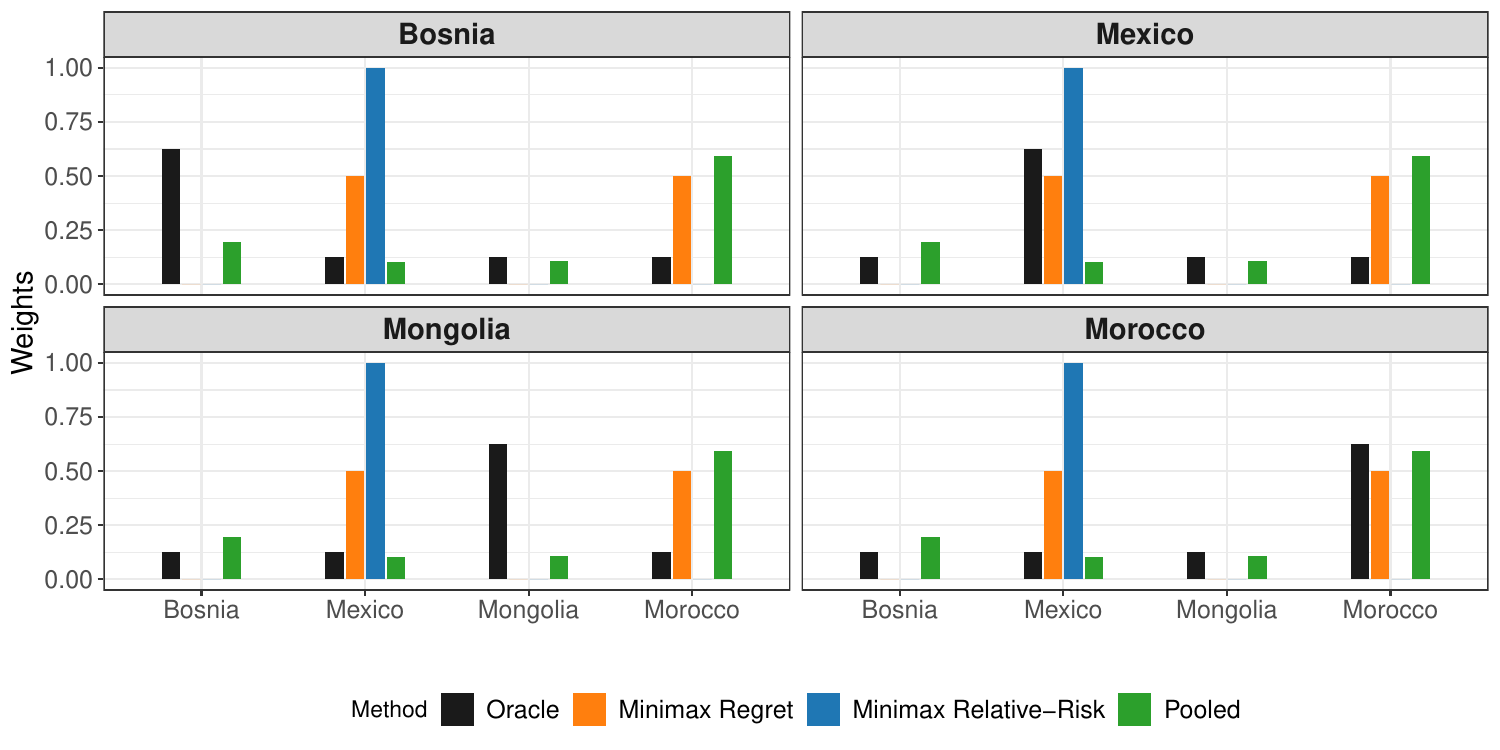}\\
    \textbf{(b) Scenario 4}\\ 
        \includegraphics[width=\textwidth]{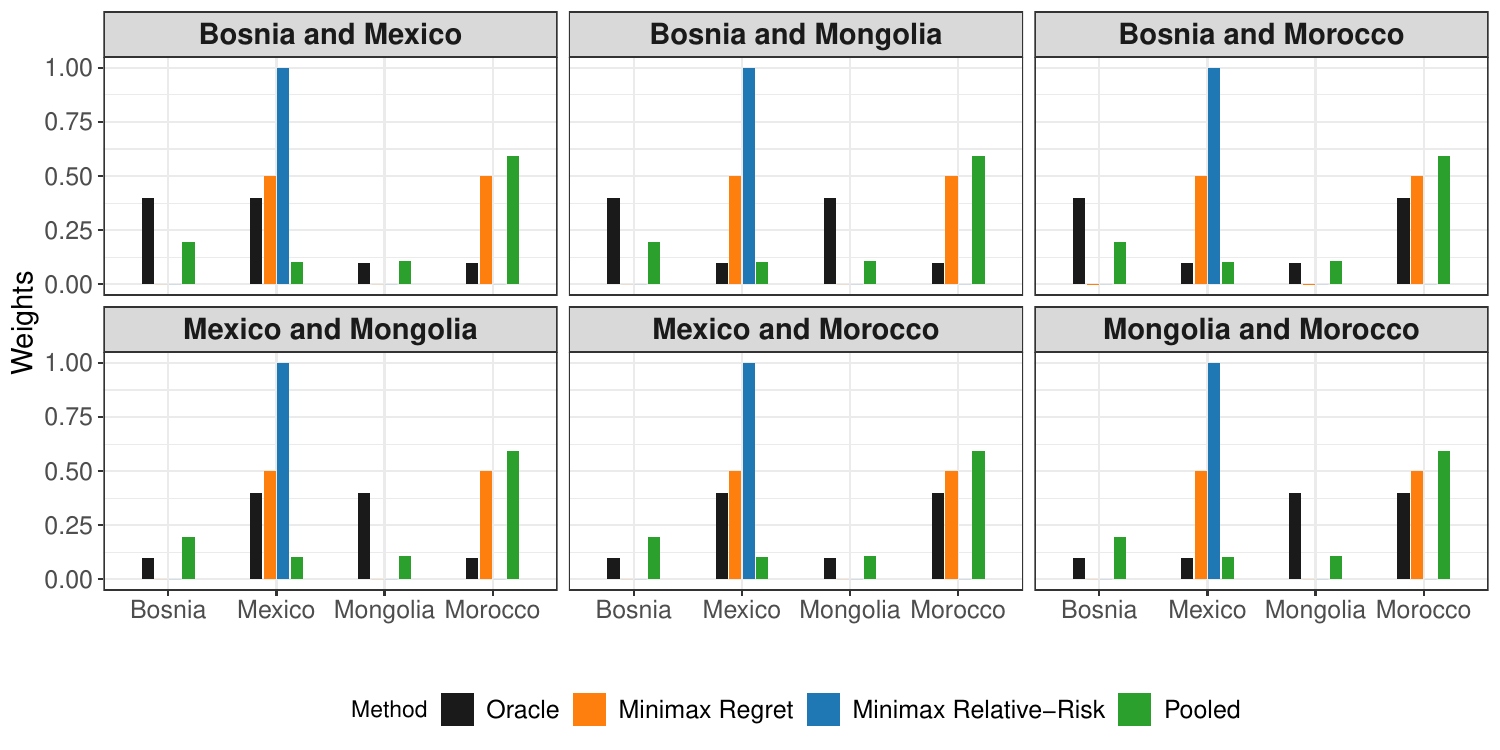}
    \caption{Visualization of the weights estimated under each approach for (a) Scenario 3 and (b) Scenario 4. For the pooled approach, the estimated weights remain fixed and match the proportions implied by the source distributions. In other words, the pooled distribution is unable to adapt to the target population, even when the target distribution is very different from the source distributions. The weights for the minimax relative-risk approach consistently assigns a weight of 1 to Mongolia, and as such, considers a very conservative, worst-case scenario. The minimax regret approach assigns a weight of 0.5 to Mongolia, and 0.5 to Morocco.}
    \label{fig:scenario_weights}
\end{figure}

\begin{figure} 
\centering 
\textbf{(a) Bosnia} \\
\includegraphics[width=0.8\textwidth]{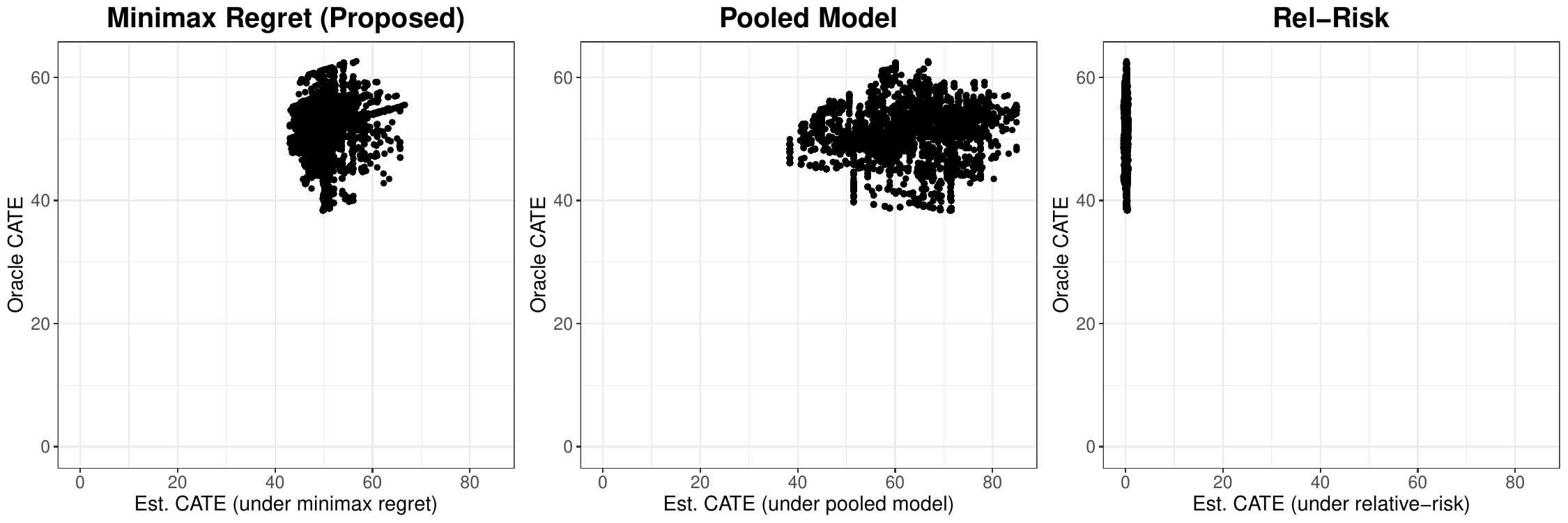}\\ 
\textbf{(b) Mexico} \\ 
\includegraphics[width=0.8\textwidth]{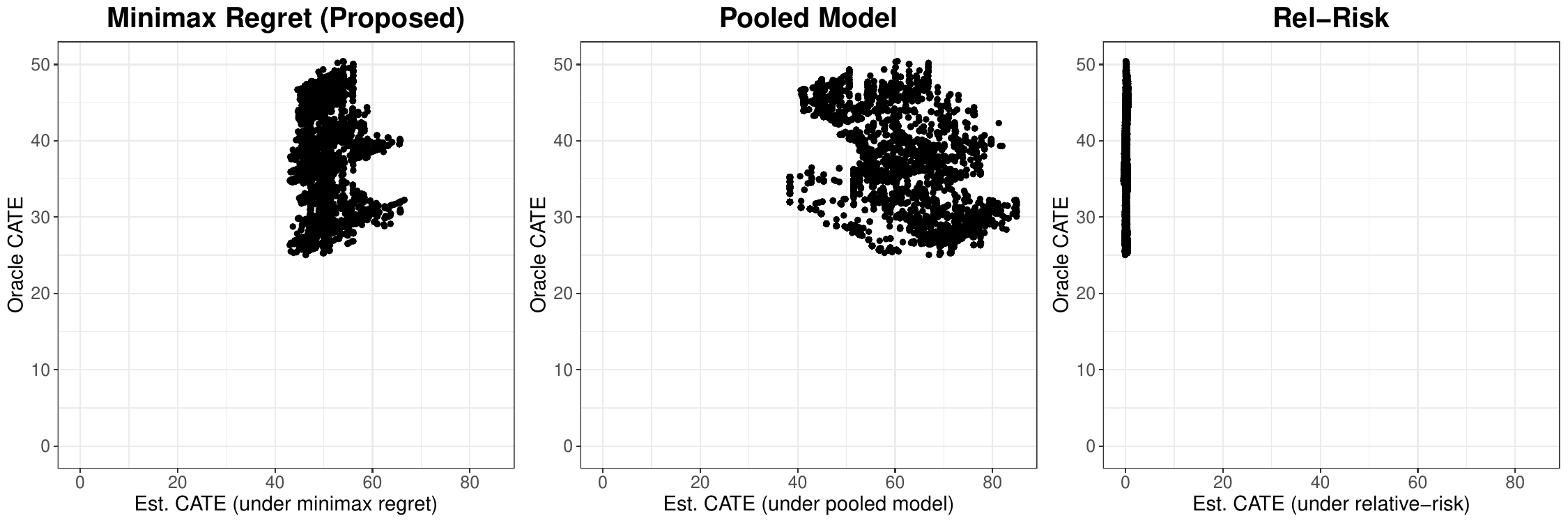}\\
\textbf{(c) Mongolia} \\ 
\includegraphics[width=0.8\textwidth]{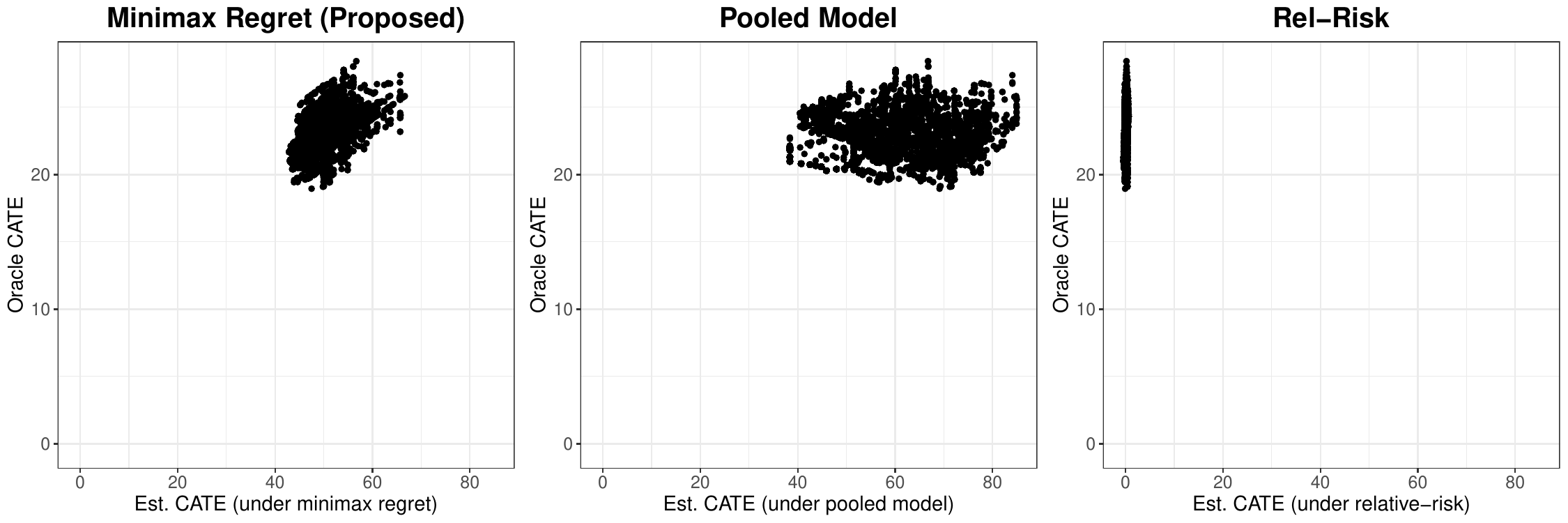}\\
\textbf{(d) Morocco} \\ 
\includegraphics[width=0.8\textwidth]{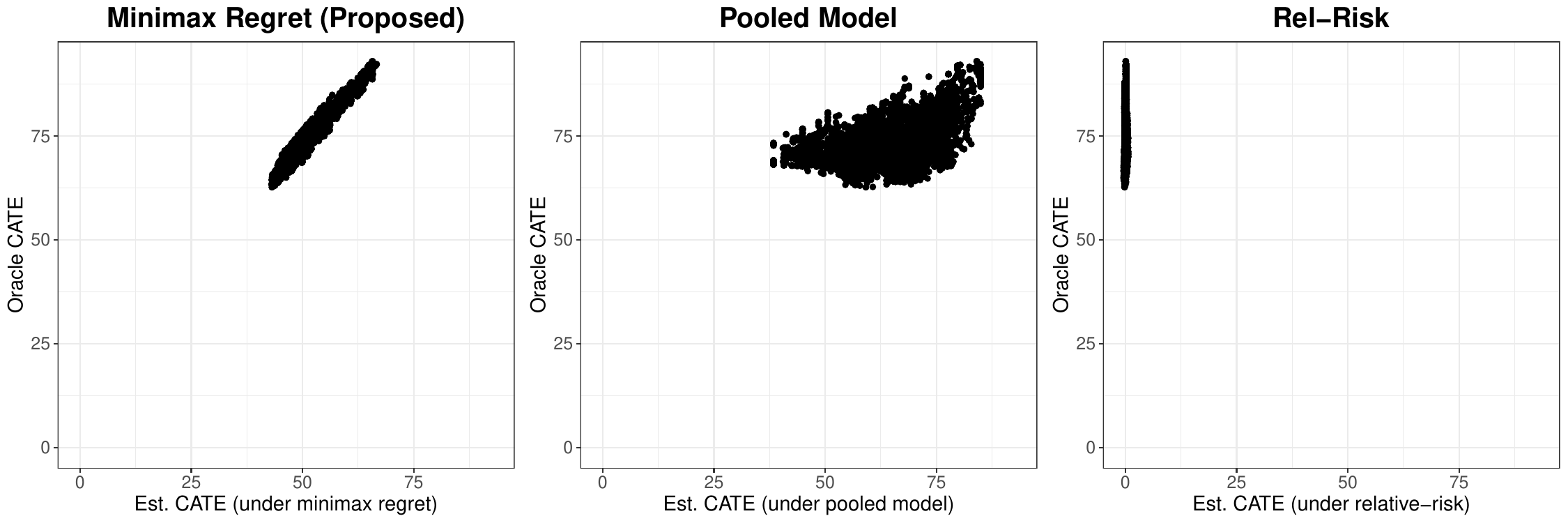}
\caption{Comparison of CATE estimates across the different methods, in Scenario 3. In general, we see that the relative-risk approach results in CATE estimates that are almost always close to zero. In contrast, the minimax regret approach results in CATE estimates that more closely proxy the true, oracle CATE.}
\end{figure}

\end{document}